\documentclass[preprint,12pt,sort&compress]{elsarticle}
\usepackage{graphicx} % Required for inserting images

\usepackage{amsmath}
\usepackage{mathtools}
\usepackage{amssymb}
\usepackage[mathscr]{eucal}
\usepackage{amscd}
\usepackage{color}
\usepackage{verbatim}
\usepackage{sistyle}
\usepackage{subfigure}
\usepackage{makecell}
\usepackage[usenames,dvipsnames]{xcolor}
\usepackage{tikz-cd}
\usepackage{algorithm}
\usepackage{xfrac}
\usepackage[permil]{overpic}
\usepackage{url}

\DeclareMathOperator{\SO3}{{{\rm SO(3)}}} 
\DeclareMathOperator{\so3}{{{\rm so(3)}}}
\DeclareMathOperator{\id}{{{id}}}
\newcommand{\f}[1]{{\boldsymbol{#1}}}
\newcommand{\baf}[1]{{\bar{\boldsymbol{#1}}}}

\newcommand{\sub}{\subset}
\newcommand{\bEq}{\begin{equation}}
\newcommand{\eEq}{\end{equation}}
\newcommand{\beq}{\begin{equation*}}
\newcommand{\eeq}{\end{equation*}}

\newcommand{\car}{\times}
\newcommand{\mto}{\mapsto}

\newcommand{\byd}{\,{\raisebox{.092ex}{\rm :}{\rm =}}\,}

\newcommand{\fR}[1]{{\mathbf{#1}}}
\newcommand{\hfR}[1]{\hat{\mathbf{#1}}}

\newcommand{\sepr}[1]{\,\,\,\,\textnormal{{#1}}\,\,\,\,}

\newcommand{\lf}{\left}
\newcommand{\rg}{\right}

\newcommand{\sS}[1]{{\scriptscriptstyle {#1}}}
\newcommand{\Tra}{^{\mathsf{\sS\!T}}}
\newcommand{\Rn}{\text{I\!R}}
\newcommand{\tif}[1]{{\widetilde{\boldsymbol{#1}}}}
\newcommand{\htif}[1]{\hat{{\widetilde{\boldsymbol{#1}}}}}

\newcommand{\bAl}{\begin{align}}

\newcommand{\Gam}{\varGamma}
\newcommand{\B}[1]{{\mathbb{#1}}}
\newcommand{\veps}{\varepsilon}

\newcommand{\del}{\delta}
\newcommand{\Del}{\Delta}
\newcommand{\vtht}{\vartheta}
\newcommand{\vTht}{\varTheta}

\newcommand{\fr}[2]{\frac{#1}{#2}\,}

\newcommand{\wti}[1]{{\widetilde{#1}}}

\newcommand{\haf}[1]{{\hat{\boldsymbol{#1}}}}

\newcommand{\chf}[1]{{\check{\f{#1}}}}
\newcommand{\alp}{\alpha}

\newcommand{\bdg}{\beq\begin{diagram}}
\newcommand{\edg}{\end{diagram}\eeq}

\newcommand{\CNinf}{\B C_{N\infty}}
\newcommand{\CMinf}{\B C_{M\infty}}
\newcommand{\CNalp}{\B C_{N\alp}}
\newcommand{\CMalp}{\B C_{M\alp}}
\newcommand{\CNz}{\B C_{N0}}
\newcommand{\CMz}{\B C_{M0}}
\newcommand{\CNb}{\bar{\B C}_{N}}
\newcommand{\CMb}{\bar{\B C}_{M}}

\newcommand{\nm}[1]{\textnormal{#1}}
\newcommand{\dde}{\frac{d}{d\veps}}

\definecolor{blu}{RGB}{0, 114, 189}
\definecolor{aran}{RGB}{217, 83, 25}
\usepackage{amssymb}
\usepackage{amsthm}

\newproof{prf}{Proof}[section]
\newproof{rmk}{Remark}[section]

\usepackage{lineno}
\usepackage[top=1in,bottom=1in,left=1in,right=1in]{geometry}

\begin{document}
\begin{frontmatter}

\title{Simulating programmable morphing of shape memory polymer beam systems with complex geometry and topology}

\author[fi]{Giulio Ferri}
\author[fi]{Enzo Marino\corref{cor1}}
\ead{enzo.marino@unifi.it}
\cortext[cor1]{Corresponding author}

\address[fi]{Department of Civil and Environmental Engineering, University of Florence\\ Via di S. Marta 3, 50139 Firenze, Italy}

\begin{abstract}
We propose a novel approach to the analysis of programmable geometrically exact shear deformable beam systems made of shape memory polymers. The proposed method combines the viscoelastic Generalized Maxwell model with the Williams, Landel and Ferry relaxation principle, enabling the reproduction of the shape memory effect of structural systems featuring complex geometry and topology. 
Very high efficiency is pursued by discretizing the differential problem in space through the isogeometric collocation (IGA-C) method. The method, in addition to the desirable attributes of isogeometric analysis (IGA), such as exactness of the geometric reconstruction of complex shapes and high-order accuracy, circumvents the need for numerical integration since it discretizes the problem in the strong form. Other distinguishing features of the proposed formulation are: 
i) $\SO3$-consistency for the linearization of the problem and for the time stepping; 
ii) minimal (finite) rotation parametrization, that means only three rotational unknowns are used; 
iii) no additional unknowns are needed to account for the rate-dependent material compared to the purely elastic case. 
Through different numerical applications involving challenging initial geometries, we show that the proposed formulation possesses all the sought attributes in terms of programmability of complex systems, geometric flexibility, and high order accuracy.
\end{abstract}


\end{frontmatter}

\section{Introduction}
In the era of smart materials and structures \cite{Kuang_etal2019,VanManen_2021,Wan_etal2024}, whose development has been dramatically boosted by the advent of additive manufacturing \cite{Subasha&Kandasubramanian2020,Sajjad_etal2024}, Shape Memory Polymers (SMPs) \cite{Behl&Lendlein2007} play a crucial role due to their potential to revolutionize a wide range of sectors. Among them, the personalized medicine may take enormous advantages of the potentialities to design patient-tailored devices \cite{Gomez_etal2014,Arif_etal2022,Fattah-alhosseini_etal2024}.

Compared to other classes of shape memory materials, such as alloys, SMPs exhibit several appealing features, such as low cost, light weight, high recoverable deformations and, especially for biomedical applications, biocompatibility and biodegradability \cite{Subasha&Kandasubramanian2020,Sajjad_etal2024,Chen_etal2024}.

SMPs have the ability to change shape, shifting from a temporary configuration to a permanent one, upon the effect of a specific stimulus. Typical stimuli are temperature gradients, magnetic fields, light radiations and humidity. In the case of thermo-responsive SMPs, the shape changing process is governed by an appropriate thermal cycle. 
The process starts with forming the object, e.g. via 3D printing, in its initial (also referred to as permanent) shape. Then, after heating it up above the glass transition temperature, $\nm T_G$, through the application of external conditions (loads or displacements), a predetermined temporary configuration is obtained. 
At this stage, the material is in a rubbery state, its stiffness is significantly reduced allowing large inelastic deformations. Lowering the temperature below $\nm{T}_G$, keeping active the external conditions, allows to ``freeze'' the deformations in the temporary shape, which is spontaneously maintained once the external conditions are removed. Once the material is heated again, raising the temperature above $\nm{T}_G$, the object recovers the permanent shape without any other external condition~\cite{Yiping_etal2006}. 

To reproduce this shape memory effect, phase transition or viscoelasticity-based models can be employed. The former are typically more suitable for semi-crystalline polymers, the latter for amorphous polymers, such as resins and PLA \cite{Nguyen_2010}. 
In phase transition models, the shape memory effect is described at the macroscopic level \cite{Lendlein_2002,Boatti_2016,Li_2023}, where the mechanical properties of the material are governed by the transition of a volume fraction of the material from glassy to rubbery phases when the temperature increases above $\nm{T}_G$. 
One of the first internal state variable constitutive model, suitable for epoxy resins at small strains, was derived in~\cite{Liu_2006}. Phase transition macroscopic models at finite strains were developed in \cite{Scalet_2015} and in \cite{Reese_2010,Boatti_2016} with applications on biomedical devices. 
Other nonlinear phase transition models have been proposed combining different rheological models and free energy expressions \cite{Chen_2008a,Chen_2008b,Xue_2022,Wang_2023}. Three-dimensional small and finite strains models with applications to cardiovascular stents have been proposed~\cite{Baghani_2012,Baghani_2014}.

On the other hand, viscoelasticity-based models assume that the shape memory effect occurs at the micro scale due to changes in the mobility of the polymer chains induced by temperature variations, which result in a modification on the relaxation properties of the material \cite{Lendlein_2002,Boatti_2016,Li_2023}. 
One of the first models of this kind assumed both viscous and elastic material properties as exponential functions of the temperature \cite{Tobushi_1997}.
A three-dimensional thermo-viscoelastic constitutive model at finite strains, using a multiplicative decomposition of the elastic and viscoelastic deformation gradient, was developed in \cite{Diani_2006}. In a continuum thermo-viscoelastic finite deformation framework, a nonlinear structural relaxation model was combined with a viscous flow rule in \cite{Nguyen_2008} for the prediction of the shape memory response. Thermo-visco-hyperelastic  models at finite strains have been proposed in \cite{Westbrook_2011,Fan_2018,Baniasadi_2020}.
Thermo-viscoelastic constitutive models have been also developed on the basis of the Williams, Landel and Ferry (WLF) relaxation principle for amorphous polymers, also known as Time-Temperature Superimposition Principle (TTSP) \cite{Williams_1955,Ferry1980,Lomellini_1992,Dagdug_1998}. 
These models, referred to as thermo-rheologically simple due to the fact that the mechanical behaviour of the material at different temperatures differs solely in the relaxation times, have been recently employed for the simulation of 4D-printed morphing devices \cite{VanManen_2021}, including cardiovascular stents \cite{Lin_2020}.

Both phase transition and viscoelasticity-based models have been mostly developed and applied to three-dimensional continua discretized with solid finite elements. However, there is a very large variety of structures, ranging from lattice metamaterials to biomedical devices, composed by complex assemblies of slender elements for which solid-based discretization would lead to an unaffordable computational cost. For such kind of applications, one-dimensional geometrically exact rod elements, endowed with appropriate constitutive models, represent a very appealing alternative. 
These beam models, in addition to the ability to reproduce the shape memory effect, must support finite deformations (displacement and rotations) in order to reproduce the large shape changing process, and must be able to reconstruct complex assemblies of arbitrarily curved elements. 
Existing shape memory beam models lack one or more the above features being planar and not able to support large deformations nor curved initial geometries, see e.g. \cite{Baghani_etal2014,Baghani&Taheri2014,Zhao_etal2021a,Zhao_etal2021b,Ghosh_etal2013,Pandit&Srinivasan2019}.

In this paper, we present a novel computational model that meets all the necessary requirements. We extend the viscoelastic formulation proposed in \cite{Ferri_etal2023,Ferri&Marino2024} to the temperature-dependent case making use of the WLF equation \cite{Williams_1955}, enabling the reproduction of the shape memory effect. High efficiency is achieved discretizing the differential problem in space through the isogeometric collocation (IGA-C) method \cite{Auricchio2010,Auricchio2012,Schillinger2013,Fahrendorf_etal2022}. IGA-C possesses all the attributes of isogeometric analysis (IGA) \cite{Hughes2005a}, such as exactness of the geometric reconstruction of complex shapes and high-order accuracy. Moreover, it bypasses the numerical integration over the elements since the method discretizes the problem in the strong form. IGA-C has been already successfully employed for a number of problems~\cite{Gomez2014,Kruse2015,Auricchio2013,Kiendl2015a,Reali2015,Kiendl2015, KiendlMarinoDeLorenzis2017,Maurin_elal2018,Maurin_etal2018b,Fahrendorf_etal2020,Torre_etal2022,Torre_etal2023,Jia_etal2025}, including geometrically exact beams~\cite{Marino2016a,Weeger_etal2017,Marino2017b,Marino2019a,Marino2019b,Ferrietal2024}. 
There is a limited number of existing works addressing the problem of geometrically exact rods with inelastic materials \cite{Linn_etal2013,Lestringant&Kochmann2020,Lestringant_etal2020} and even less using IGA-based methods \cite{Weeger_etal2022,Ferri_etal2023,Ferri&Marino2024,Ferrietal2024}.
The present contribution, making use of the temperature-dependent Generalized Maxwell model, is apparently the first one able to reproduce the full shape memory effect (programming and recovery) of complex assemblies of geometrically exact rods. In addition to the discretization technique, other distinguishing features of our formulation are: i) $\SO3$-consistent linearization and time stepping; ii) minimal (finite) rotation parametrization, that means only three unknowns are used in contrast to other approaches (e.g., those based on quaternions), iii) no additional unknowns are needed to account for the rate-dependent material compared to the purely elastic case. 

The outline of the paper is as follows. In Section 2, we introduce the theoretical background of geometrically exact temperature-dependent viscoelastic beams, presenting the concept of polymers morphing  through temperature changes. 
In Section 3, we present the discretization of the problem, namely time discretization, linearization of the governing equations, and space discretization. In Section 4 we show the expected capabilities of the formulation through numerical applications, including challenging stent-like structures. Finally, in Section 5, the main conclusions of the work are drawn.

\section{Geometrically exact, temperature-dependent viscoelastic beams}
\subsection{A brief review of geometrically exact beams}
Let $(t,s) \mto \f c(t,s) \in \Rn^3$ be a space curve representing the axis of a beam. Assuming a quasi-static deformation for which inertial effects can be neglected, the governing equations in the local form \cite{Simo1985} are 
\bAl
\f{n},_{s} + \, \baf n & = \f 0\,, \label{eq:spa_f} \\ 
\f{m},_{s} + \, \f{c},_{s}\car \, \f{n} +\baf m & = \f 0\,, \label{eq:spa_m}
\end{align}
valid for any $t\in[0,T]\sub\Rn$ and $s\in [0, L]\sub\Rn$, $L$ and $T$ being the upper bounds of the space and time domains, respectively.
In the above equations, $\f{n}$ and $\f{m}$ denote the internal forces and couples per unit length, respectively; $\baf{n}$ and $\baf{m}$ are the external distributed forces and couples; $(\cdot),_s$ denotes the partial derivative with respect to the spatial coordinate $s$. 
The boundary conditions in the spatial setting are written as follows
\bEq
\f \eta = \baf \eta_c \sepr{or} \f{n}  = \baf n_c \sepr{with}	s = \{0,L\}\,, t\in[0,T]\,,\label{eq:bcseta_or_n}\\
\eEq
\bEq
\f \vtht = \baf \vtht_c \sepr{or} \f m = \baf m_c  \sepr{with} s = \{0,L\}\,, t\in[0,T]\,,\label{eq:bcstht_or_m}\\
\eEq
where $\baf\eta_c$ and $\baf\vtht_c$ are the spatial displacements and rotations imposed to any of the beam ends, $\baf n_c$ and $\baf m_c$ are the external concentrated forces and moments applied to any of the beam ends. 

Eqs.~\eqref{eq:spa_f} and \eqref{eq:spa_m} can be pulled back to the material form as follows
\bAl
\tif K \f N +\f N,_s+\fR R\Tra \baf n =  \f 0\,,  \label{mat_f} \\ 
\tif K \f M +\f M,_s+ \lf( \fR R\Tra \f{c},_{s}\rg) \car \, \f N  + \fR R\Tra \baf m =  \f 0\,, \label{mat_m}
\end{align}
where $(t,s)\mto \fR R(t,s) \in \SO3$ is a curve in rotation group describing the (rigid) rotation of the beam cross over time and space. Namely, $\fR R(t,s)$ rigidly maps cross sections at $t$ and $s$ from the material to the current configuration. 
Note that to obtain the material form of the governing equatons, Eqs.~\eqref{mat_f} and \eqref{mat_m}, we exploited the orthogonality of $\fR R$, $\fR R\fR R\Tra = \fR R\Tra \fR R = \f\id$, and the definition of the (material) beam curvature $(t,s)\mto \tif K(s) \byd \fR R\Tra \fR R,_{s}\in\so3$\footnote{With the symbol $\sim$ we mark elements of $\so3$, that is the set of $3 \times 3$ skew-symmetric matrices. In this context, they are used to represent curvature matrices and infinitesimal incremental rotations. Furthermore, we recall that for any skew-symmetric matrix $\tif a\in\so3$,  $\f a = \textnormal{axial}(\tif a)$ indicates the axial vector of $\tif a$ such that $\tif a \f h= \f a \car \f h$, for any $\f h\in \Rn^3$, where $\car$ is the cross product.}. 
Moreover, we used the definitions of internal forces and couples per unit length in the material form, respectively given by $\f N = \fR R\Tra \f n$ and $\f M = \fR R\Tra \f m$. Similar transformations apply to the boundary conditions. 

\subsection{Programming morphing of polymers through temperature-depending viscoelasticity}
\begin{figure}
\centering
\includegraphics[width=0.7\textwidth]{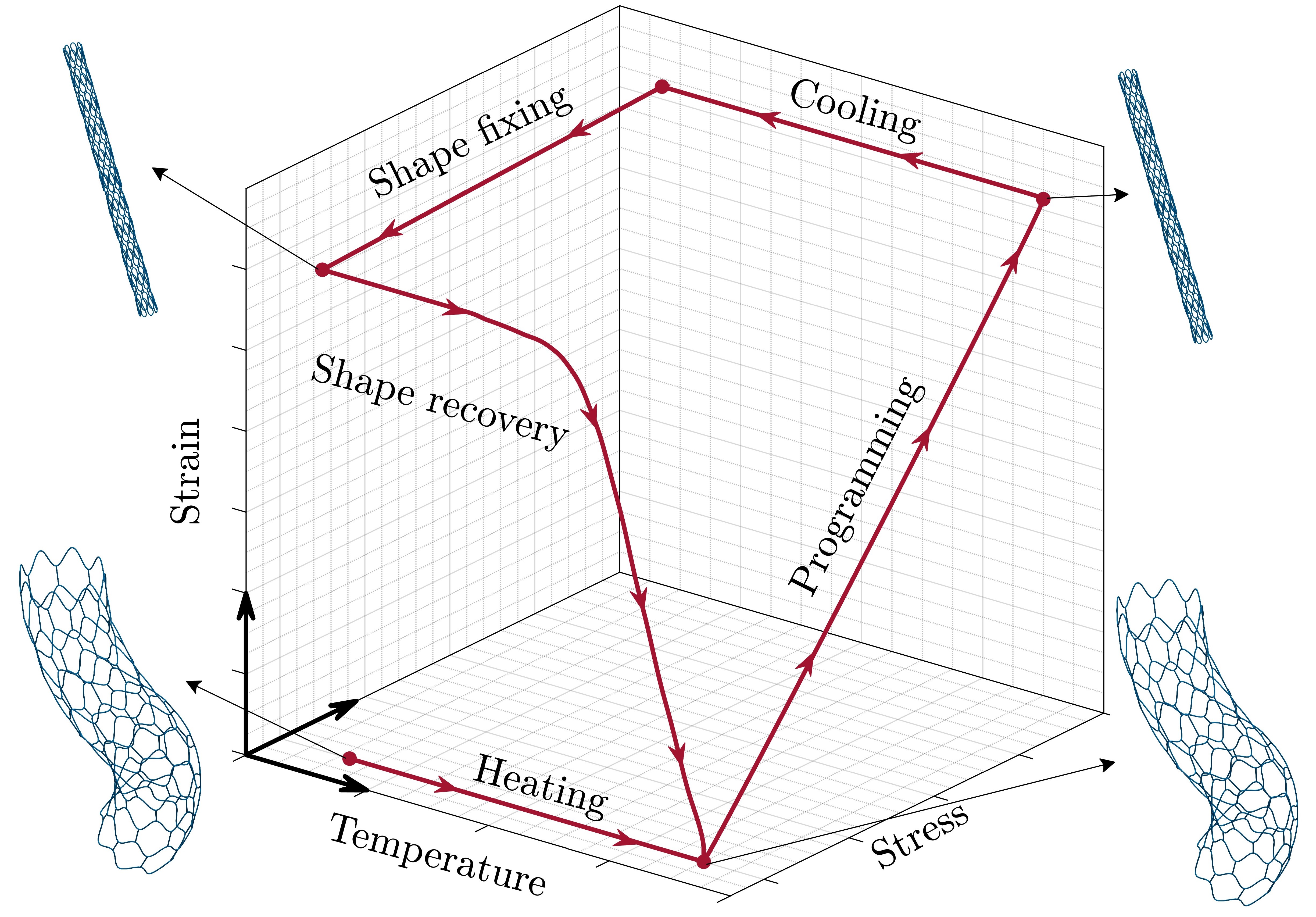}
\caption{The shape memory cycle in SMP.\label{fig:SMC}}
\end{figure}

As mentioned above, the programming phase of SMPs starts after increasing the temperature above the glass transition temperature $\nm T_G$. In this rubbery state, loads are applied and the material undergoes (mostly) irreversible, viscous deformations. When the predetermined shape is reached, the temperature is reduced below $\nm T_G$. This cooling process fixes the temporary shape, the material becomes glassy and exhibits an elastic behaviour. Upon re-heating the material above the glass transition, the initial shape can be totally or partially recovered. A schematic representation of this shape memory cycle is shown in Figure~\ref{fig:SMC}.

\subsubsection{The generalized Maxwell model for temperature-dependent geometrically exact viscoelastic beams}
The thermo-viscoelastic behaviour of amorphous polymers can be described making use of the TTSP \cite{Williams_1955,Ferry1980}, through the shift factor $a_\nm{T}$
\bEq
a_\nm{T}(\nm T) =-\fr{C_1(\nm{T}-\nm{T}_{ref})}{C_2+\nm{T}-\nm{T}_{ref}}\,
\eEq
where $\nm T$ is the temperature, $\nm{T}_{ref}$ is a reference temperature, and $C_1$ and $C_2$ are empirical material constants. 
Generally, $\nm{T}_{ref}$ is set to the glass transition temperature, $\nm{T}_G$. 
The shift factor controls the change of the relaxation times of Maxwell elements with the temperature as follows \cite{Williams_1955,Lomellini_1992,Dagdug_1998}
\bEq
\tau(\nm{T})=\tau({\nm{T}_G})10^{a_\nm{T}}=\tau({\nm{T}_G})10^{-C_1(\nm{T}-\nm{T}_{G})/(C_2+\nm{T}-\nm{T}_G)}\,. \label{eq:at}
\eEq

\begin{figure}
\centering
\includegraphics[width=0.75\textwidth]{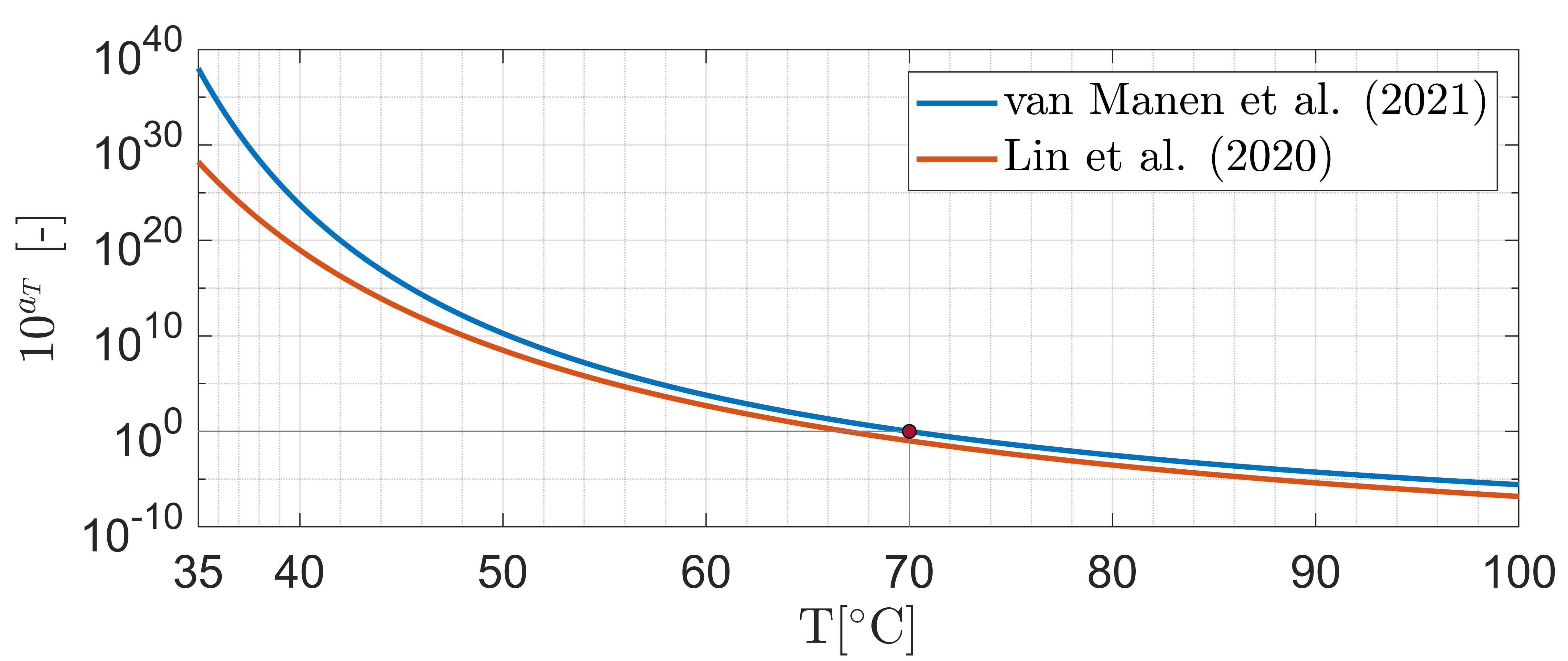}
\caption{Shift factor $a_{\nm T}$ as a function of temperature \cite{vanManen_2021s,Lin_2020}\label{fig:aT_PLA}.}
\end{figure}

Examples of variation of the shift factor, $a_{\rm T}$, with temperature for PLA materials, with parameters given in Table~\ref{tab:aT_PLA}, are shown in Figure~\ref{fig:aT_PLA}.
It is noted that when
\begin{itemize}
    \item $\nm{T}<\nm{T}_G\Rightarrow a_\nm{T}>>1\Rightarrow \tau(\nm{T})>>\tau({\nm{T}_G})$: the relaxation time is so high such that the material tends to have mainly an elastic behaviour;
    \item $\nm{T}=\nm{T}_G \Rightarrow  a_\nm{T}=1 \Rightarrow \tau(\nm{T})=\tau({\nm{T}_G})$;
    \item $\nm{T}>\nm{T}_G\Rightarrow a_\nm{T}<<1\Rightarrow \tau(\nm{T})<<\tau({\nm{T}_G})$: the relaxation time is so small such that the material has a dominant viscous behaviour. 
\end{itemize}

\begin{table}
\centering
	\begin{tabular}{ccccc}
    	
$$&  $C_1~[-]$	& $C_2~[\SI{}{\nm{K}}] $ & $\nm{T}_G~[\SI{}{^\circ{}\nm{C}}] $\\
	\hline
Van Manen et al. (2021)&\SI{14.59}{} &\SI{48.43}{}  &\SI{70.0}{} \\
Lin et al. (2020)&\SI{17.44}{} &\SI{51.60}{}  &\SI{66.9}{}\\
	\hline
	\end{tabular}
		\caption{Empirical parameters for the characterization of the shift factor, $a_\nm{T}$, for two PLA materials.}\label{tab:aT_PLA}
\end{table}

Making use of Eq.~\eqref{eq:at}, the thermal effect on the geometrically exact viscous strain rates associated with the $\alp$th Maxwell element \cite{Ferri_etal2023,Ferri&Marino2024} is accounted as follows
\bEq\label{eq:evol_f_m_thermal}
\dot{\f\Gam}_{N\alp}=\fr {1} {\tau_\alp(\nm T)}(\f\Gam_N-\f\Gam_{N\alp})\,,
\sepr{}
\dot{\f K}_{M\alp}=\fr {1} {\tau_\alp(\nm T)}(\f K_M-\f {K}_{M\alp})\,, 
\eEq
where
\begin{align}
\f\Gam_N & = \f \Gam - \f \Gam_0 =  \fR R\Tra \f c,_s - \fR R_0\Tra \f c_{0,s}\,,\label{eq:Gam_N}\\
\f K_M & =\rm axial (\tif K - \tif K_0) = \f K - \f K_0\,,\label{eq:Gam_M}
\end{align}
are the total strains, whereas 
$\f\Gam_{N\alp} $ and $\f {K}_{M\alp}$ are the viscous strains of the $\alp$th Maxwell element~\cite{Ferri_etal2023}.
In Eqs.~\eqref{eq:Gam_N} and \eqref{eq:Gam_M}, $s \mto \f c_0(s) \in \Rn^3$ is the initial configuration of the beam axis, $s \mto \fR R_0(s) \in \SO3$ is the rotation of the beam cross section in the initial configuration, and $s \mto \tif K_0 \in \so3$ is the initial curvature of the beam. A proper representation of these quantities allows to address any arbitrarily curved initial geometry~\cite{Marino_etal2020,Ignesti_etal2023}.

\section{Discretization and linearization of the problem}
\subsection{Time discretization and integration scheme}
We employ the trapezoidal rule for the integration of the time-discretized version of the viscous strain rates. At $t^{n+1}$, being $h$ the time step size, Eq.~\eqref{eq:evol_f_m_thermal} reads as follows
\bEq\label{eq:evol_f_thermal_disc}
{\f\Gam}^{n+1}_{N\alp}={\f\Gam}^{n}_{N\alp} + \fr {h}{2}\left[\fr{1}{\tau^n_\alp}(\f\Gam^n_N-\f\Gam^n_{N\alp})+\fr{1}{\tau^{n+1}_\alp}(\f\Gam^{n+1}_N-\f\Gam^{n+1}_{N\alp})\right],
\eEq
which, after some rearrangements, becomes
\bEq
{\f\Gam}^{n+1}_{N\alp}=\fr{h}{2\tau^{n+1}_\alp+h}\f\Gam^{n+1}_N + \fr{\tau^{n+1}_\alp}{2\tau^{n+1}_\alp+h}\f{\varPsi}^n_{\Gam\alp}\,,
\eEq
where we have set 
\bEq\label{eq:evol_varPhi_thermal_1}
\f{\varPsi}^n_{\Gam\alp}=\fr{h}{\tau^{n}_\alp}(\f\Gam^{n}_N)+\fr{2\tau^{n}_\alp-h}{\tau^{n}_\alp}{\f\Gam}^{n}_{N\alp}\,. 
\eEq

Similarly, the vector representing bending and torsional viscous strains for the $\alp$th Maxwell element is written as
\bEq\label{eq:evol_m_thermal_disc}
{\f K}^{n+1}_{M\alp}=\fr{h}{2\tau^{n+1}_\alp+h}\f{K}^{n+1}_M+\fr{\tau^{n+1}_\alp}{2\tau^{n+1}_\alp+h}\f{\varPsi}^n_{K\alp}\,,
\eEq
with 
\bEq\label{eq:evol_varPhi_thermal_2}
\f{\varPsi}^n_{K\alp}=\fr{h}{\tau^{n}_\alp}\f{K}^{n}_M+\fr{2\tau^{n}_\alp-h}{\tau^{n}_\alp}{\f{K}}^{n}_{M\alp}\,. 
\eEq

Importantly, we observe that $\f{\varPsi}^n_{\f{\Gam}\alp}$ and $\f{\varPsi}^n_{\f{K}\alp}$ are quantities computed solely using information known from the previous time step.
$\tau^n_\alp$ and $\tau^{n+1}_\alp$ are the relaxation times of the $\alp$th Maxwell element corresponding to the temperature at time $n$ and $n+1$, respectively. Namely, $\tau^n_\alp = \tau({\rm T}(t^n))$ and $\tau^{n+1}_\alp = \tau({\rm T}(t^{n+1}))$. 

Making use of Eqs.~\eqref{eq:evol_f_thermal_disc} and~\eqref{eq:evol_m_thermal_disc}, the internal stresses \cite{Ferri_etal2023,Ferri&Marino2024} at time $t^{n+1}$ become 
\begin{align}
\f N^{n+1} & = \CNinf \f\Gam_N^{n+1} +\sum_{\alp=1}^{m}\CNalp(\f\Gam_N^{n+1}-\f\Gam^{n+1}_{N\alp}),\\
\f M^{n+1} & = \CMinf \f K_M^{n+1} + \sum_{\alp=1}^{m}\CMalp(\f K_M^{n+1}-\f K^{n+1}_{M\alp})\,, 
\end{align}
which, after some manipulation, become
\begin{align}
\f N^{n+1} & = \CNb^{n+1}\f\Gam^{n+1}_N-\sum_{\alp=1}^{m}\CNalp\left(\fr{\tau^{n+1}_\alp}{2\tau^{n+1}_\alp+h}\right)\f{\varPsi}^n_{\Gam\alp}\,,\label{eq:N_constitutive_disc}\\
\f M^{n+1} & = \CMb^{n+1}\f{K}^{n+1}_M-\sum_{\alp=1}^{m}\CMalp\left(\fr{\tau^{n+1}_\alp}{2\tau^{n+1}_\alp+h}\right)\f{\varPsi}^n_{K \alp}\,,\label{eq:M_constitutive_disc}
\end{align}
where we have defined 
\beq
\CNb^{n+1}=\CNz-\sum_{\alp=1}^{m}\CNalp\left(\fr{h}{2\tau^{n+1}_\alp+h}\right)
\sepr{and}
\CMb^{n+1}=\CMz-\sum_{\alp=1}^{m}\CMalp\left(\fr{h}{2\tau^{n+1}_\alp+h}\right)\,,
\eeq
being $\CNz = \CNinf + \sum_{\alp=1}^m \CNalp$ and $\CMz =  \CMinf + \sum_{\alp=1}^m \CMalp$ the instantaneous elasticity tensors. 

Exploiting Eqs.~\eqref{eq:N_constitutive_disc} and~\eqref{eq:M_constitutive_disc}, the time-discretized version of the governing equations, Eqs.~\eqref{mat_f} and~\eqref{mat_m}, turn into   
\begin{gather}
 \tif K^{n+1}\CNb^{n+1}\f\Gam^{n+1}_N -\tif K^{n+1}\sum_{\alp=1}^{m}\CNalp\left(\fr{\tau^{n+1}_\alp}{2\tau^{n+1}_\alp+h}\right)\f{\varPsi}^n_{\Gam\alp}+\nonumber\\
 +\CNb^{n+1}\f\Gam^{n+1}_{N,s}-\sum_{\alp=1}^{m}\CNalp\left(\fr{\tau^{n+1}_\alp}{2\tau^{n+1}_\alp+h}\right)\f{\varPsi}^n_{\Gam\alp},_s+\fR {R\Tra}^{n+1} \baf n^{n+1} =  \f 0\,,\label{eq:gov_N_disc}
\end{gather}
and
\begin{gather}
  \tif K^{n+1}\CMb^{n+1}\f{K}^{n+1}_M-\tif K^{n+1}\sum_{\alp=1}^{m}\CMalp\left(\fr{\tau^{n+1}_\alp}{2\tau^{n+1}_\alp+h}\right)\f{\varPsi}^n_{K \alp}+\CMb^{n+1}\f{K}^{n+1}_{M,s}+\nonumber\\
  -\sum_{\alp=1}^{m}\CMalp\left(\fr{\tau^{n+1}_\alp}{2\tau^{n+1}_\alp+h}\right)\f{\varPsi}^n_{K \alp,s}+\fR {R\Tra}^{n+1} \baf m^{n+1}+\nonumber\\
   +\lf( \fR {R\Tra}^{n+1} \f{c},^{n+1}_{s}\rg) \car \, \lf[  \CNb^{n+1}\f\Gam^{n+1}_N-\sum_{\alp=1}^{m}\CNalp\left(\fr{\tau^{n+1}_\alp}{2\tau^{n+1}_\alp+h}\right)\f{\varPsi}^n_{\Gam\alp} \rg] = \f 0\,.\label{eq:gov_M_disc}
\end{gather}
Note that in the above equations we have assumed that both temperature and relaxation times are constant over the beam length, namely $\nm{T},_s^{n+1} = 0$ and $\tau^{n+1}_{\alp,s} = 0$. 

With similar arguments, the Neumann boundary conditions take the following form
\begin{gather}
\CNb^{n+1}\f\Gam^{n+1}_N-\sum_{\alp=1}^{m}\CNalp\left(\fr{\tau^{n+1}_\alp}{2\tau^{n+1}_\alp+h}\right)\f{\varPsi}^n_{\Gam\alp}=\fR {R\Tra}^{n+1} \baf n_c^{n+1}\,,  \label{BC_N_disc}
\end{gather}
\begin{gather}
\CMb^{n+1}\f{K}^{n+1}_M-\sum_{\alp=1}^{m}\CMalp\left(\fr{\tau^{n+1}_\alp}{2\tau^{n+1}_\alp+h}\right)\f{\varPsi}^n_{K \alp}=\fR {R\Tra}^{n+1} \baf m_c^{n+1}\,. \label{BC_M_disc}
\end{gather}

\subsection{Linearization of the governing equations}
In the formulation we present here, we assume to know the temperature at each time step considering it as an external action that influences the mechanical properties of the viscoelastic material. 
The $\SO3$-consistent linearization leads to the following form of the governing equations, where the unknown fields $\del\f\vTht^{n+1}$ and $\del\f\eta^{n+1}$, which are the incremental rotations and displacements, respectively,  evaluated at $t^{n+1}$ appear.
\begin{gather}
[\CNb^{n+1}\wti{(\hat{\fR R}\Tra{^{n+1}}\f{c},_s^{n+1})}-\wti{(\CNb^{n+1}\haf\Gam^{n+1}_N)}+\nonumber\\
+\sum_{\alp=1}^{m}\left(\fr{\tau^{n+1}_\alp}{2\tau^{n+1}_\alp+h}\right)\wti{(\CNalp\f{\varPsi}^n_{\Gam\alp})} ]\del\f\vTht,_s^{n+1} +  [\htif {K}^{n+1}\CNb^{n+1}\wti{(\hat{\fR R}\Tra{^{n+1}}\f{c},_s^{n+1})}+ \nonumber\\
-\CNb^{n+1}\wti{(\htif {K}^{n+1}\hat{\fR R}\Tra{^{n+1}}\f{c},_s^{n+1})}+\CNb^{n+1}\wti{(\hat{\fR R}{\Tra{^{n+1}}}\hat{\f{c}},_{ss}^{n+1})}+\wti{(\hat{\fR R}{\Tra{^{n+1}}}\baf n^{n+1})} +                                         \nonumber\\
-\wti{(\CNb^{n+1}\haf\Gam^{n+1}_N)}\htif {K}^{n+1} +\sum_{\alp=1}^{m}\left(\fr{\tau^{n+1}_\alp}{2\tau^{n+1}_\alp+h}\right)\wti{(\CNalp\f{\varPsi}^n_{\Gam\alp})}\htif {K}^{n+1}] \del\f\vTht^{n+1}+\nonumber\\
+\CNb^{n+1}\hat{\fR R}{\Tra{^{n+1}}}\del\f\eta,_{ss}^{n+1}+[\htif {K}^{n+1}\CNb^{n+1}\hat{\fR R}{\Tra{^{n+1}}}-\CNb^{n+1}\htif {K}^{n+1}\hat{\fR R}{\Tra{^{n+1}}}]\del\f\eta,_{s}^{n+1}+\hfR F^{n+1} = \f 0\,, \label{gov_N_disc_lin}
\end{gather}

\begin{gather}
\CMb^{n+1}\del\f\vTht,_{ss}^{n+1}+[\CMb^{n+1}\htif {K}^{n+1}+\htif {K}^{n+1}\CMb^{n+1}-\wti{(\CMb^{n+1}\haf{K}^{n+1}_M)}+ \nonumber \\
+\sum_{\alp=1}^{m}\left(\fr{\tau^{n+1}_\alp}{2\tau^{n+1}_\alp+h}\right)\wti{(\CMalp\f{\varPsi}^n_{K \alp})}]\del\f\vTht,_{s}^{n+1}+[\htif{K}^{n+1}\CMb^{n+1}\htif {K}^{n+1}-\wti{(\CMb^{n+1}\haf{K}^{n+1}_M)}\htif{K}^{n+1} + \nonumber\\ 
+\sum_{\alp=1}^{m}\left(\fr{\tau^{n+1}_\alp}{2\tau^{n+1}_\alp+h}\right)\wti{(\CMalp\f{\varPsi}^n_{K \alp})}\htif{K}^{n+1}+\CMb^{n+1}\haf{K},_s^{n+1}+\wti{(\fR {R\Tra}^{n+1} \baf m^{n+1})}+\nonumber\\ 
+(\wti{(\hat{\fR R}\Tra{^{n+1}}\haf{c},_s^{n+1})}\CNb^{n+1}-\wti{(\CNb^{n+1}\haf\Gam^{n+1})}+\nonumber\\ 
+\sum_{\alp=1}^{m}\left(\fr{\tau^{n+1}_\alp}{2\tau^{n+1}_\alp+h}\right)\wti{(\CNalp\f{\varPsi}^n_{\Gam\alp})} ) 
\wti{(\hat{\fR R}\Tra{^{n+1}}\haf{c},_s^{n+1})}]\del\f\vTht^{n+1}+\nonumber\\ 
+[\wti{(\hat{\fR R}\Tra{^{n+1}}\haf{c},_s^{n+1})}\CNb^{n+1}-\wti{(\CNb^{n+1}\haf\Gam^{n+1})}+\nonumber\\ 
+\sum_{\alp=1}^{m}\left(\fr{\tau^{n+1}_\alp}{2\tau^{n+1}_\alp+h}\right)\wti{(\CNalp\f{\varPsi}^n_{\Gam\alp})}] \hat{\fR R}\Tra{^{n+1}}\del\f\eta,_{s}^{n+1} + \hfR T^{n+1} = \f 0\,,\label{gov_M_disc_lin}
\end{gather}
where we have set 
\begin{gather}
\hfR F^{n+1} = \htif K^{n+1}\CNb^{n+1}\haf\Gam^{n+1}_N -\htif K^{n+1}\sum_{\alp=1}^{m}\CNalp\left(\fr{\tau^{n+1}_\alp}{2\tau^{n+1}_\alp+h}\right)\f{\varPsi}^n_{\Gam\alp} + \nonumber\\
 +\CNb^{n+1}\haf\Gam^{n+1}_{N,s}-\sum_{\alp=1}^{m}\CNalp\left(\fr{\tau^{n+1}_\alp}{2\tau^{n+1}_\alp+h}\right)\f{\varPsi}^n_{\Gam\alp},_s+\hat{\fR R}\Tra{^{n+1}} \baf n^{n+1} \,,
\end{gather}
and
\begin{gather}
\hfR T^{n+1} =     \htif K^{n+1}\CMb^{n+1}\haf{K}^{n+1}_M-\htif K^{n+1}\sum_{\alp=1}^{m}\CMalp\left(\fr{\tau^{n+1}_\alp}{2\tau^{n+1}_\alp+h}\right)\f{\varPsi}^n_{K \alp}+\CMb^{n+1}\haf{K}^{n+1}_{M,s} + \nonumber\\
  -\sum_{\alp=1}^{m}\CMalp\left(\fr{\tau^{n+1}_\alp}{2\tau^{n+1}_\alp+h}\right)\f{\varPsi}^n_{K \alp,s}+\hat{\fR R}\Tra{^{n+1}} \baf m^{n+1}+\nonumber\\
   +\lf( \hat{\fR R}\Tra{^{n+1}} \haf{c},^{n+1}_{s}\rg) \car \, \lf[  -\sum_{\alp=1}^{m}\CNalp\left(\fr{\tau^{n+1}_\alp}{2\tau^{n+1}_\alp+h}\right)\f{\varPsi}^n_{\Gam\alp} \rg]\,.
\end{gather}

Similarly, Neumann BCs are linearized as follows
\begin{gather}
\lf[\CNb^{n+1}\wti{( \hat{\fR R}\Tra{^{n+1}} \haf{c},^{n+1}_{s})}-\wti{( \hat{\fR R}\Tra{^{n+1}} \baf n_c^{n+1})}\rg]\del\f\vTht^{n+1}+\CNb^{n+1} \hat{\fR R}\Tra{^{n+1}}\del\f\eta,_{s}^{n+1} = \nonumber \\ 
=\hat{\fR R}\Tra{^{n+1}} \baf n_c^{n+1}-\CNb^{n+1}\haf\Gam^{n+1}_N+\sum_{\alp=1}^{m}\CNalp\left(\fr{\tau^{n+1}_\alp}{2\tau^{n+1}_\alp+h}\right)\f{\varPsi}^n_{\Gam\alp}\,, \label{neu_N_disc_lin}
\end{gather}

\begin{gather}
\CMb^{n+1}\del\f\vTht,_s^{n+1}+[\CMb^{n+1}\htif{K}^{n+1}-\wti{( \hat{\fR R}\Tra{^{n+1}} \baf m_c^{n+1})}]\del\f\vTht^{n+1} = \nonumber \\ 
=\hat{\fR R}\Tra{^{n+1}} \baf m_c^{n+1}+\sum_{\alp=1}^{m}\CMalp\left(\fr{\tau^{n+1}_\alp}{2\tau^{n+1}_\alp+h}\right)\f{\varPsi}^n_{K \alp}-\CMb^{n+1}\haf{K}^{n+1}_M\,. \label{neu_M_disc_lin}
\end{gather}

Note that for the rotations, the above linearization is based on the material incremental vector $\del\f\vTht = \rm axial (\del\tif\vTht)$, such that $\fR R_\veps = \fR R\exp(\veps\del\tif\vTht))$, with $\veps\in \Rn$ and being $\exp:\so3\to\SO3$ the exponential map for $\SO3$ \cite{Argyris1982}. From this this multiplicative composition rule, the directional derivatives of the rotation matrix and its space derivative are given by \cite{Marino2019b}
\begin{align}\label{direct_R}
&\dde(\fR R_\veps)_{\veps=0}=\dde(\fR R\exp(\veps\del\tif\vTht))_{\veps=0}=\fR R\del\tif\vTht\,,\\
&\dde(\fR R\Tra_\veps)_{\veps=0}=\dde(\exp(-\veps\del\tif\vTht)\fR R\Tra)_{\veps=0}=-\del\tif\vTht\fR R\Tra\,,\\
&\dde(\fR R_{\veps,s})_{\veps=0}=\fR R \tif{K}\del\tif\vTht+\fR R\del\tif\vTht,_s\,,\\
&\dde(\fR R\Tra_{\veps,s})_{\veps=0}=\del\tif\vTht\tif{K}\fR R\Tra -\del\tif\vTht,_s\fR R\Tra\,.
\end{align}
For the translational field, the standard additive composition rule leads to 
\begin{align}
&\dde(\f c_\veps)_{\veps=0}=\dde(\f c +\veps\del\f\eta)_{\veps=0}=\del\f\eta\,,\\
&\dde(\f c_{\veps,s})_{\veps=0}=\del\f\eta,_s\,,\\
&\dde(\f c_{\veps,ss})_{\veps=0}=\del\f\eta,_ss\,.\label{direct_c}
\end{align}
Exploiting the directional derivatives of the above key kinematic quantities, it possible to obtain all the linearizations needed in Eqs.~\eqref{gov_N_disc_lin}-\eqref{neu_M_disc_lin}\cite{Marino2019b}.

\subsection{Space discretization of the linearized equations, inremental update and time stepping}
Following the IGA paradigm~\cite{Hughes2005a,Cottrell2009}, the unknowns $\del\f\vTht^{n+1}$ and $\del\f\eta^{n+1}$ at time $t^{n+1}$ appearing in the linearized governing equations
(Eqs.~\eqref{gov_N_disc_lin}, \eqref{gov_M_disc_lin} and~\eqref{neu_N_disc_lin}, \eqref{neu_M_disc_lin}) are spatially discretized by using NURBS basis functions $R_{j,p}$ with $j = 1,\ldots,\rm n$ of degree $p$ as follows
\bAl
\del\f\vTht^{n+1} (u) &= \sum_{j= 1}^{\rm n} R_{j,p}(u) \del \chf\vTht^{n+1}_j\sepr{with} u\in [0,\,1]\,,\label{eq:Thtu}\\
\del\f\eta^{n+1} (u) &= \sum_{j= 1}^{\rm n} R_{j,p}(u)\del \chf\eta^{n+1}_j\sepr{with} u\in [0,\,1]\,,\label{eq:etau}
\end{align}
where $\del \chf\vTht^{n+1}_j$ and $\del\chf\eta^{n+1}_j$ are the ($2\car3\car \rm n$) unknowns of our system to be solved at each iteration. They represent the $j$th incremental control rotation and translation vectors. The same basis functions are used to discretize the beam centroid line
\bEq
\f c^{n+1} (u) = \sum_{j= 1}^{\rm n} R_{j,p}(u) \chf p^{n+1}_j\sepr{with} u\in [0,\,1]\,, \label{eq:cu}
\eEq
where $\chf p^{n+1}_j$ are the control points of the beam centreline.
We substitute Eqs.~\eqref{eq:Thtu}-\eqref{eq:cu} in~\eqref{gov_N_disc_lin}-\eqref{neu_M_disc_lin}, and collocate field equations and boundary conditions at the internal collocation points (Greville abscissae \cite{Auricchio2010}) and at the boundaries ($u=0$ and $u=1$), respectively. In this way, we build a square system, which is solved for $\del \chf\vTht^{n+1}_j$ and $\del\chf\eta^{n+1}_j$ at each iteration of the Newton-Raphson algorithm. 
The computed primal variables are then used to update the configuration of the centroid, as well as the total and the viscous strain measures \cite{Marino2019b,Ferri_etal2023}. For example, at the $k$th iteration of the Newton-Raphson algorithm, we have
\begin{align}
\f c^{n+1,k+1} (u) & = \sum_{j= 1}^{\rm n} R_{j,p}(u) \chf p^{n+1,k+1}_j=\sum_{j= 1}^{\rm n} R_{j,p}(u) (\chf p^{n+1,k}_j+\del\chf\eta^{n+1,k}_j)\,,\\
\fR R^{n+1,k+1} (u) & = \fR R^{n+1,k}(u)\exp{(\del\tif\vTht^{n+1,k} (u))}\,,
\end{align}
with $\del\f\vTht^{n+1,k} (u) = \sum_{j= 1}^{\rm n} R_{j,p}(u) \del\chf\vTht^{n+1,k} _j$. Note that the same multiplicative and additive rules employed in Eqs.~\eqref{direct_R}-\eqref{direct_c} need to be used for a consistent update.

Consistently with the time integration scheme (Eqs.~\eqref{eq:evol_f_thermal_disc}-\eqref{eq:evol_m_thermal_disc}), the viscous strains are updated as follows 
\begin{align}
{\f\Gam}^{n+1,k+1}_{N\alp}=\fr{h}{2\tau^{n+1}_\alp+h}\f\Gam^{n+1,k+1}_N + \fr{\tau^{n+1}_\alp}{2\tau^{n+1}_\alp+h}\f{\varPsi}^n_{\Gam\alp}\,,\\
{\f K}^{n+1,k+1}_{M\alp}=\fr{h}{2\tau^{n+1}_\alp+h}\f{K}^{n+1,k+1}_M+\fr{\tau^{n+1}_\alp}{2\tau^{n+1}_\alp+h}\f{\varPsi}^n_{K\alp}\,, 
\end{align}
where we remark that $\f{\varPsi}^n_{\Gam\alp}$ and $\f{\varPsi}^n_{K\alp}$ are not updated during the iterations of Newton-Raphson algorithm, but only during the time stepping passing from $t^{n}$ to $t^{n+1}$.

\section{Numerical results}
In this section, we present a series of numerical applications with increasing complexity aimed at testing all the key features of the proposed formulation. 

We start with a circular arch under concentrated couples. Then we address the morphing of a straight cantilever beam, followed by two cases of cardiovascular stent-like structures for which we analyse the programming and shape recovery phases. 

\subsection{Circular arch under simultaneous change of loads and temperature in time}\label{subsec:circ_arch}
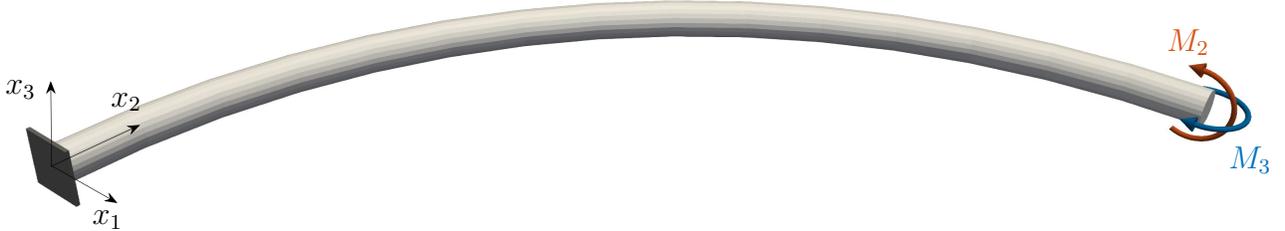
\begin{figure}[h]
\centering
\begin{overpic}[width=1\textwidth]{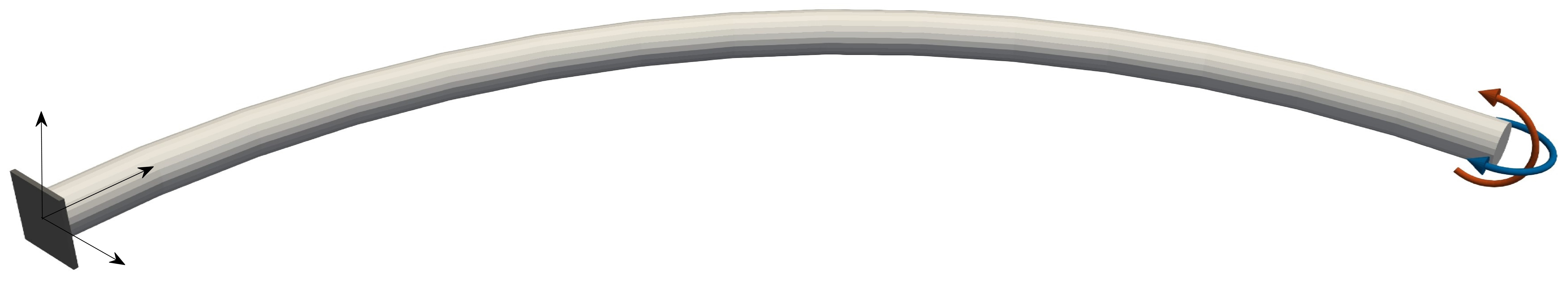}
\put(925,135){\textcolor{aran}{$M_2$}} \put(975,40){\textcolor{blu}{$M_3$}}
\put(75,90){$x_2$}  \put(60,-5){$x_1$}  \put(-10,100){$x_3$}
\end{overpic}
\caption{Circular arch under tip load: geometry and applied couples.\label{fig:circular_arch_geom}}
\end{figure}

We investigate a $90^\circ$-circular arch of radius $\SI{1}{m}$, clamped at one end, and loaded by concentrated couples at the tip of the beam (see Figure~\ref{fig:circular_arch_geom}). The arch has a circular cross-section of diameter $d=\SI{0.05}{m}$. Thermal and viscoelastic material properties are similar to those used in \cite{VanManen_2021} and are reported in Table~\ref{tab:aT_PLA} (first row) and~\ref{tab:PLA}, respectively. For all the Maxwell elements, the Poisson's ratio is kept constant to $\nu=0.33$. 

\begin{table}
\centering
	\begin{tabular}{ccccccc}
$\alp$&$E_\alp$ [$\SI{}{MPa}$] 	& $\tau_{G,\alp}$ [$\SI{}{s}$]&   &$\alp$ &$E_\alp$ [$\SI{}{MPa}$]&$\tau_{G,\alp}$ [$\SI{}{s}$]\\
	\hline
$\infty$&\SI{80.59}{} &-& & $8$&\SI{474.56}{} &\SI{1e-3}{} \\
$1$&\SI{20.12}{} &\SI{1e-10}{}& &$9$&\SI{449.43}{} &\SI{1e-2}{}\\
$2$&\SI{50.31}{} &\SI{1e-9}{}& &$10$&\SI{237.98}{} &\SI{1e-1}{}\\
$3$&\SI{81.37}{} &\SI{1e-8}{}& &$11$&\SI{114.16}{} &\SI{1e0}{}\\
$4$&\SI{97.02}{} &\SI{1e-7}{}& &$12$&\SI{51.82}{} &\SI{1e1}{}\\
$5$&\SI{173.70}{} &\SI{1e-6}{}& &$13$&\SI{29.98}{} &\SI{1e2}{}\\
$6$&\SI{225.60}{} &\SI{1e-5}{}& &$14$&\SI{14.40}{} &\SI{1e3}{}\\
$7$&\SI{292.64}{} &\SI{1e-4}{}& &$15$&\SI{0.72}{} &\SI{1e5}{}\\
	\hline
	\end{tabular}
		\caption{Viscoelastic properties of PLA modelled with 15 Maxwell elements \cite{vanManen_2021s}.}\label{tab:PLA}
\end{table}
 
\begin{figure}
\centering
\subfigure[Tip displacement along $x_1$.\label{fig:u1_circle_2_1}]
{\includegraphics[width=.40\textwidth]{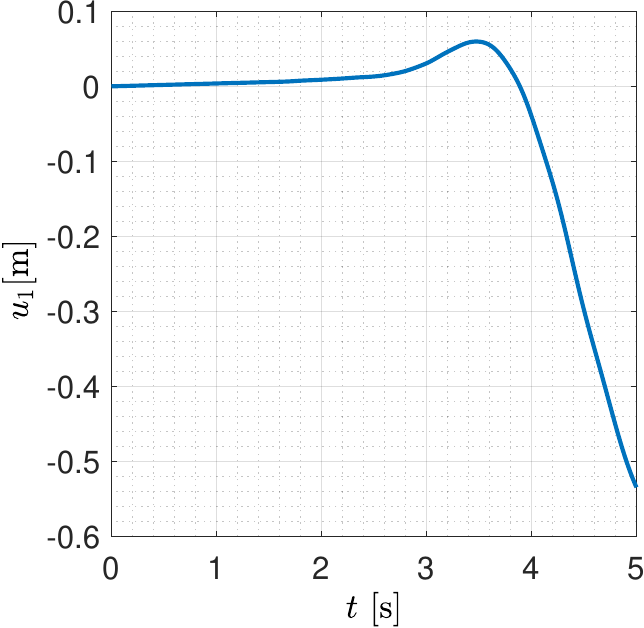}}
\hspace{0.1\textwidth}
\subfigure[Tip displacement along $x_2$.\label{fig:u2_circle_2_1}]
{\includegraphics[width=.40\textwidth]{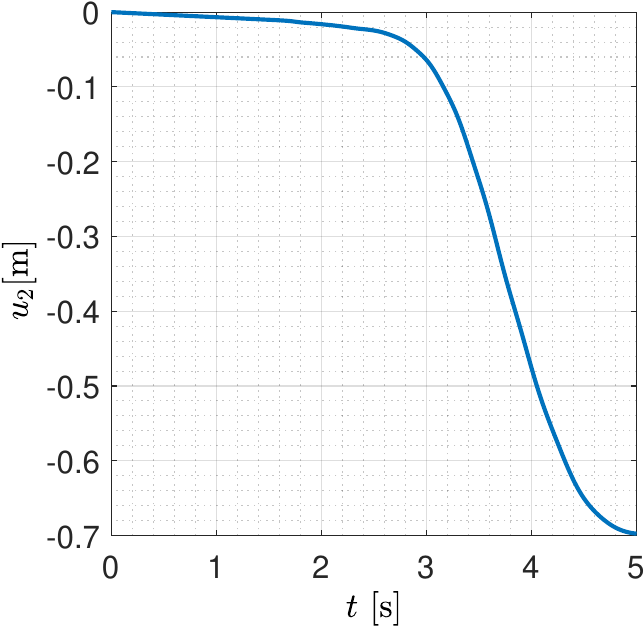}}
\subfigure[Tip displacement along $x_3$.\label{fig:u3_circle_2_1}]
{\includegraphics[width=.40\textwidth]{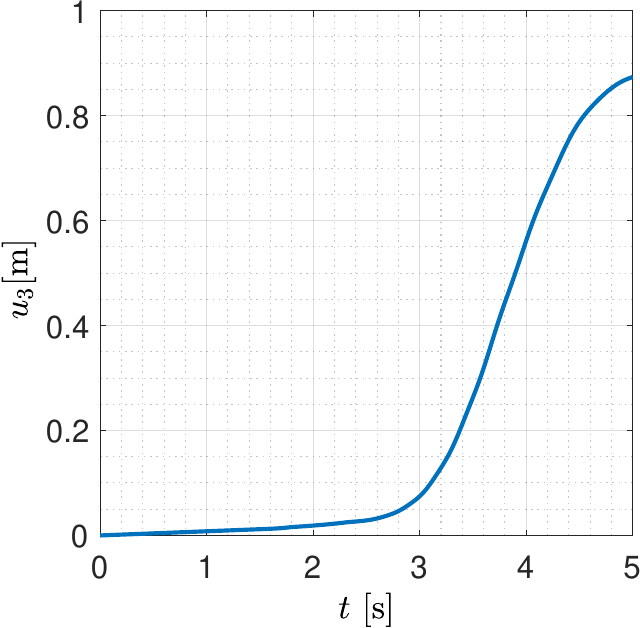}}
\hspace{0.1\textwidth}
\subfigure[Deformed shape at $t=\SI{5}{s}$.\label{fig:deformed_circle_2_1}]
{\begin{overpic}[width=0.40\textwidth]{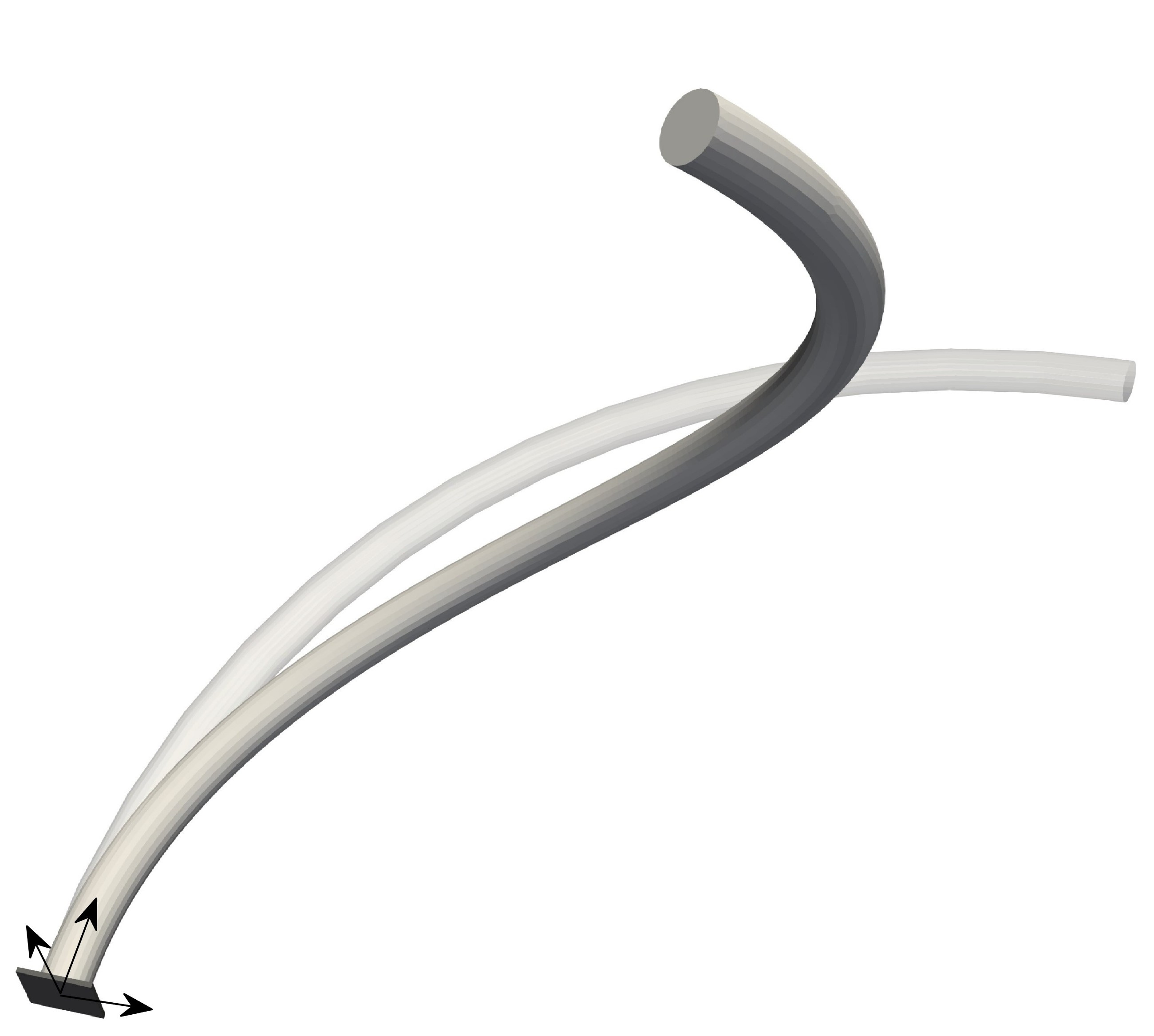}
\put(10,135){$x_2$}  \put(140,-15){$x_1$}  \put(-60,75){$x_3$}
\end{overpic}}
\caption{Circular arch: tip displacements under concentrated couples, $M_2$ and $M_3$.}\label{fig:circle_2_1}
\end{figure}

The simulation time is $T=\SI{5}{s}$, with $h=\SI{0.001}{s}$. Both temperature and loads are assumed to vary linearly during the simulation, from $\SI{31.5}{^\circ\nm{C}}$ to $\SI{90}{^\circ\nm{C}}$, and from $\SI{0}{Nm}$ to $\SI{25}{Nm}$, respectively. Tip displacement time histories are shown in Figure~\ref{fig:circle_2_1}. 
As expected, the progressive drop of the mechanical properties of the structure occurring from $t=\SI{2.5}{s}$ on is caused by the increase of temperature above the glass transition value $\nm{T}_G = \SI{70}{^{\circ}\nm C}$.

\begin{figure}
\centering
\includegraphics[width=0.6\textwidth]{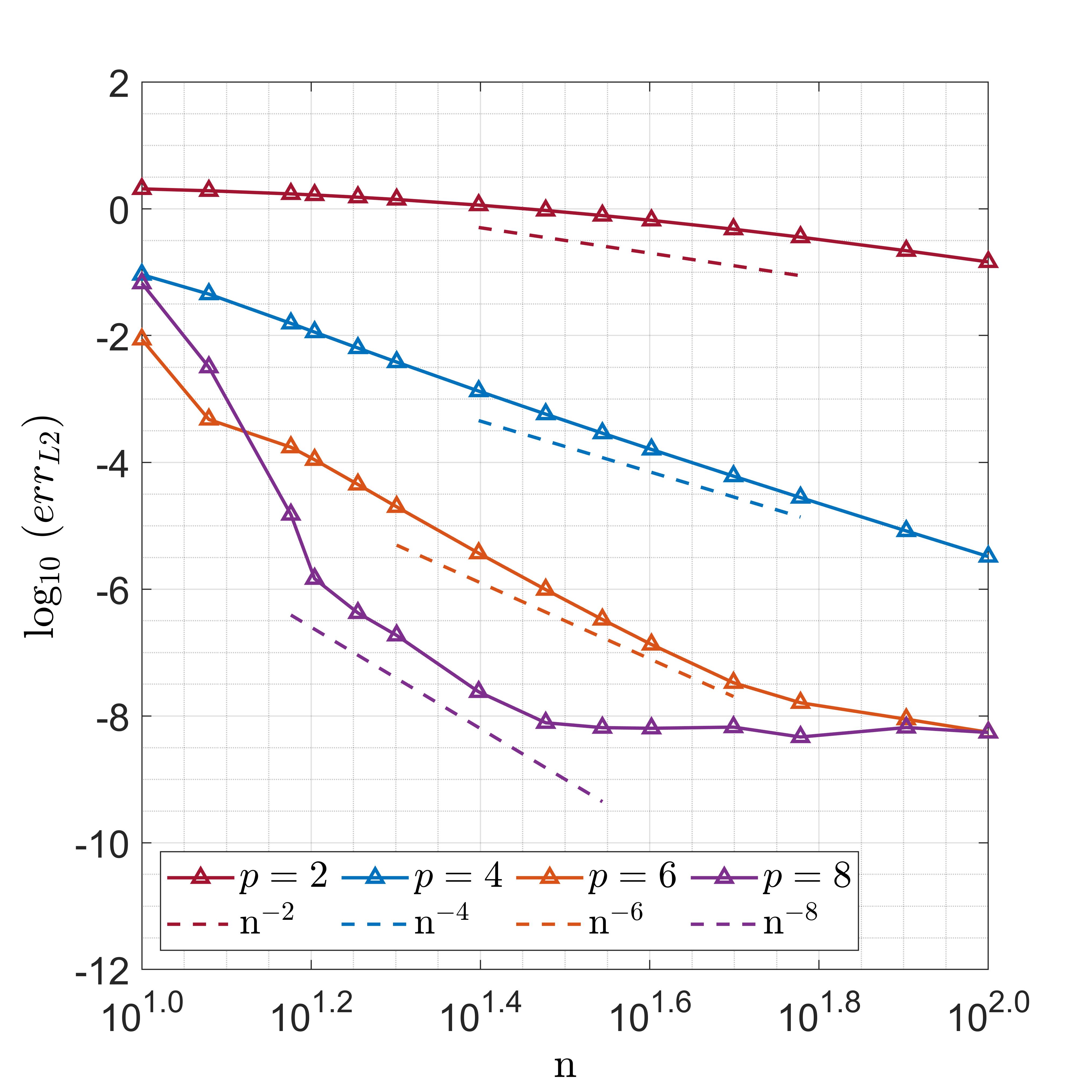}
\caption{Circular Arc: spatial convergence plots for $p=2,\ldots, 8$ vs. the number of collocation points $\rm n$.\label{fig:conv_circ}}
\end{figure}

Among the appealing attributes of IGA-C, there is the high-order space accuracy. To verify that this property is preserved in the proposed formulation, in Figure~\ref{fig:conv_circ} we show the convergence rates of the $L_2$-norm of the error for different number of collocation points, $\rm n$, and degrees of the B-Splines basis functions, $p$. 
The error is computed at $t=\SI{3}{s}$ over a grid of 9 points as $err_{L_2}=||\f{u}^r-\f{u}^h||/||\f{u}^r||$, where $\f{u}^h$ is the approximated displacement, while $\f{u}^r$ is the reference solution, obtained with $p=8$ and $\rm n=150$. Note that at the time when the error is evaluated the arch has undergone already very large deformations. 
Excellent rates (solid lines), in agreement with the expected ones (dashed lines), are achieved. 
The Newton-Raphson iteration tolerance prevents to reach accuracy below $\SI{10^{-8}}{}$. A super-convergent rate can be observed for $p=8$ and coarse discretization ($\rm n <16$). It is remarkable that that with $p=8$ an accuracy of $\SI{10^{-6}}{}$ is reached with just 16 collocation points, proving the utility of the higher-order formulation.

\subsection{Morphing of a cantilever beam}
\begin{figure}
\centering
\subfigure[Geometry and applied force and couple.\label{fig:cantilever_beam_geom}]
{\begin{overpic}[width=0.8\textwidth]{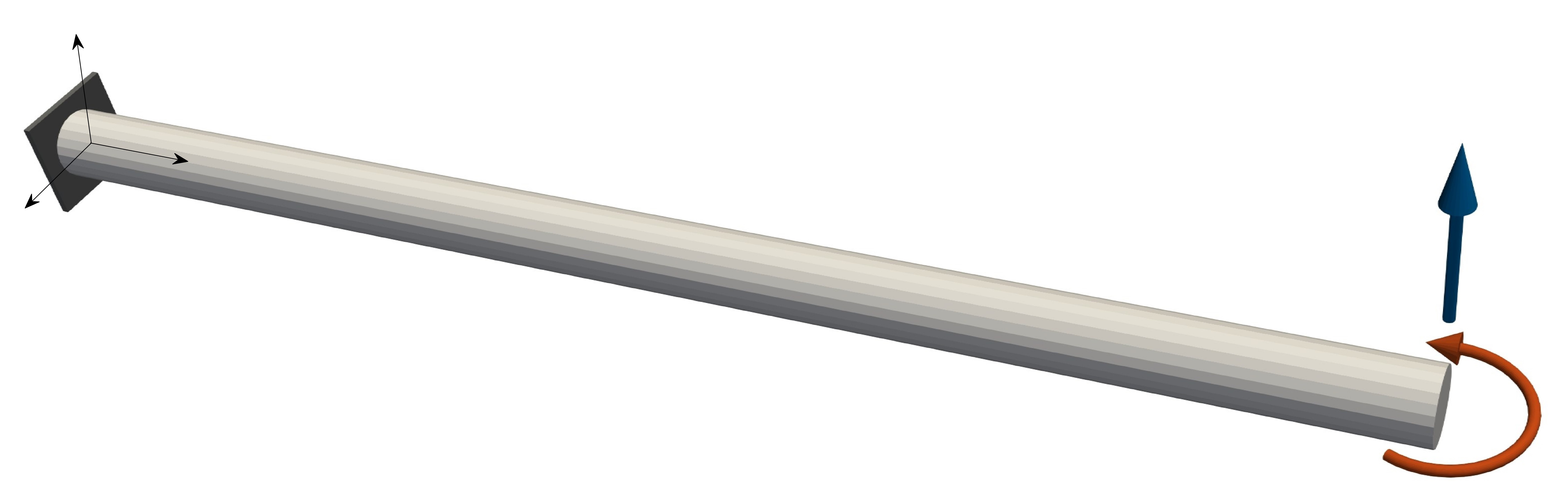}
\put(875,200){\textcolor{blu}{$F_3$}} \put(950,110){\textcolor{aran}{$M_3$}}
\put(0,290){$x_3$}  \put(120,235){$x_2$}  \put(-20,200){$x_1$}
\end{overpic}
}
\subfigure[Temperature time history.\label{fig:Temp_cantilever_1_2}]
{\includegraphics[width=.47\textwidth]{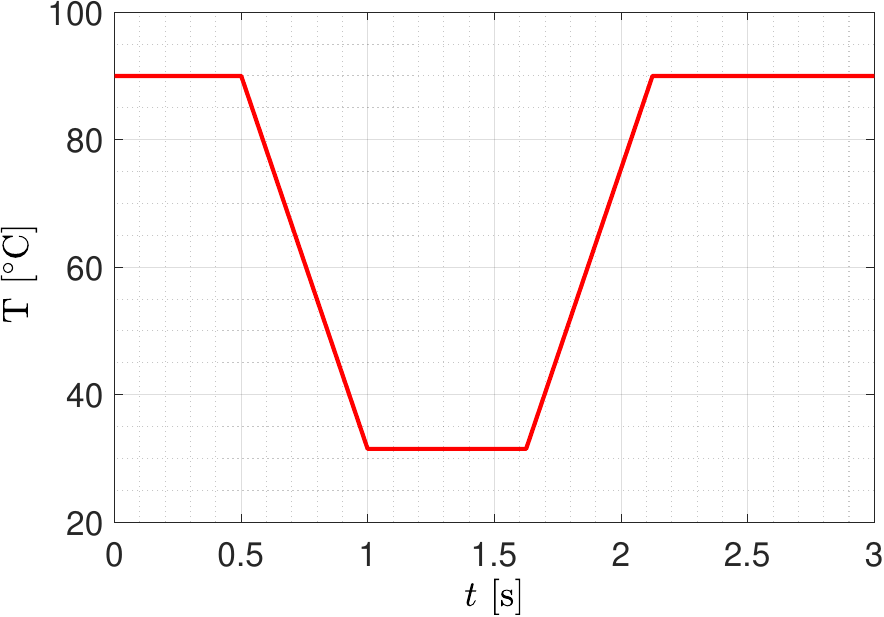}}
\subfigure[Load time history.\label{fig:Load_cantilever_1_2}]
{\includegraphics[width=.50\textwidth]{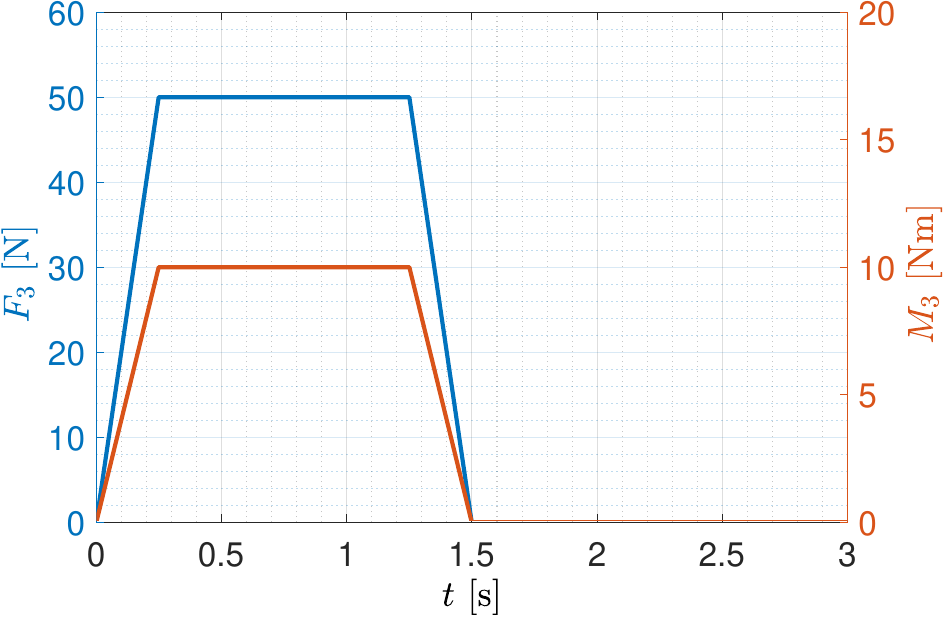}}
\caption{Morphing of a cantilever beam: beam geometry, temperature and load time histories applied during the shape-changing process}\label{fig:cantilever_beam}
\end{figure}

This numerical experiment concerns a straight $\SI{1}{m}$-long cantilever beam placed along the $x_2$-axis, subjected to a vertical tip force, $F_3$, and a concentrated couple, $M_3$. See Figure~\ref{fig:cantilever_beam}. 
The structure features the same cross-section and the same material properties adopted for the circular arch discussed in Section~\ref{subsec:circ_arch}. 
Temperature, as well as $F_3$ and $M_3$, vary as shown in Figure~\ref{fig:Temp_cantilever_1_2} and~\ref{fig:Load_cantilever_1_2}, respectively. 
These conditions reproduce a morphing process where the temporary shape is reached at $t=\SI{0.5}{s}$ and fixed (loads removed) at $t=\SI{1}{s}$, while the shape recovery process starts at $t=\SI{1.625}{s}$.

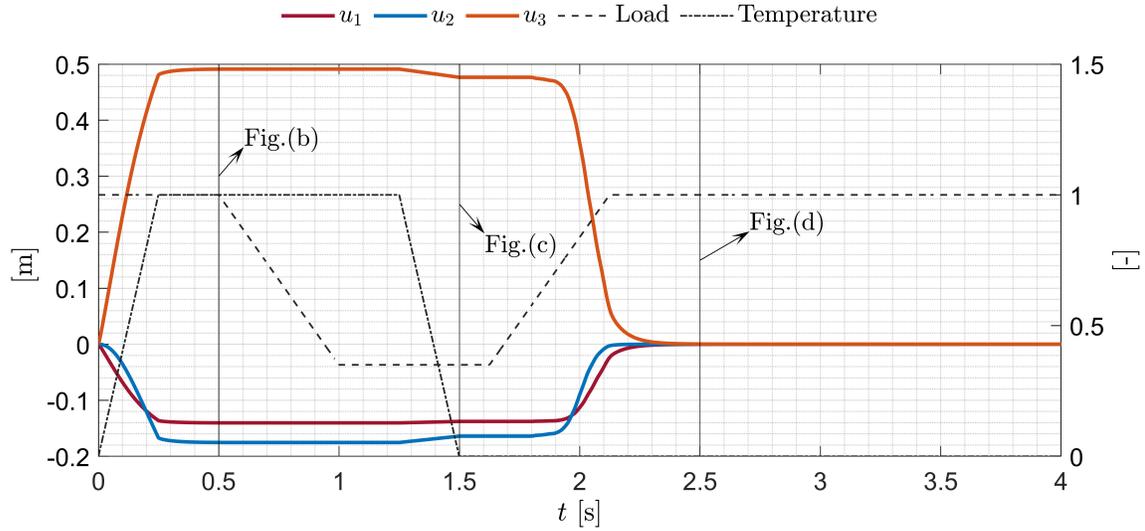
\begin{figure}
\centering
\subfigure[Tip displacements, loads and temperature functions time histories.\label{fig:u_cantilever_1_1}]
{\includegraphics[width=1\textwidth]{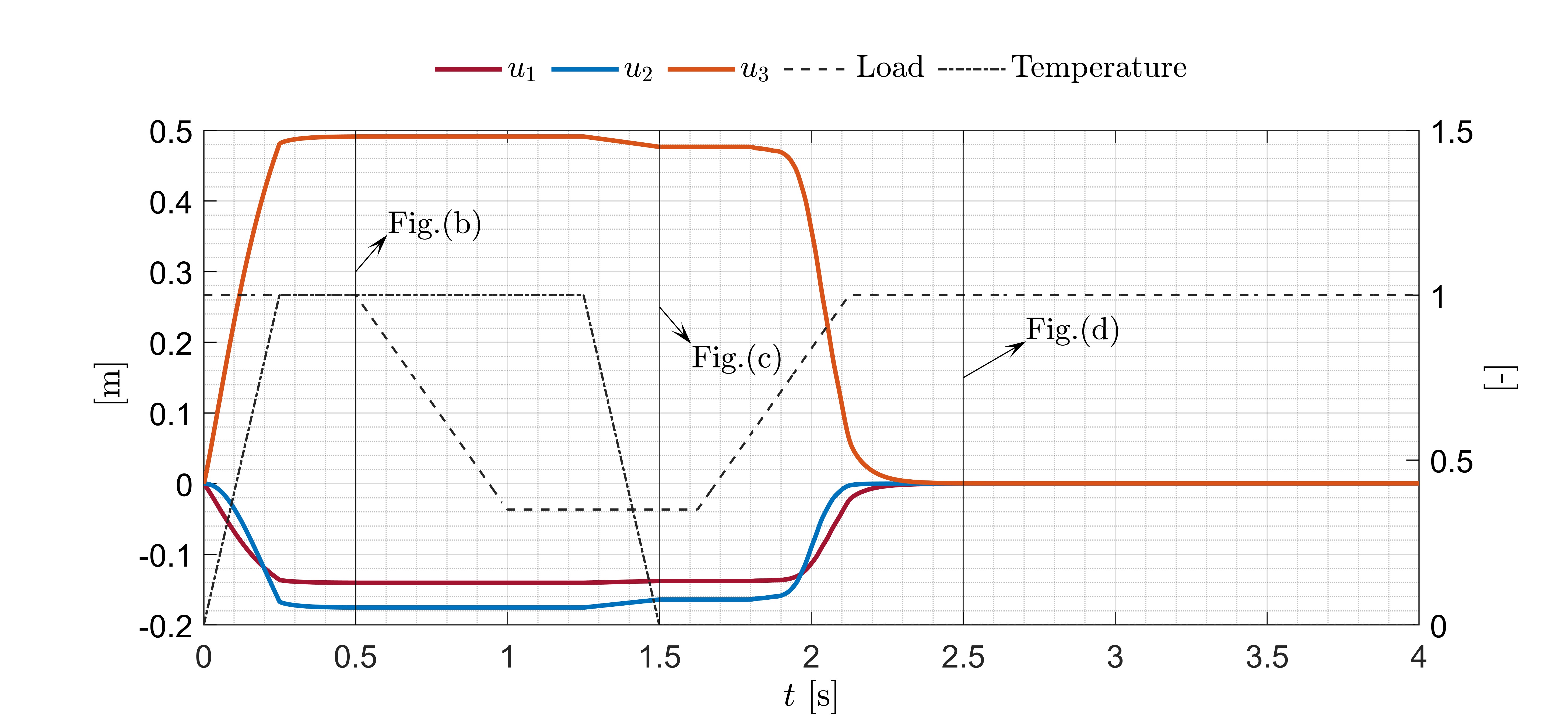}}
\subfigure[$t=\SI{0.5}{s}$: $\nm T=\SI{90}{^\circ \nm C}$, $F_3=\SI{50}{N}$ and $M_3=\SI{10}{Nm}$.\label{fig:mensola_1_2_05}]
{
\begin{overpic}[width=0.25\textwidth]{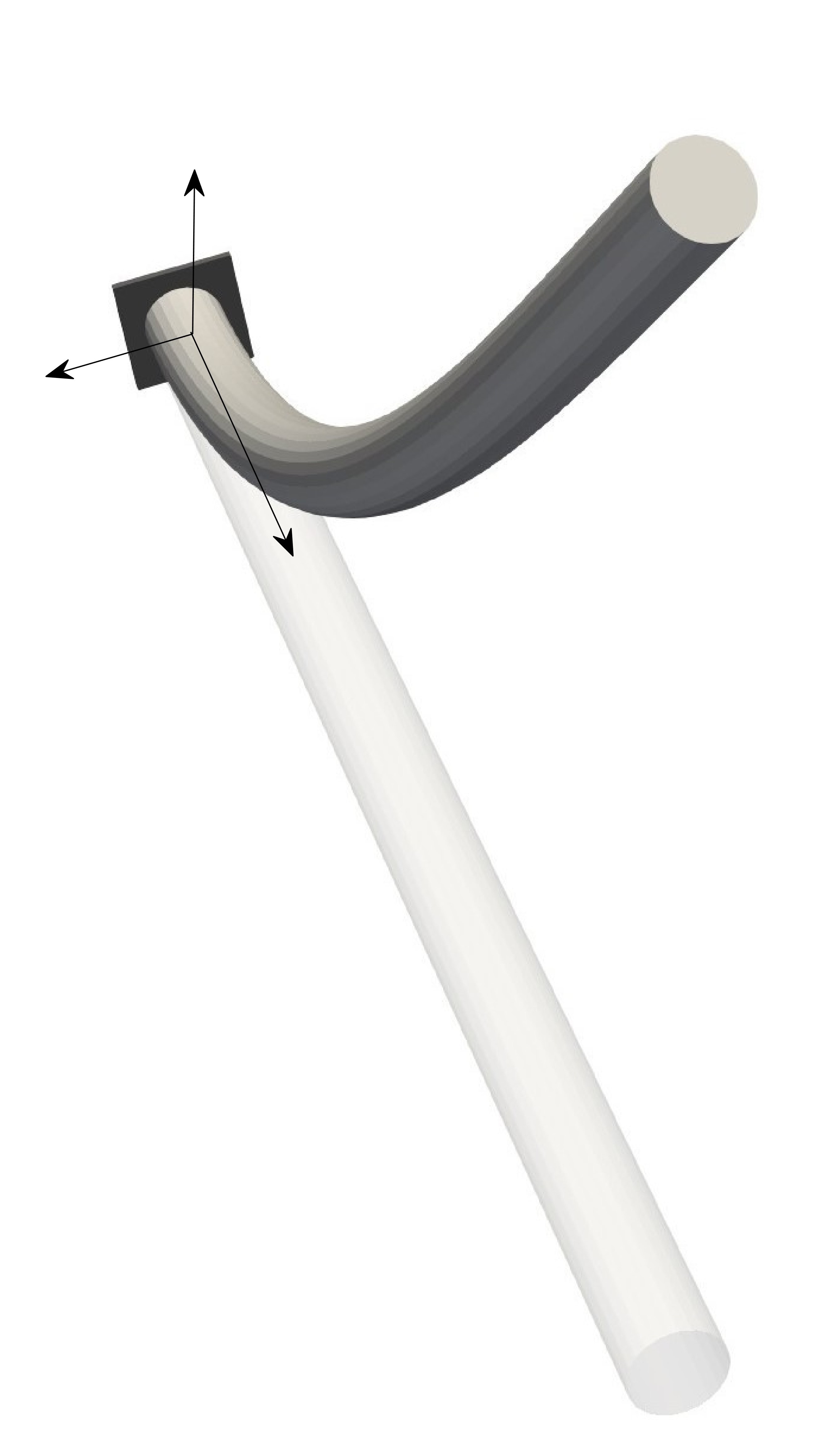}
\put(50,875){$x_3$}  \put(110,600){$x_2$}  \put(-20,700){$x_1$}
\end{overpic}
}\hspace{0.5cm}
\subfigure[$t=\SI{1.5}{s}$: $\nm T=\SI{31.5}{^\circ \nm C}$, $F_3=\SI{0}{N}$ and $M_3=\SI{0}{Nm}$.\label{fig:mensola_1_2_15}]
{
\begin{overpic}[width=0.25\textwidth]{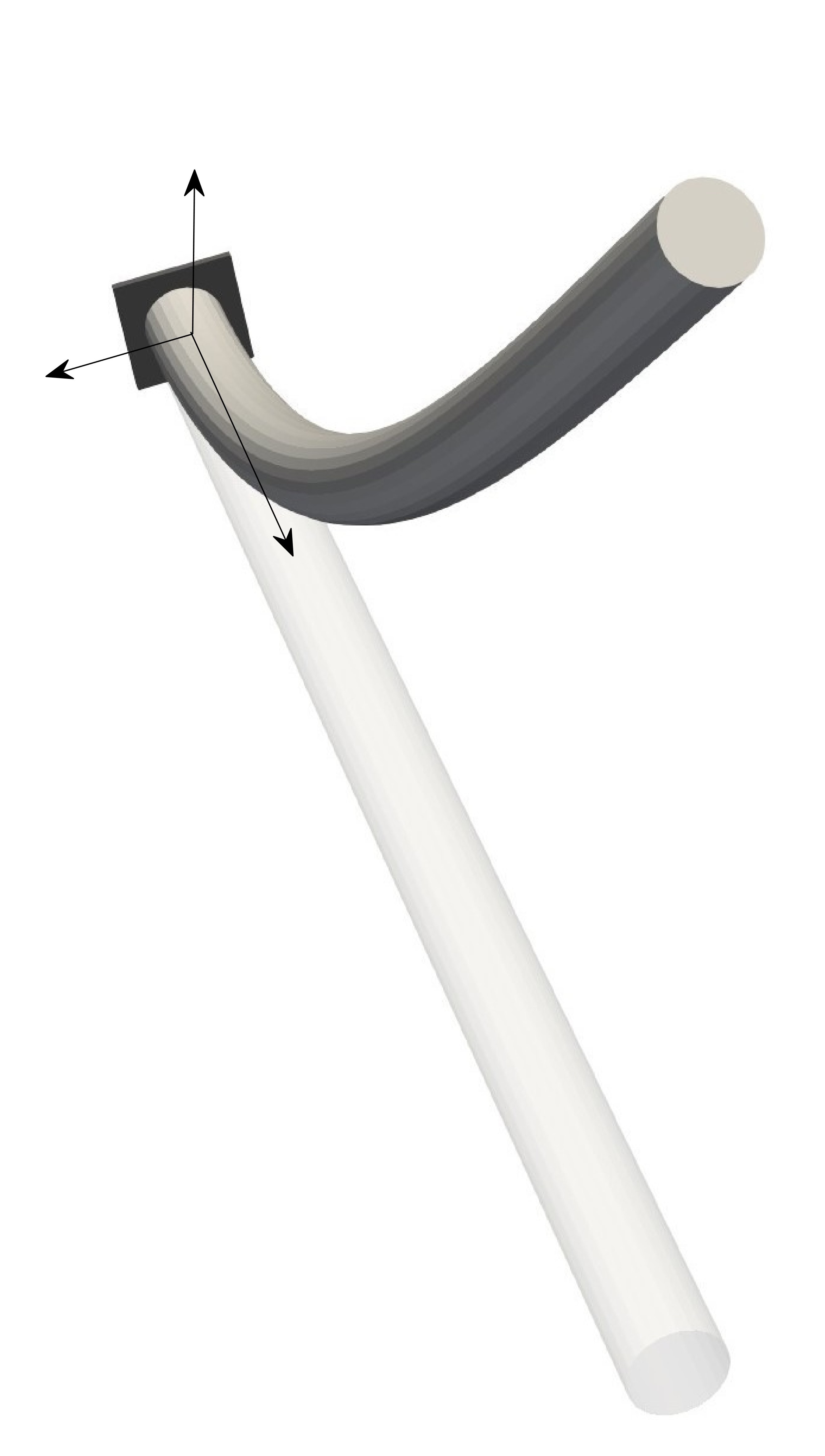}
\put(50,875){$x_3$}  \put(110,600){$x_2$}  \put(-20,700){$x_1$}
\end{overpic}
}\hspace{0.5cm}
\subfigure[$t=\SI{2.5}{s}$: $\nm T=\SI{90}{^\circ \nm C}$, $F_3=\SI{0}{N}$ and $M_3=\SI{0}{Nm}$.\label{fig:mensola_1_2_25}]
{
\begin{overpic}[width=0.25\textwidth]{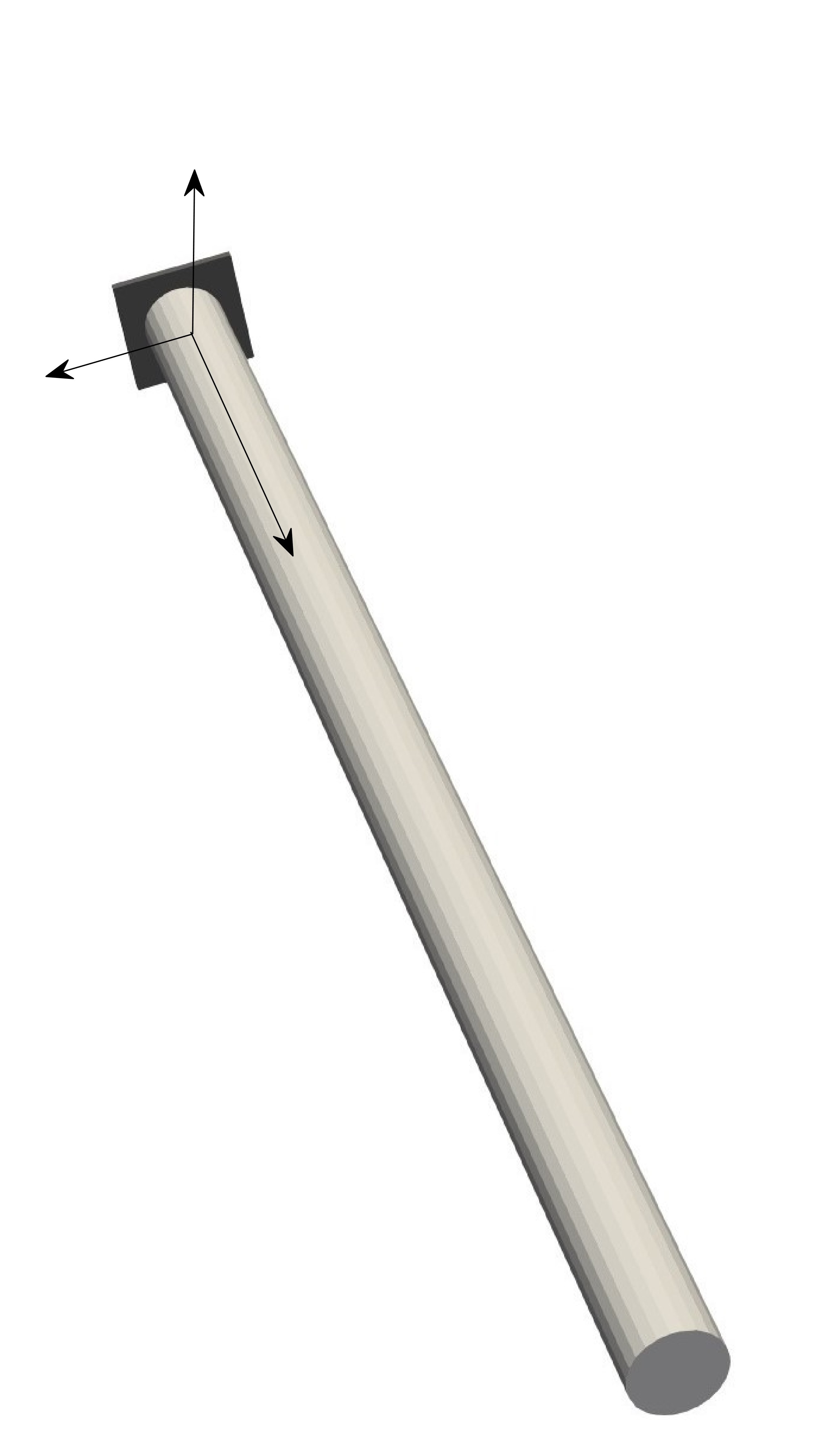}
\put(50,875){$x_3$}  \put(110,600){$x_2$}  \put(-20,700){$x_1$}
\end{overpic}
}
\caption{Morphing of a cantilever beam with tip force, $F_3$, and couple, $M_1$: displacement time histories and snapshots of the deformed configurations.}\label{fig:cantilever_1_2}
\end{figure}

Results are presented in Figure~\ref{fig:cantilever_1_2}. The time history of the tip displacement components (solid lines) is shown in Figure~\ref{fig:u_cantilever_1_1}, where the left $y$-axis refers to the magnitude of the displacements, while the right axis refers to the loads and temperature variation functions (dashed and dashed-dot black lines, respectively), that are defined as $\nm{T}(t)/\nm{max}\{\nm{T}(t)\}$ and  $F_3(t)/\nm{max}\{F_3(t)\}=M_1(t)/\nm{max}\{M_1(t)\}$. 
In Figure~\ref{fig:cantilever_1_2} we also provide three snapshots of the deformed beam: 
$i)$ $t=\SI{0.5}{s}$, Figure~\ref{fig:mensola_1_2_05}; 
$ii)$ $t=\SI{1.5}{s}$, Figure~\ref{fig:mensola_1_2_15}, where loads are completely removed and the (temporary) shape is fixed;
$iii)$ $t=\SI{2.5}{s}$, Figure~\ref{fig:mensola_1_2_25}, where the beam initial shape is fully recovered by increasing the temperature to the initial value. 
It is noted how the model reproduces the full morphing process. 

\subsection{Programming and shape recovery of cardiovascular stent-like structures}\label{sec:programming_stent}
We present two numerical applications of the proposed model to stent-like structures whose geometry has been inspired by~\cite{Zaccaria2020}. 
Stents are complex shape memory devices whose function is, once implanted in the occluded vessel, to reopen the lumen restoring the proper blood flow.
Normally, existing devices have a cylindrical geometry with a straight axis. Our main challenge (see the related project IGA4Stent \cite{MarinoIGA4Stent_webpage} for additional details) is to design programmable devices such that they not only expand radially to reopen the vase, but also morph to self-adapt to the patient-specific vessel morphology. These anatomies are often characterized by high tortuosity, raising a strong need for personalisation. 
To address this problem, we present a second numerical test of a stent-like system with a curved permanent shape. We demonstrate that with the proposed computational scheme, the full shape programming and recovery process can effectively be reproduced. 
We remark that, since the scope of this test is limited to a proof of concept showing just the morphing process and not to simulate real cardiovascular stents, the device geometry is upscaled with respect to real devices and the material has the same PLA properties presented in Table~\ref{tab:PLA} \cite{VanManen_2021}, which are clearly not suitable for realistic applications due to the high glass transition temperature. 

\begin{figure}
\centering
\subfigure[Two-patch wire.\label{fig:wire}]
{\includegraphics[width=0.21\textwidth]{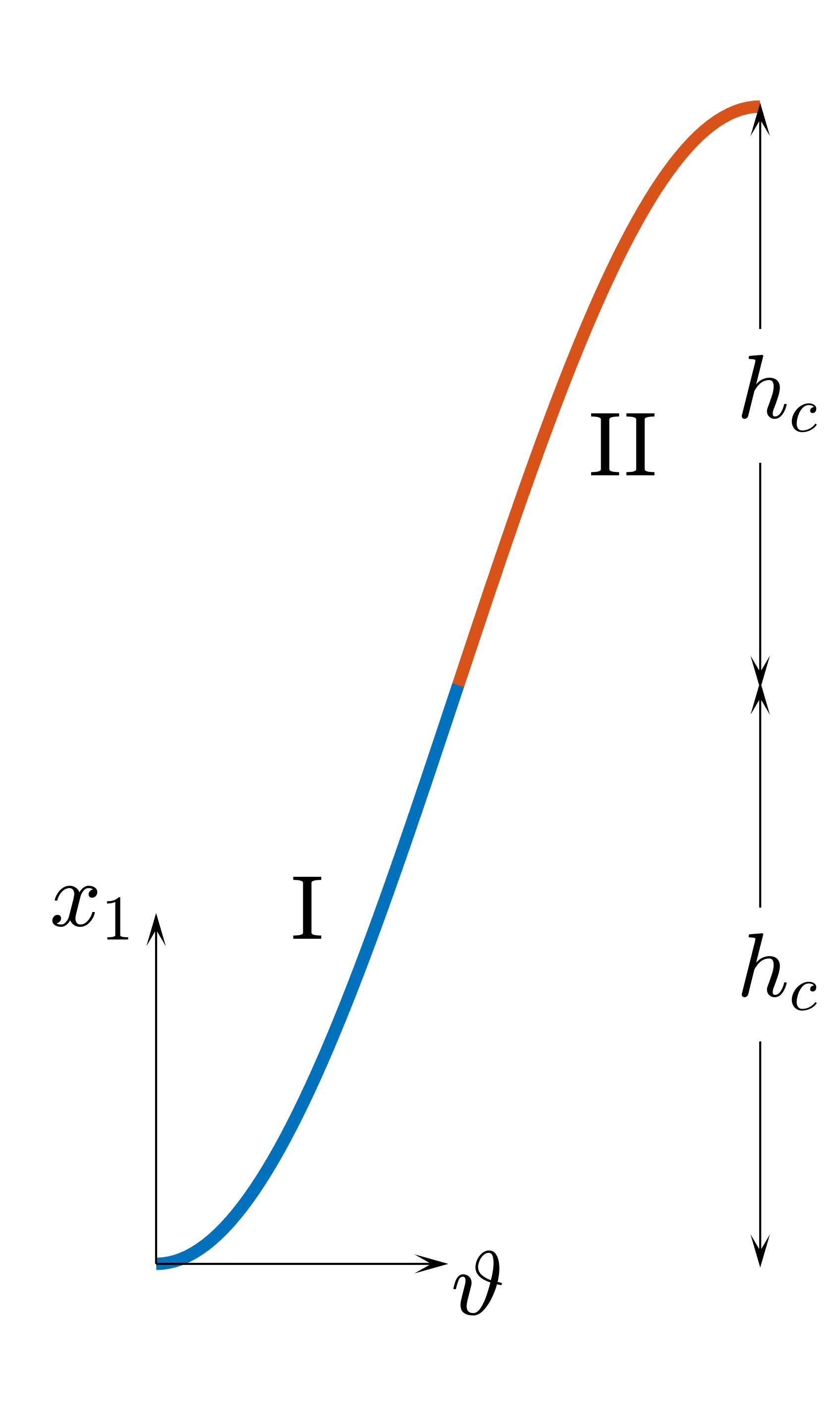}}
\subfigure[Reference crown in the ($\vartheta,x_1$) plane.\label{fig:straight_crown}]
{\includegraphics[width=0.5\textwidth]{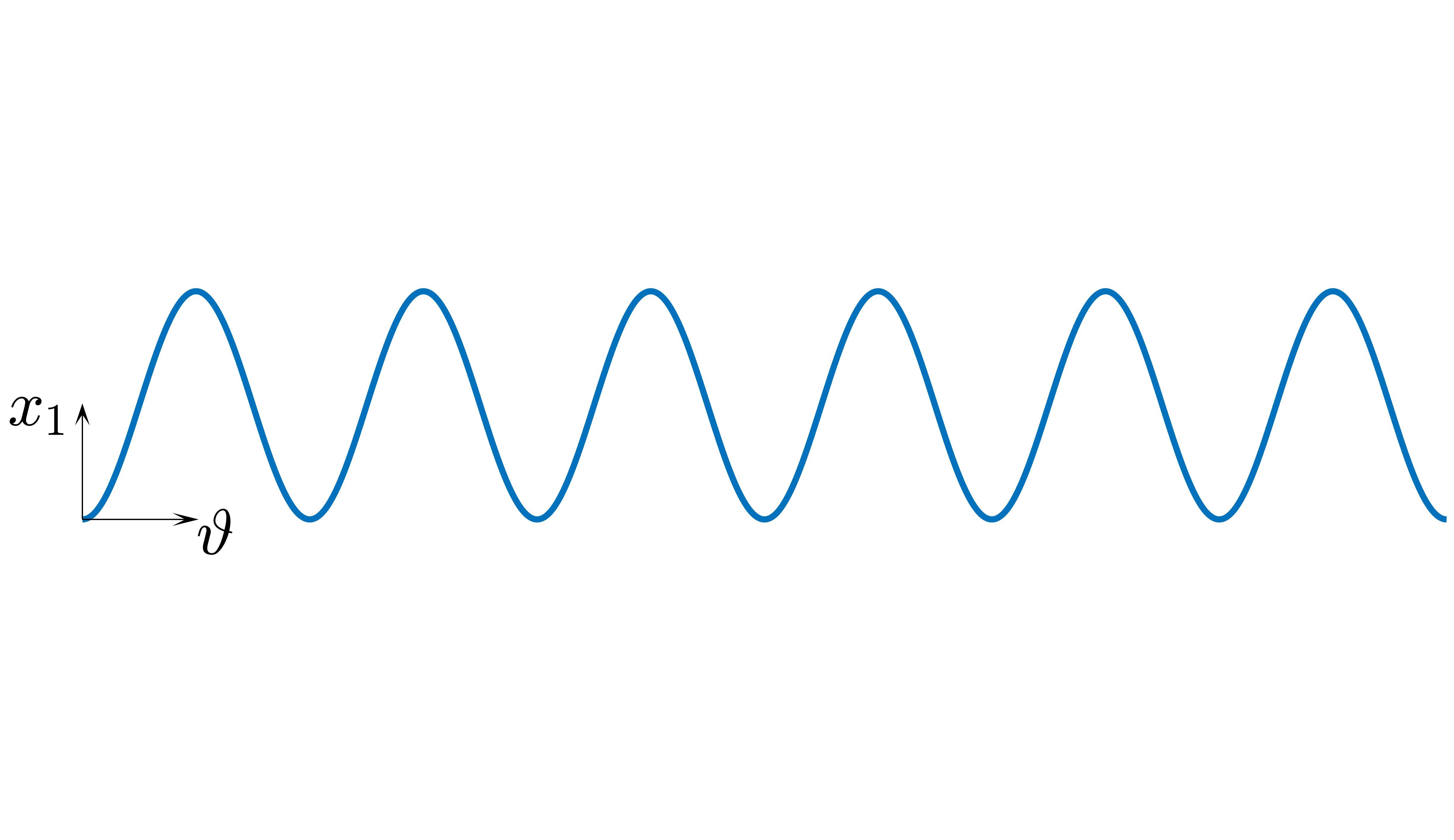}}
\subfigure[Reference crown.\label{fig:crown_cyl}]
{\includegraphics[width=0.25\textwidth]{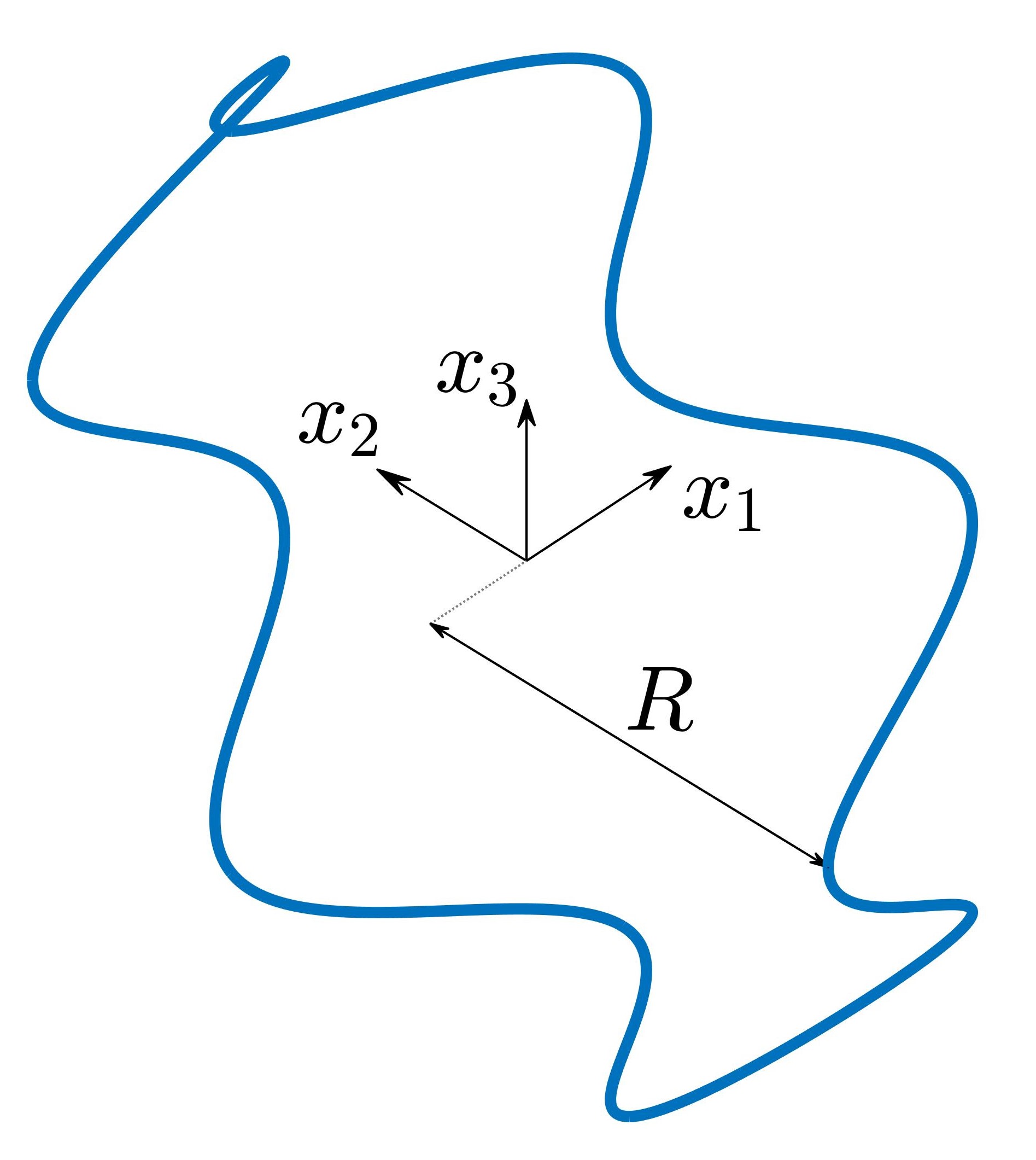}}
\caption{Parametric construction of the reference crown of radius $R$ and height $2h_c$, starting from a two-patch, sinusoidal wire.}\label{fig:parametric_crown}
\end{figure}

For both numerical tests, the initial geometry is constructed parametrically starting from an elementary wire, repeated in a Cartesian reference system ($\vartheta,x_1$), and then projected in cylindrical coordinates to form a reference stent crown in three-dimensional space $(x_1,x_2,x_3)$. 
Additional crowns are then created by duplicating the initial one along the stent axis at a distance $d_c$, and twisted of an angle $\pi/2$. The full geometry is finally obtained connecting crests and troughs of the crowns with bridges. The full process is summarized in Figure~\ref{fig:parametric_crown}.

\subsubsection{Radial morphing of a straight device}\label{radial_morph}
\begin{figure}
\centering
\includegraphics[width=0.75\textwidth]{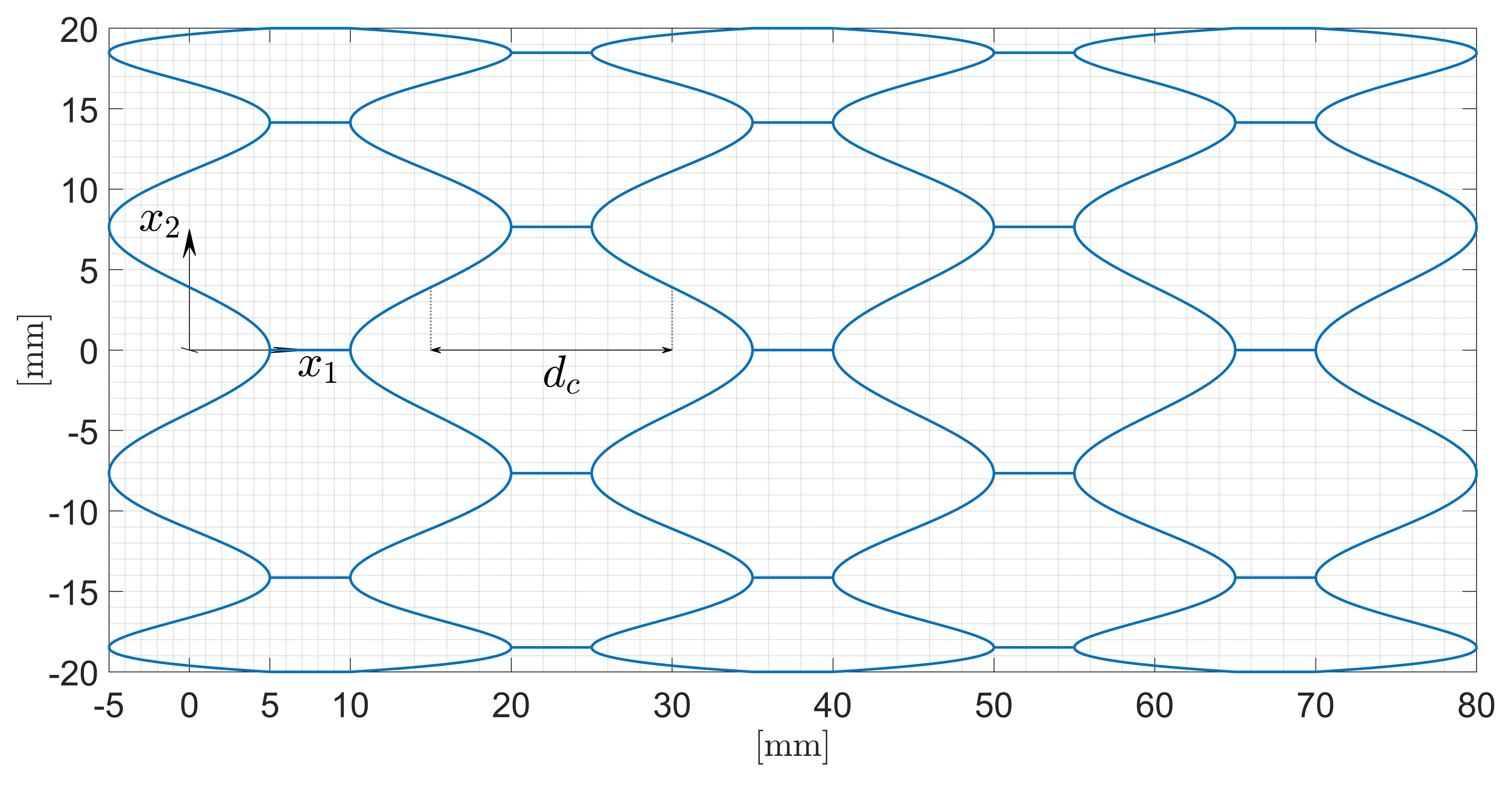}
\caption{Plane view ($x_1,x_2$) of the straight device.}\label{fig:plane_straight_stent}
\end{figure}

In this numerical application, we consider a stent-like tube made of 6 crowns of radius $R=\SI{20}{mm}$ and height $2h_c=\SI{10}{mm}$, spaced along the $x_1$-axis of $d_c=\SI{15}{mm}$ (see Figure~\ref{fig:plane_straight_stent}). 
Each wire has a circular cross section of diameter $d=\SI{0.6}{mm}$.
Both patches composing the single wire (see Figure \ref{fig:wire}) and the rectilinear bridges are discretized with B-Spline basis functions with $p=6$ and $\nm{n}=20$. 

We investigate the morphing of the device imposing a radial contraction, $\Delta r$, of $\SI{15}{mm}$. This condition is enforced through Dirichlet BCs (Eq.~\eqref{eq:bcseta_or_n}) at the nodes where different patches are connected.

\begin{figure}
\centering
\subfigure[3D view.\label{fig:1_4_3D}]
{\includegraphics[width=0.7\textwidth]{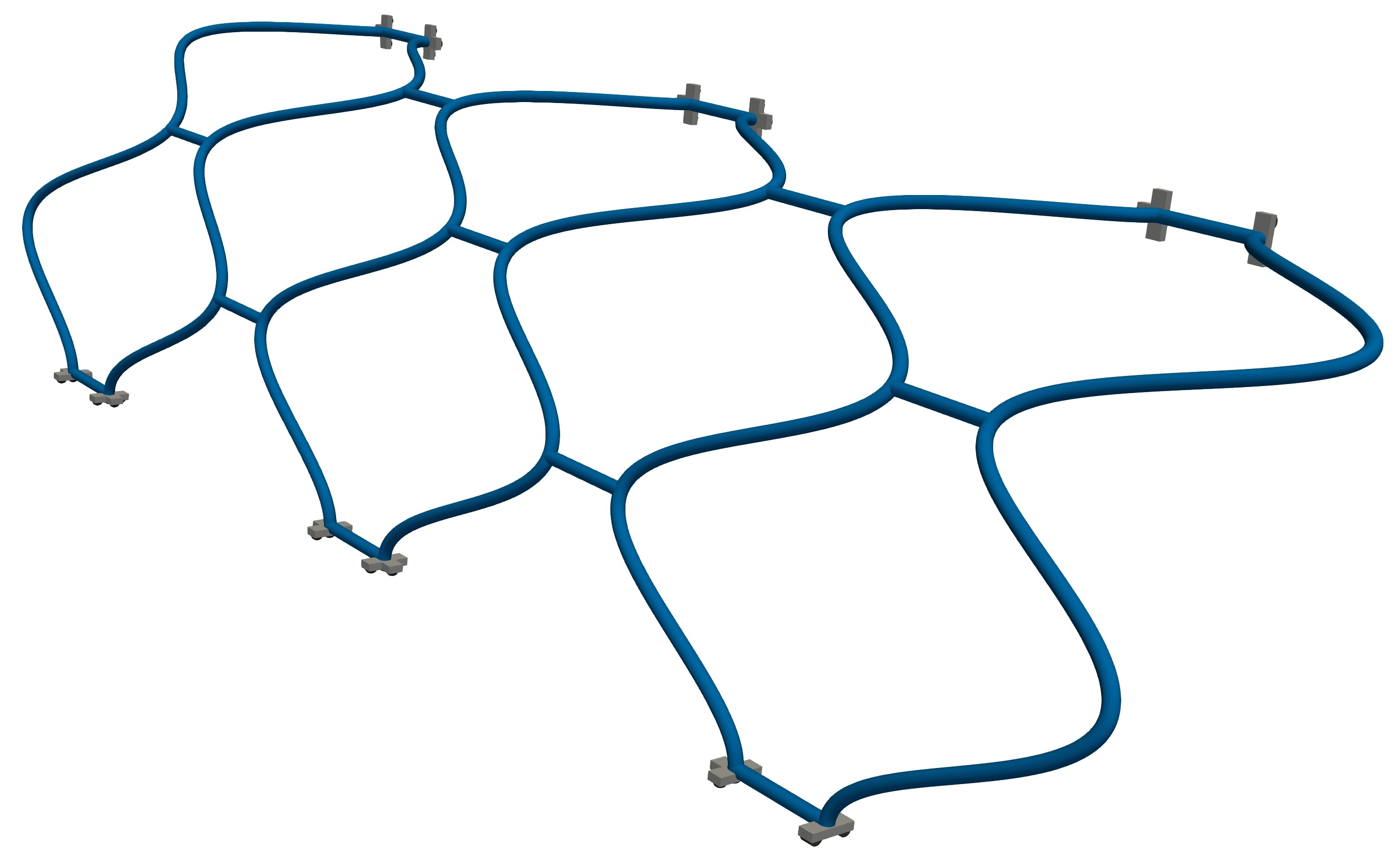}}
\subfigure[Plane ($x_1,x_2$) view.\label{fig:1_4_xy}]
{\includegraphics[width=0.77\textwidth]{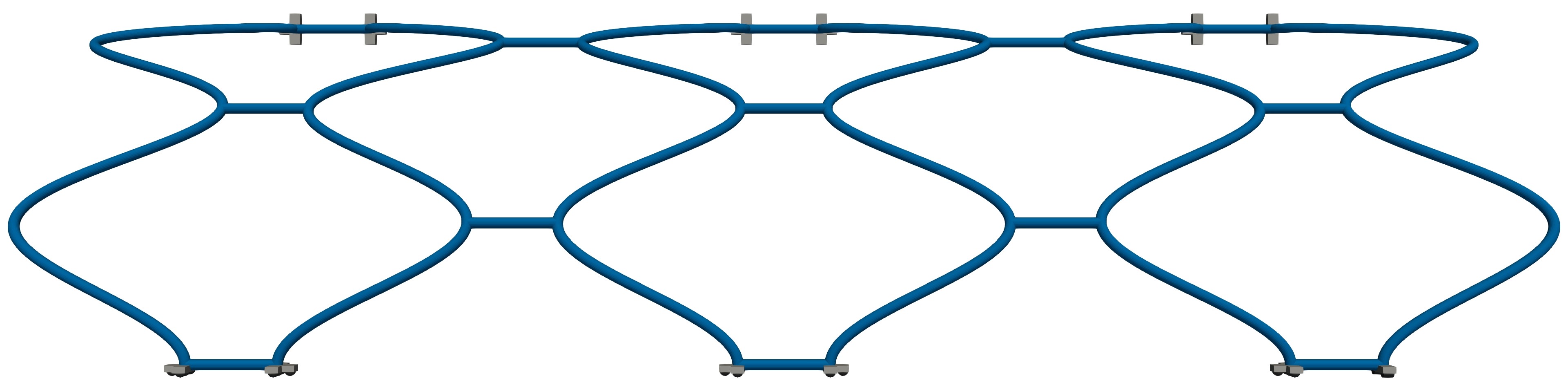}}
\subfigure[Plane ($x_2,x_3$) view.\label{fig:1_4_yz}]
{\includegraphics[width=0.215\textwidth]{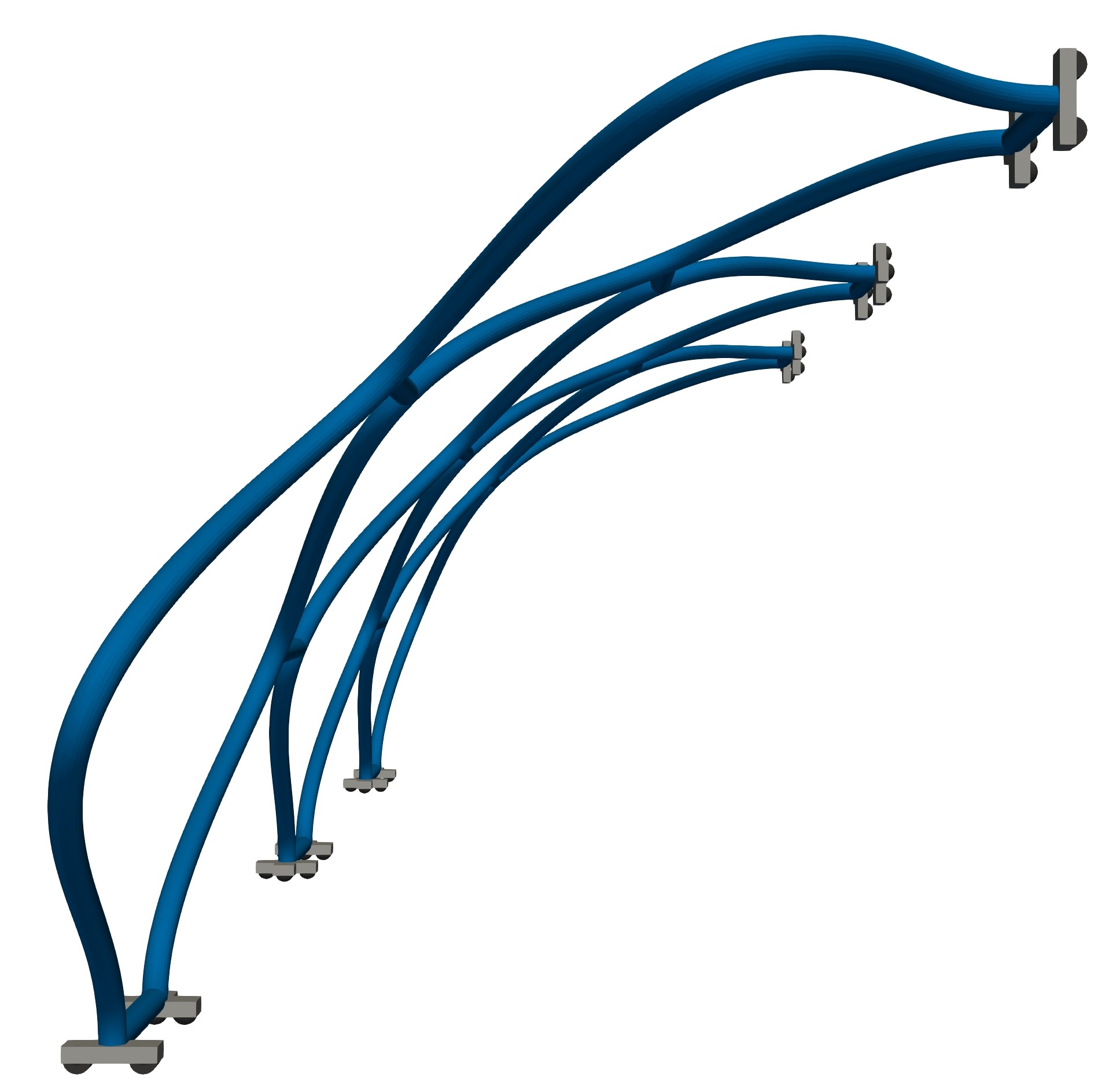}}
\caption{One-quarter of the simulated structure.\label{fig:1_4_stent}} 
\end{figure}

Given the axial-symmetry of the problem, we analyse only one-quarter of the structure, enforcing symmetry constraints on the ($x_1$,$x_2$) and ($x_1$,$x_3$) planes (see Figure~\ref{fig:1_4_stent}). 
The simulation time is set to $\SI{3.25}{s}$, with time step size $h=\SI{2.5e-3}{s}$.

\begin{figure}
\centering
\includegraphics[width=0.75\textwidth]{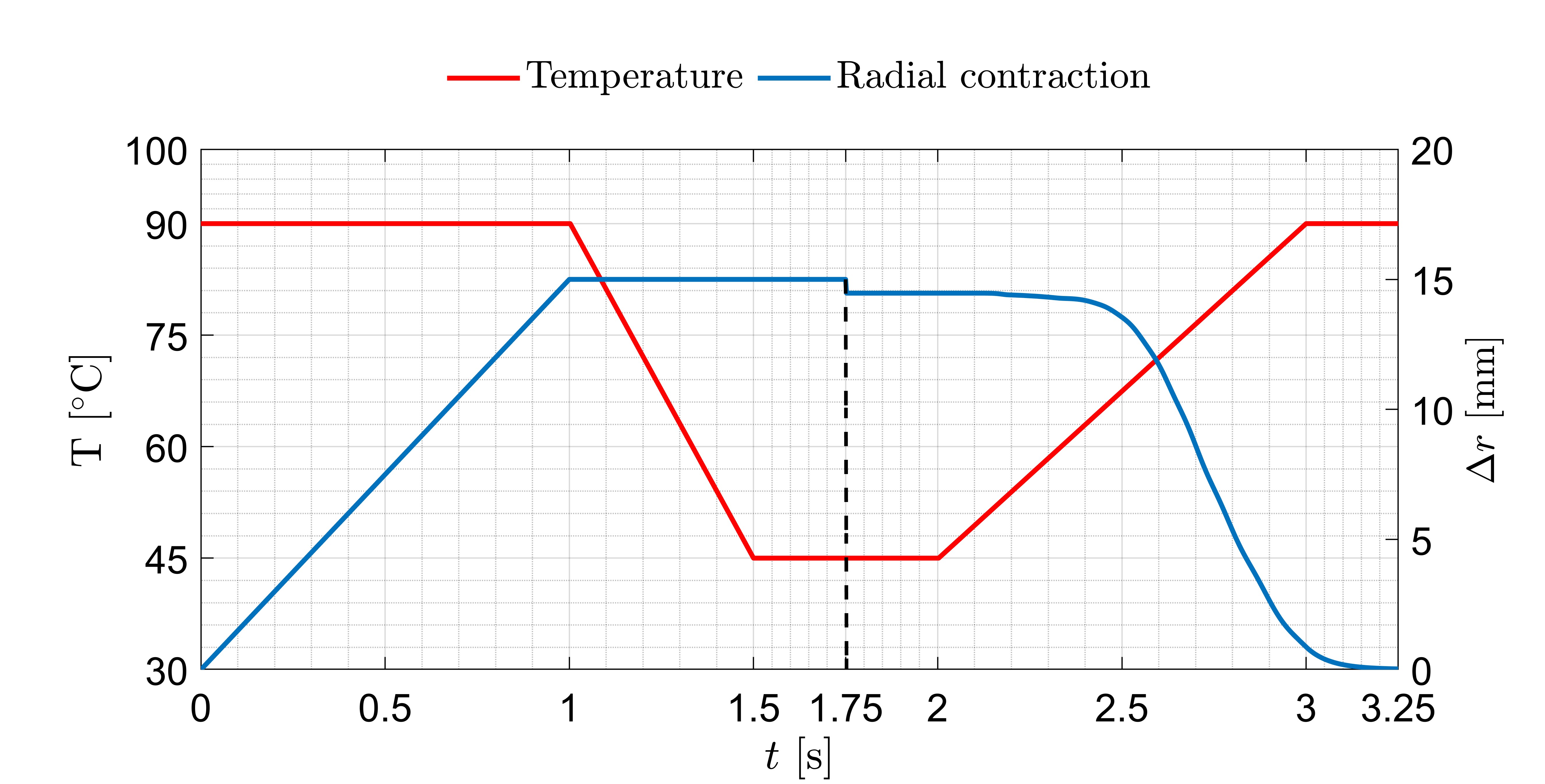}
\caption{Time histories of temperature (red solid line) and radial contraction, $\Delta r$ (blue solid line).}\label{fig:load_straight_stent}
\end{figure}

Temperature and radial contraction time histories are shown in Figure~\ref{fig:load_straight_stent}. The radius variation $\Delta r$ is applied with a linear ramp from $\SI{0}{s}$ to $\SI{1}{s}$ and then kept constant during the shape-fixing phase, where the temperature is reduced from $\SI{90}{^\circ C}$ to $\SI{45}{^\circ C}$, up to $\SI{1.75}{s}$. After this time, the imposed displacements are removed by switching from Dirichlet to free-end Neumann BCs at the interface nodes. Since mainly viscous deformations have been stored during this process, and relaxation times are significantly increased, the structure exhibits an expected small elastic snap during the BCs switch. 
When the system is heated again with a temperature increase reaching $\SI{90}{^\circ C}$, the device fully recovers the programmed shape ($\Del r = 0$, see Figure~\ref{fig:load_straight_stent}). 

\begin{figure}
\centering
\includegraphics[width=0.475\textwidth]{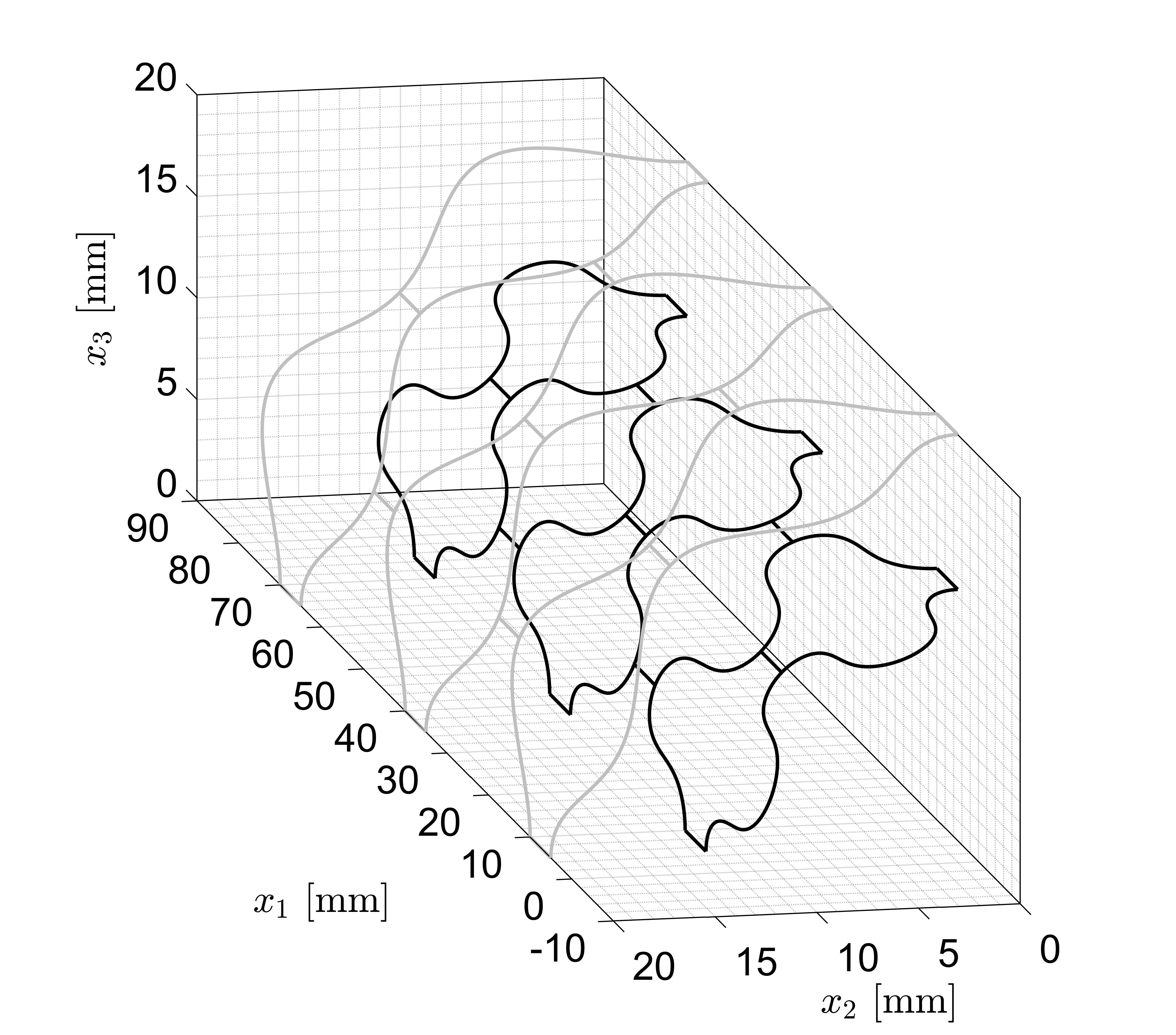}
\hspace{0.025\textwidth}
\includegraphics[width=0.475\textwidth]{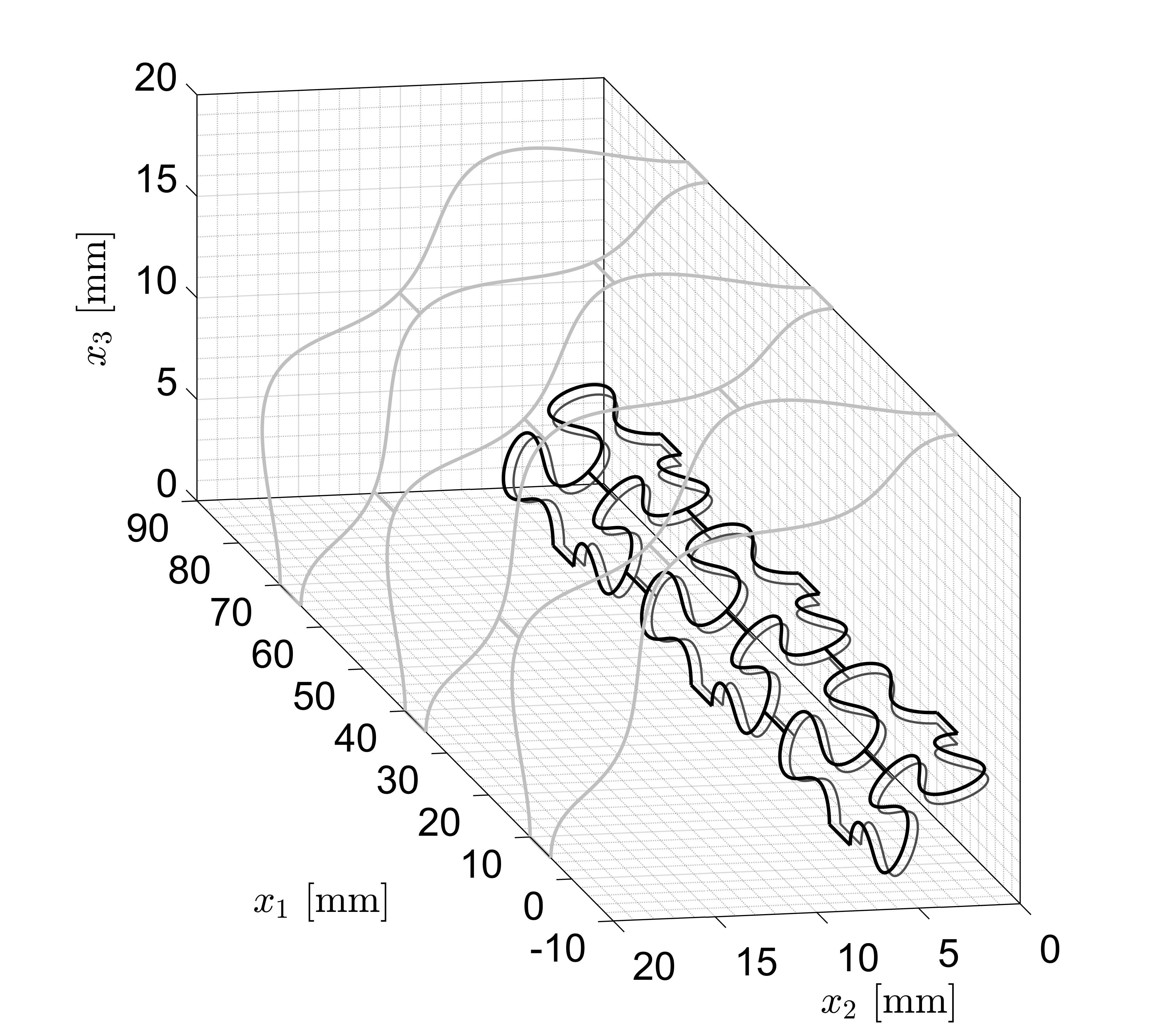}
\subfigure[3D (top) and planar ($x_1,x_3$) (bottom) views at $t=\SI{0.5}{s}$. \label{fig:05_Stent_config}]{
\includegraphics[width=0.475\textwidth]{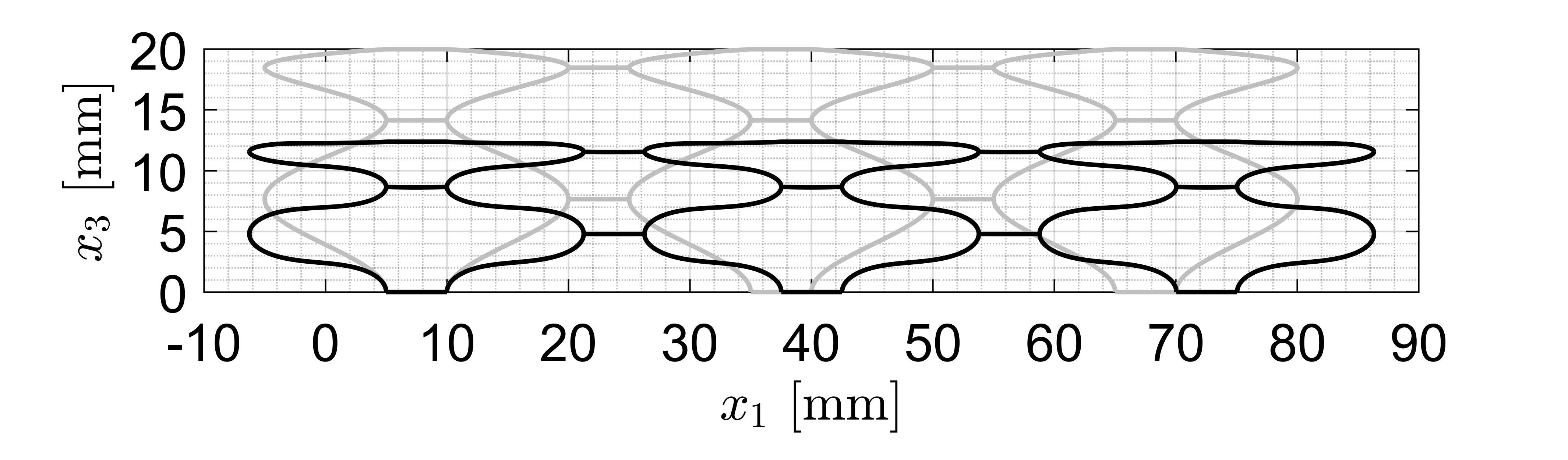}}
\hspace{0.025\textwidth}
\subfigure[3D (top) and planar ($x_1,x_3$) (bottom) views at $t=\SI{1.75}{s}$ (dark grey) and $t=\SI{1.7525}{s}$ (black). \label{fig:175_Stent_config}]{\includegraphics[width=0.475\textwidth]{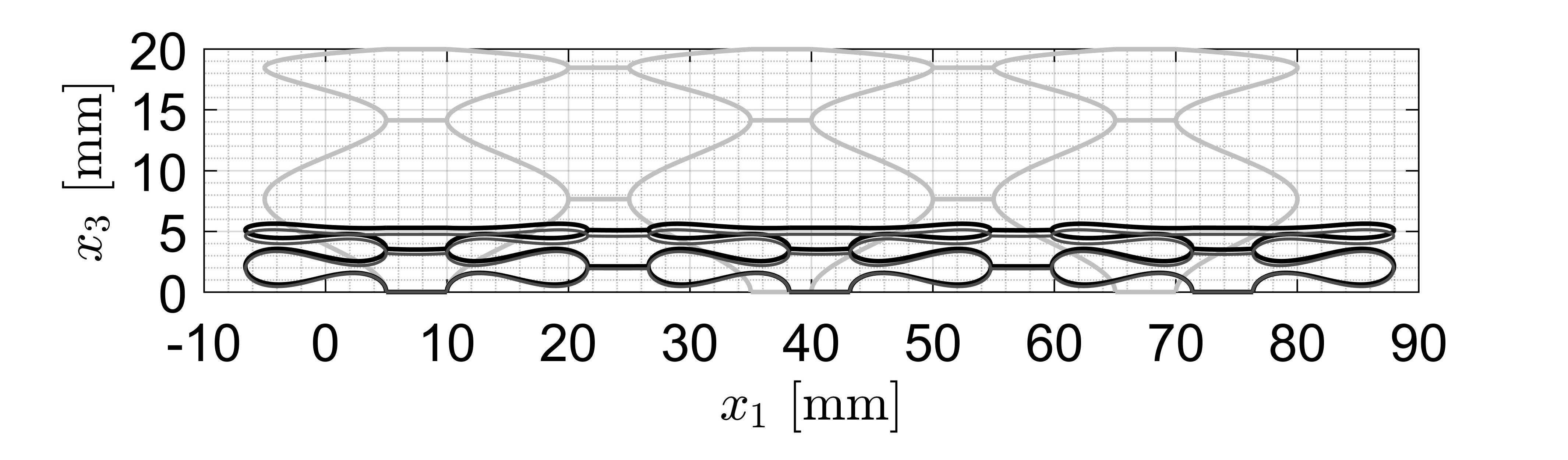}}
\includegraphics[width=0.475\textwidth]{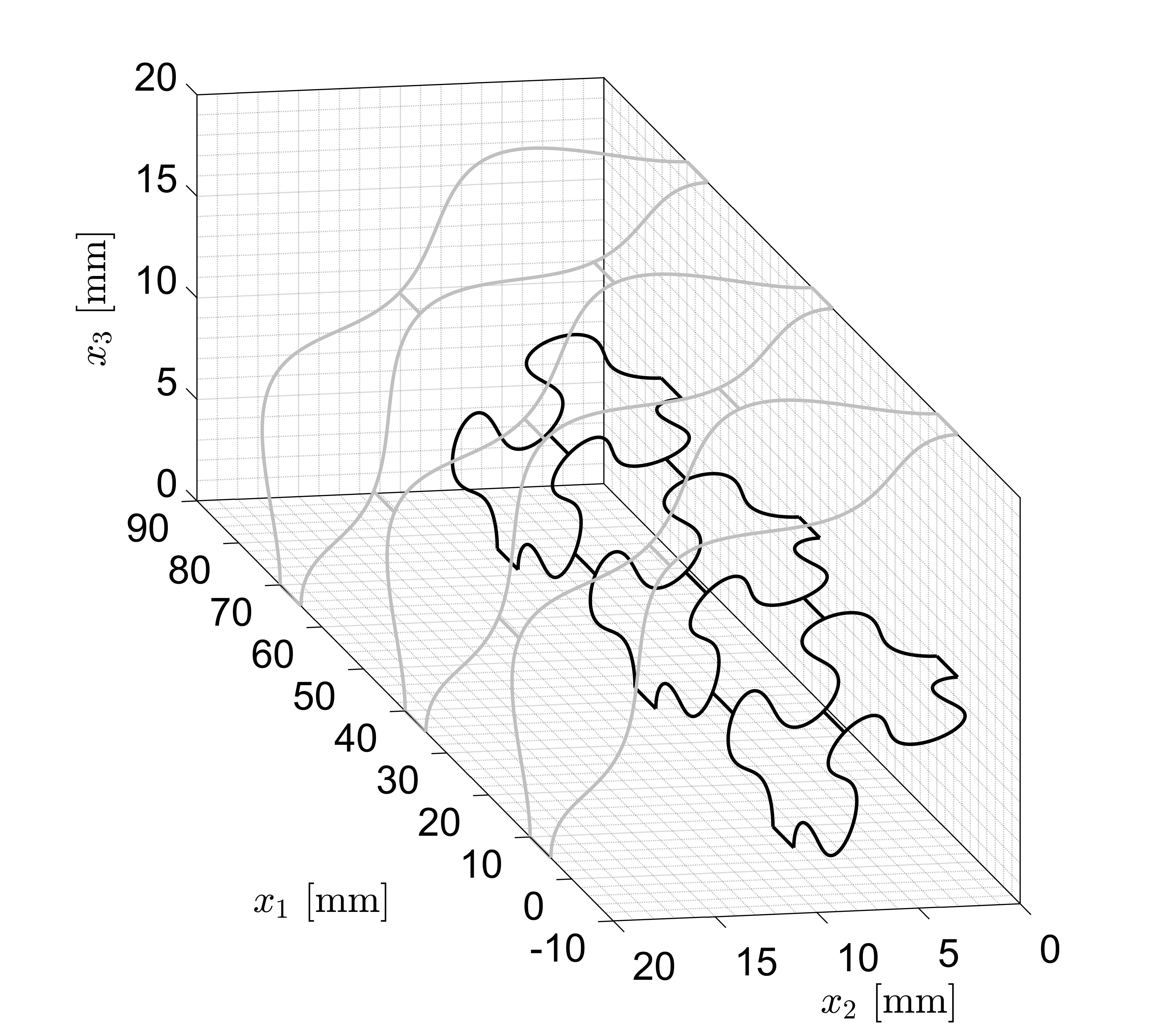}
\includegraphics[width=0.475\textwidth]{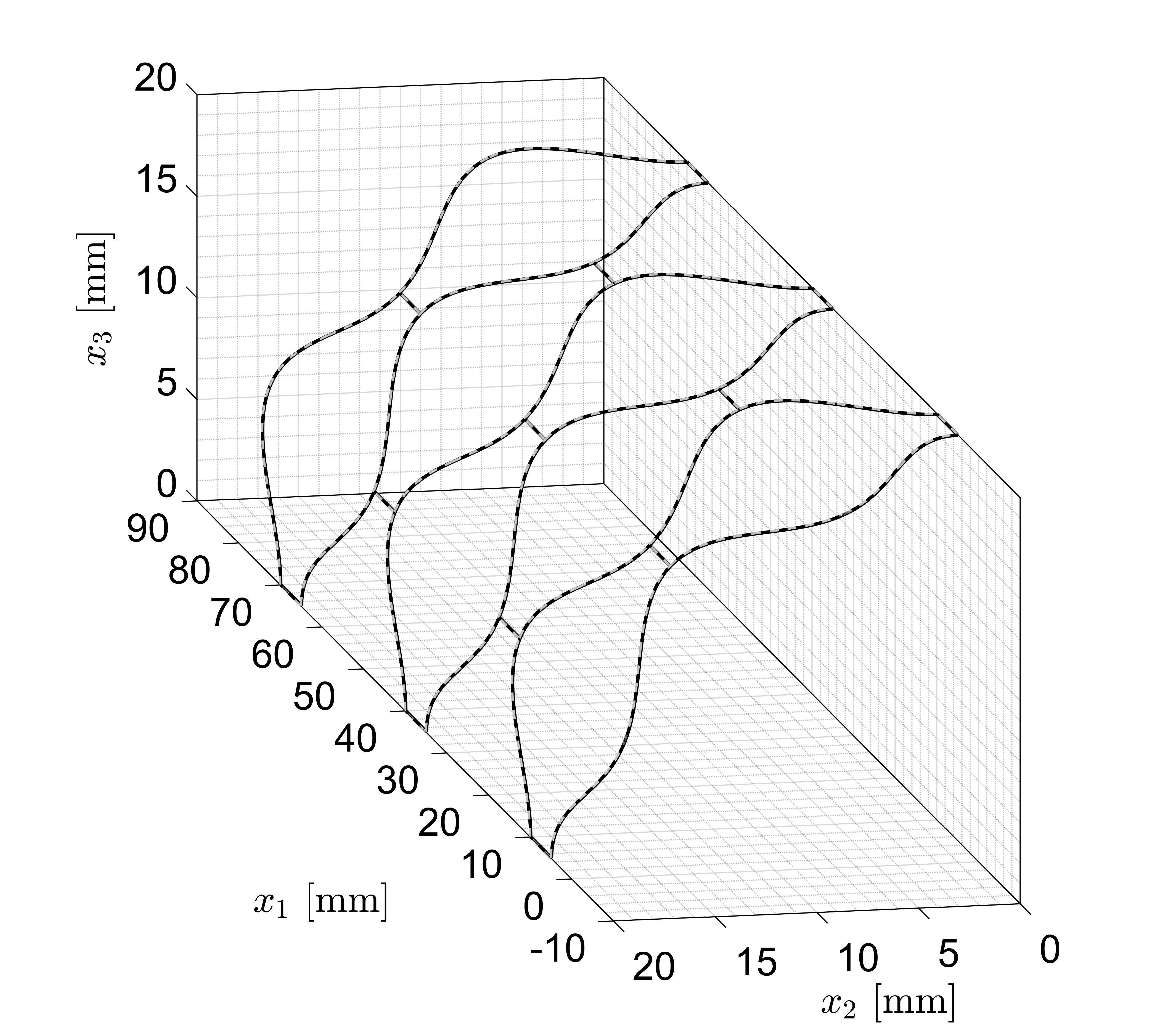}
\subfigure[3D (top) and planar ($x_1,x_3$) (bottom) views at $t=\SI{2.6}{s}$. \label{fig:27_Stent_config}]{
\includegraphics[width=0.475\textwidth]{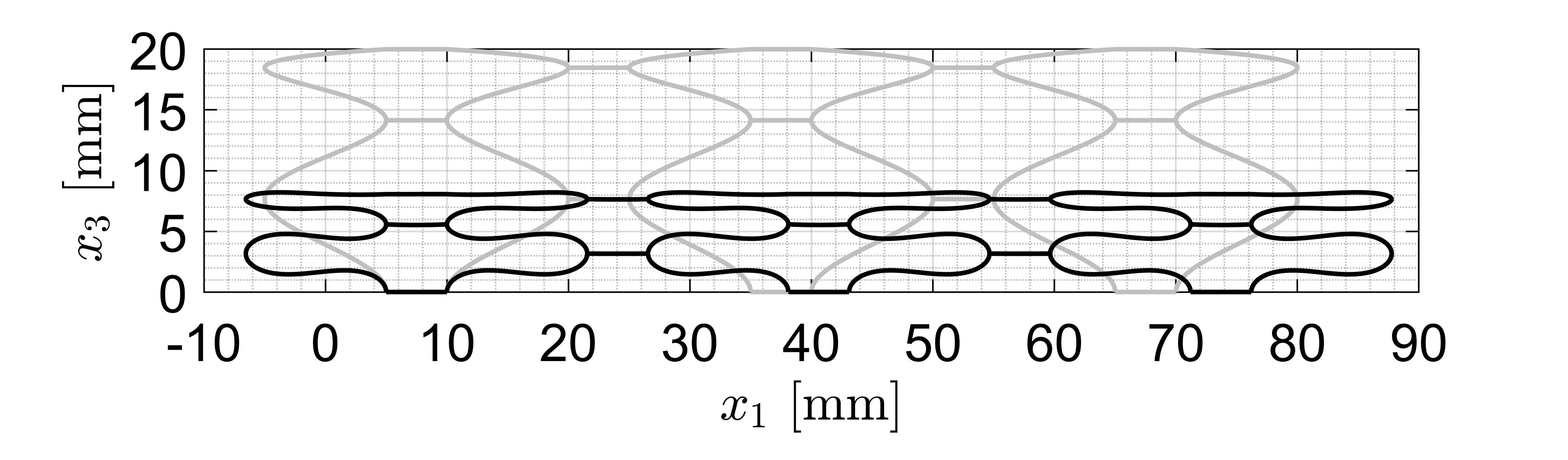}}
\hspace{0.025\textwidth}
\subfigure[3D (top) and planar ($x_1,x_3$) (bottom) views at $t=\SI{3.25}{s}$. \label{fig:325_Stent_config}]{\includegraphics[width=0.475\textwidth]{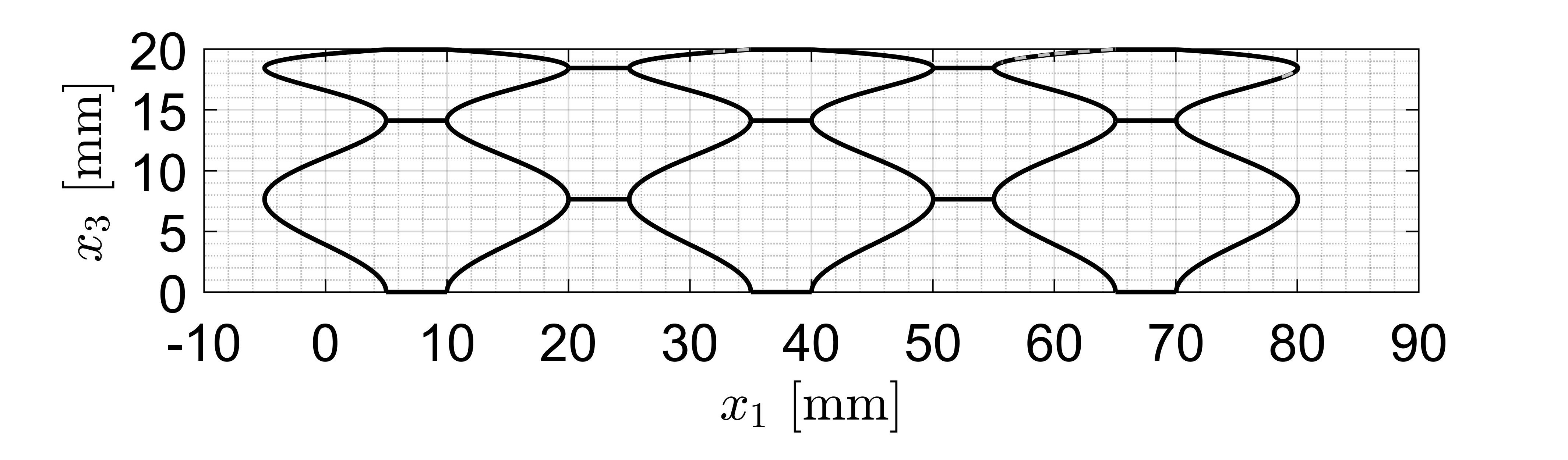}}
\caption{Snapshots of significant deformed configurations.}\label{fig:straight_stent_def}
\end{figure}

\begin{figure}
\centering
\subfigure[$t=\SI{0.0}{s}$, $\nm{T}=\SI{90}{^\circ \nm{C}}$.\label{fig:00_Stent_ext}]{\includegraphics[width=0.3\textwidth]{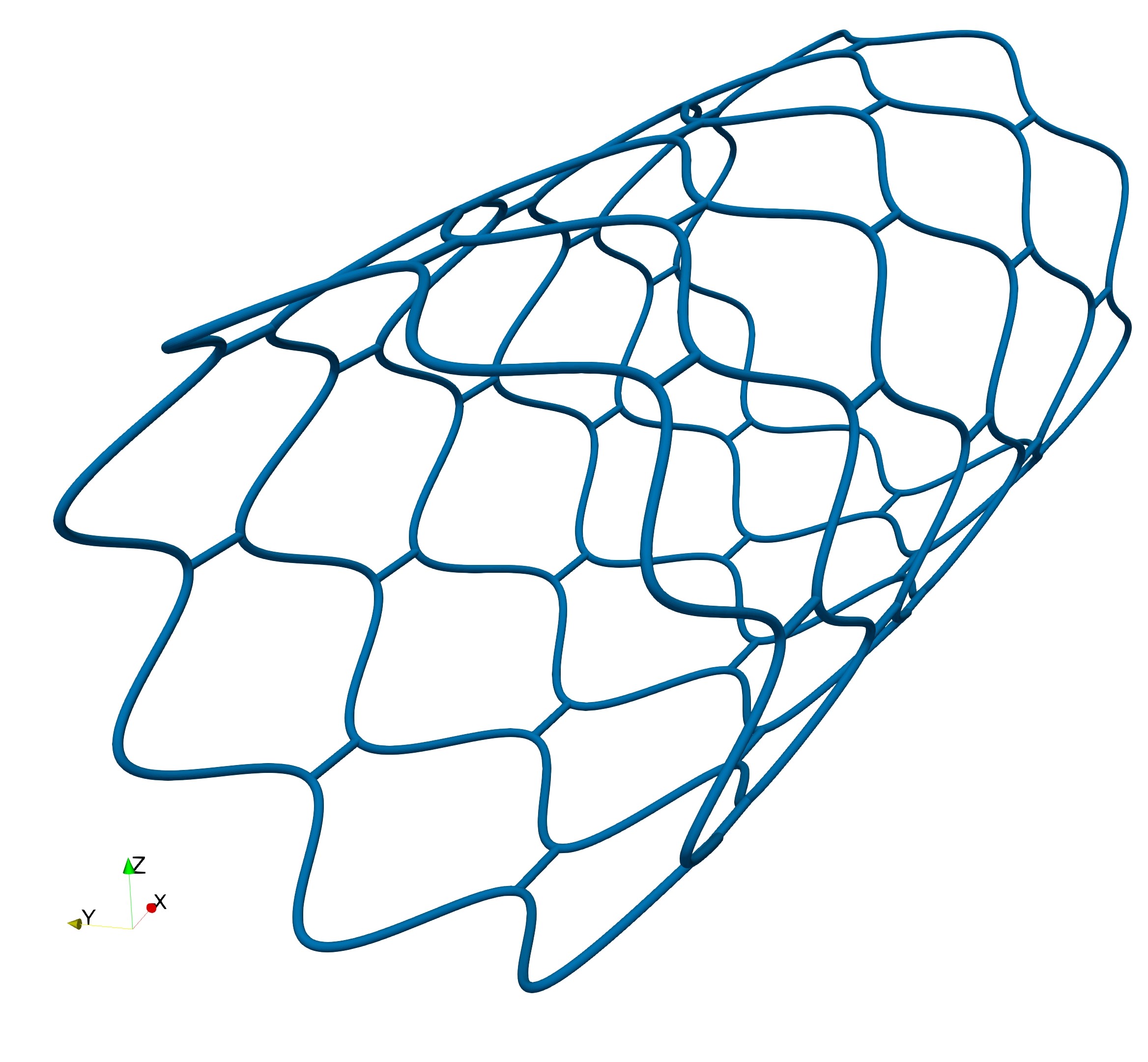}}
\hspace{0.03\textwidth}
\subfigure[$t=\SI{0.5}{s}$, $\nm{T}=\SI{90}{^\circ \nm{C}}$.\label{fig:05_Stent_ext}]{\includegraphics[width=0.3\textwidth]{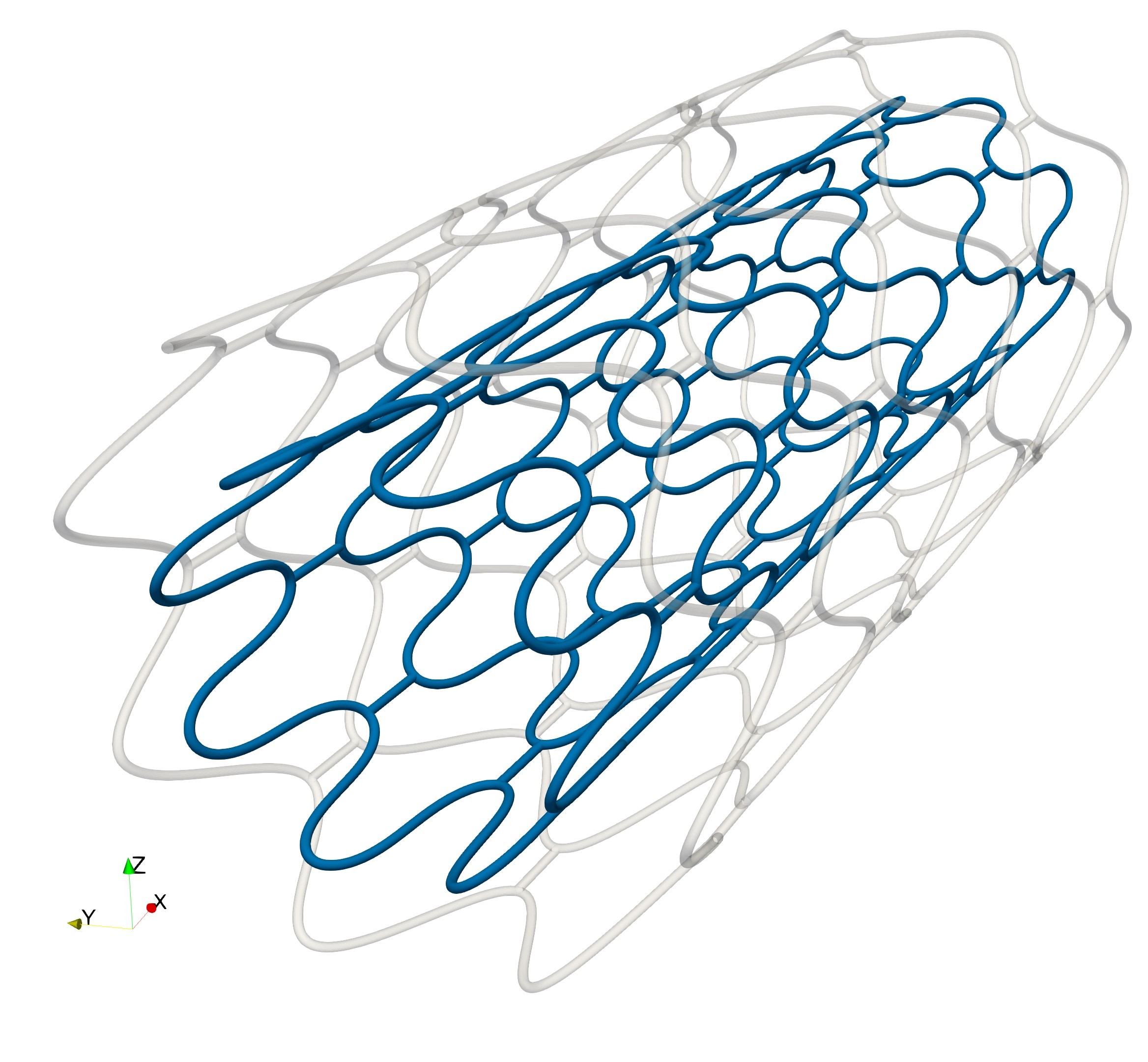}}
\hspace{0.03\textwidth}
\subfigure[$t=\SI{1.75}{s}$, $\nm{T}=\SI{45}{^\circ \nm{C}}$. \label{fig:175_Stent_ext}]{\includegraphics[width=0.3\textwidth]{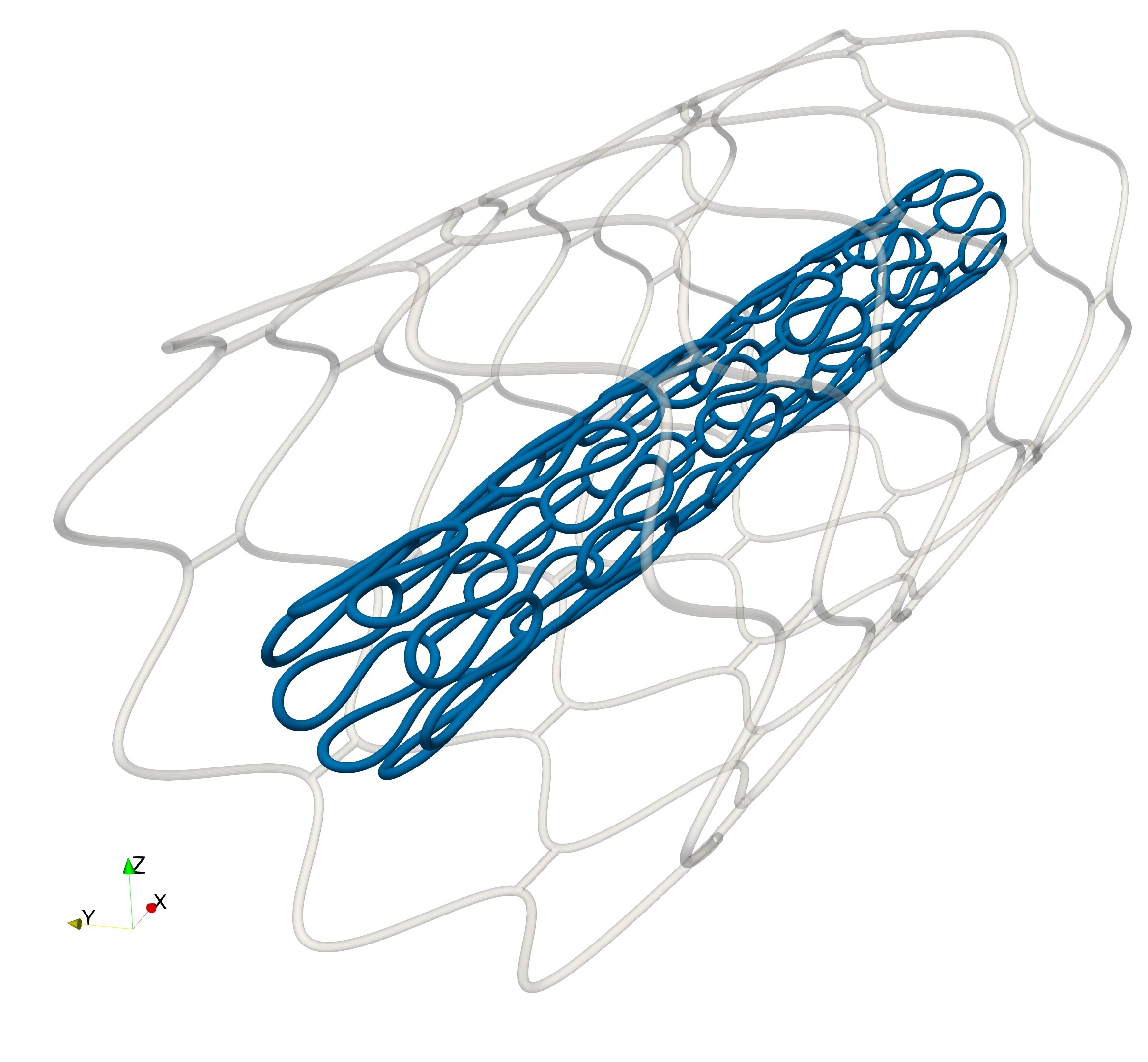}}
\subfigure[$t=\SI{1.7525}{s}$, $\nm{T}=\SI{45}{^\circ \nm{C}}$.\label{fig:17525_Stent_ext}]{\includegraphics[width=0.3\textwidth]{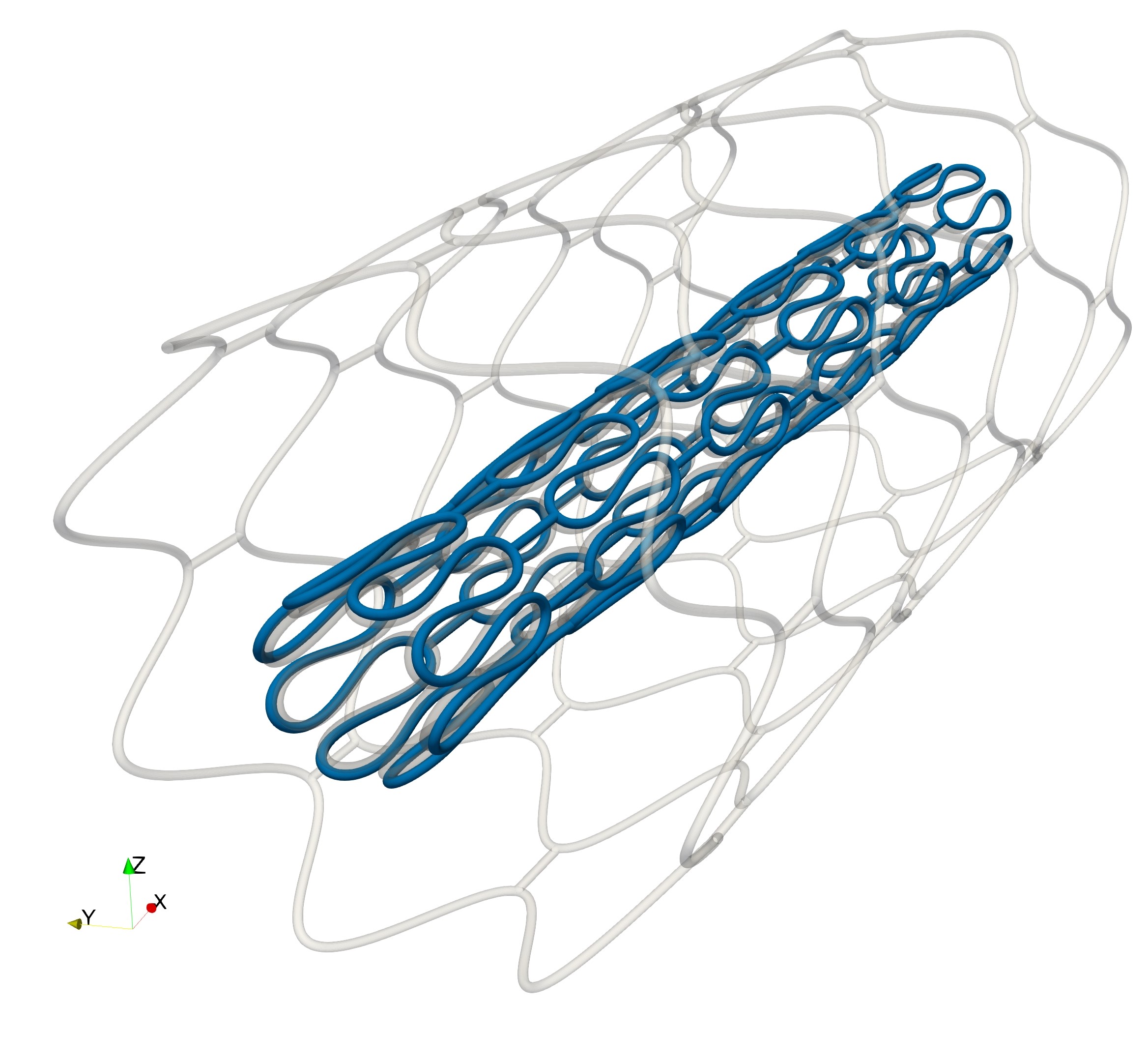}}
\hspace{0.03\textwidth}
\subfigure[$t=\SI{2.6}{s}$, $\nm{T}=\SI{71.96}{^\circ \nm{C}}$.\label{fig:260_Stent_ext}]{\includegraphics[width=0.3\textwidth]{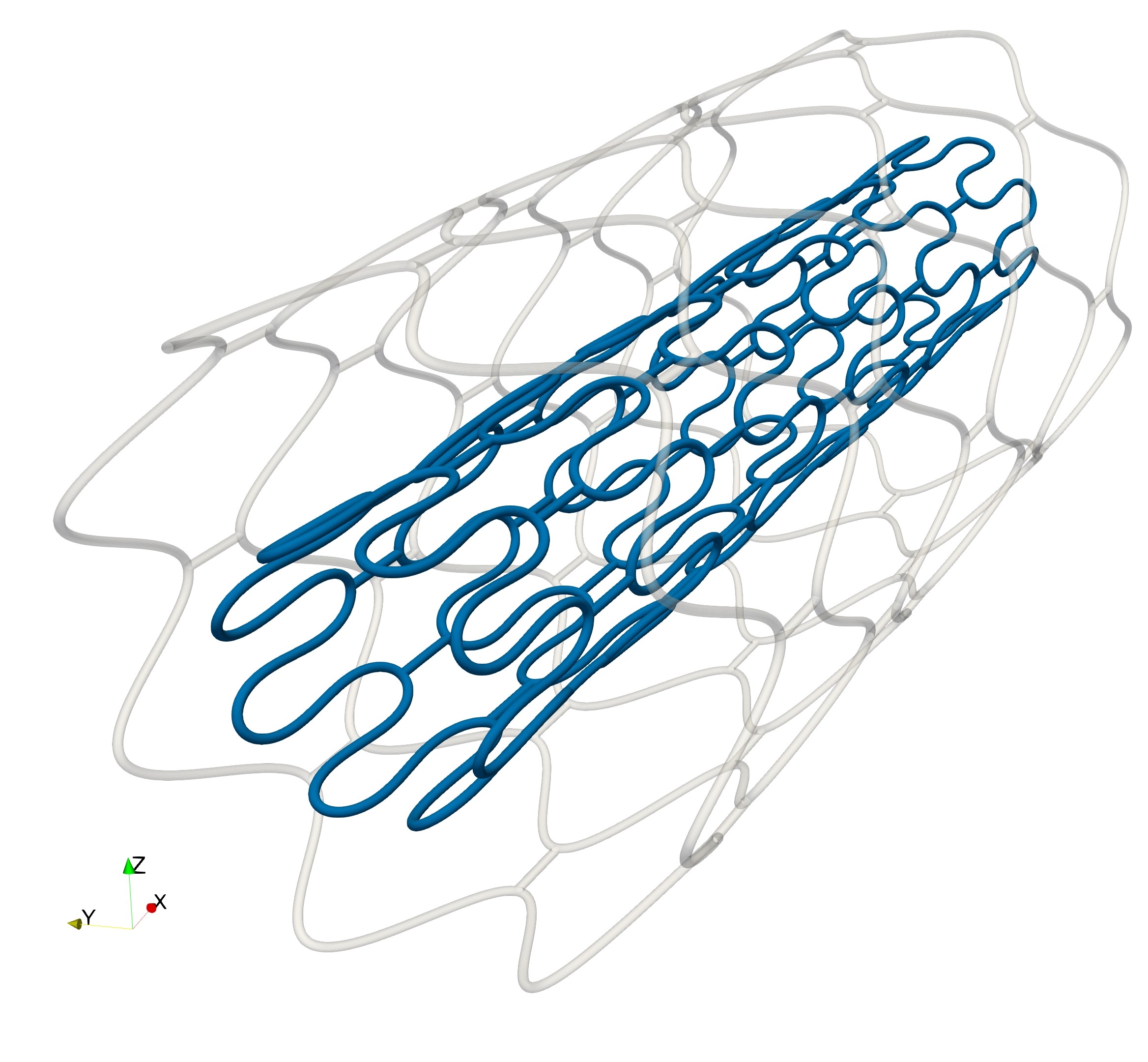}}
\hspace{0.03\textwidth}
\subfigure[$t=\SI{3.25}{s}$, $\nm{T}=\SI{90}{^\circ \nm{C}}$.\label{fig:325_Stent_ext}]{\includegraphics[width=0.3\textwidth]{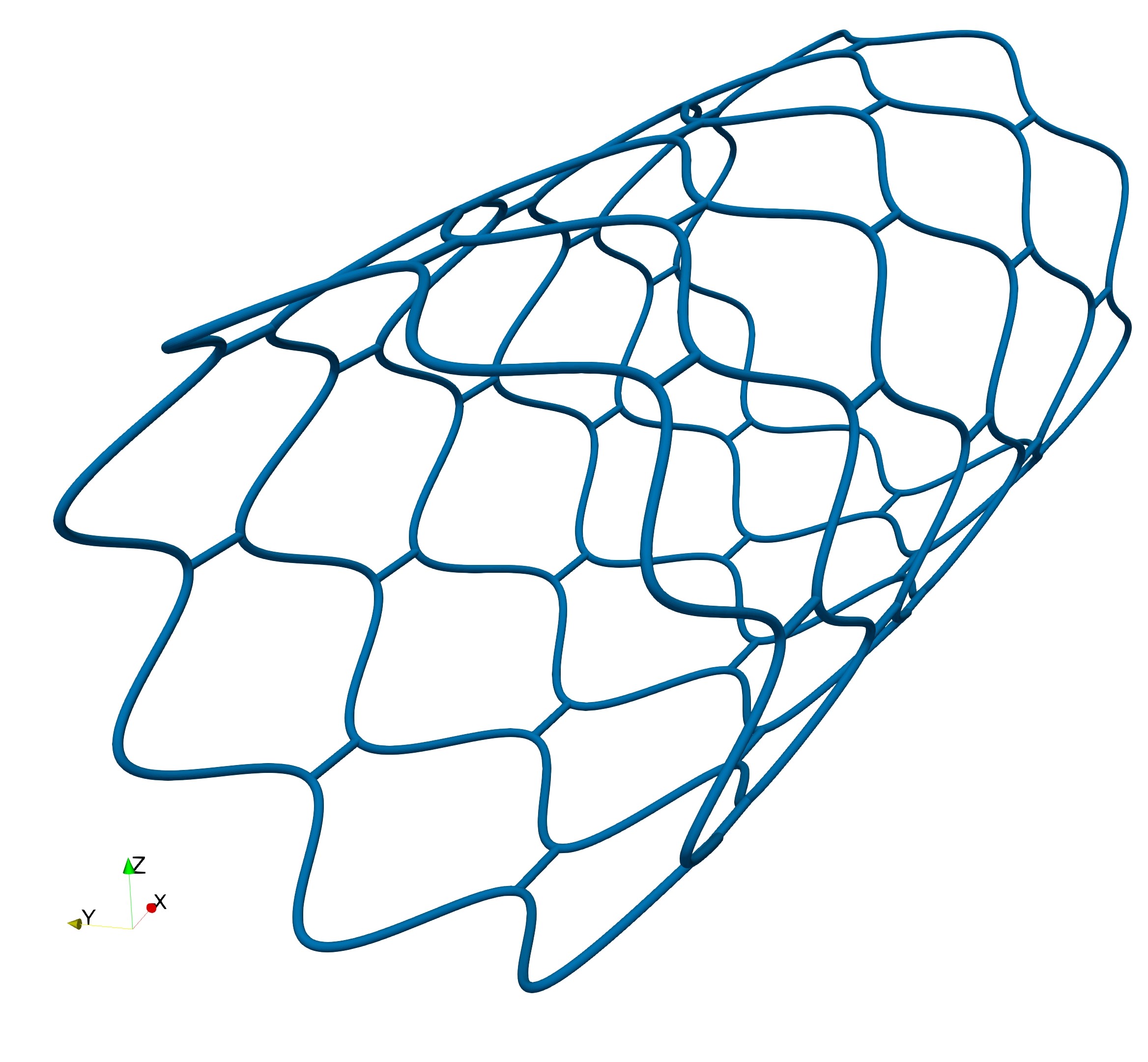}}
\caption{Straight device: three-dimensional views of the deformed configurations.}\label{fig:straight_stent_ext}
\end{figure}

In Figure~\ref{fig:straight_stent_def}, we show some relevant deformed configurations, at $t=\SI{0.5}{s}$, $\SI{1.75}{s}$, $\SI{1.7525}{s}$, $\SI{2.6}{s}$ and $\SI{3.25}{s}$. 
Light gray solid lines refer to the initial configuration, whereas black solid lines to the current one. The shape-fixing process is illustrated in Figures~\ref{fig:05_Stent_config}, $t=~\SI{0.5}{s}$,  and \ref{fig:175_Stent_config}, $t=~\SI{1.7525}{s}$. Here it is clearly noticeable the above-mentioned elastic snap due to the load removal, occurring from $t=~\SI{1.75}{s}$ to $t=~\SI{1.7525}{s}$ (see Figure~\ref{fig:175_Stent_config}). 
Finally, shape-recovery is shown in Figures~\ref{fig:27_Stent_config}, $t=~\SI{2.6}{s}$,  and \ref{fig:325_Stent_config}, $t=~\SI{3.25}{s}$, where, under the thermal action only, the structure returns to its initial shape.  
We also plot extruded three-dimensional views of the deformed full structure in Figure~\ref{fig:straight_stent_ext}.  

\subsubsection{Complex morphing of a device a with curved target geometry}\label{curved_stent}
The last numerical application concerns a curved stent-like device. The capability to simulate the morphing of complex assemblies of curved elements arranged to build a tubular system with a curved centreline is a crucial step towards the design of patient-tailored devices \cite{MarinoIGA4Stent_webpage}.

To reconstruct the device target shape (ideally provided by specific medical images), we first parametrize the stent axis, $\f c_{stent}(s)$, and the associated stent rotation matrix, $\fR R_ {stent}(s)$. This task crucially relies on the robust geometry representation strategy proposed in \cite{Ignesti_etal2023}, where the curved beam axis in the present case is the centreline of the entire stent-like tubular structure.   
Then, we place the crowns $\f P_{c}$, individually built as described in the previous section, along the (curved) stent axis to obtain their spatial placement as $\f P_{stent}(s)=\f c_{stent}(s)+\fR R_ {stent}(s)\f P_{c}$. 
This strategy for the geometric reconstruction allows to accurately design devices tailored to patient's vessel with any complexity level in terms of tortuosity since the operator $\fR R_ {stent}(s)$ and associated quantities, such as curvatures, are always well defined \cite{Ignesti_etal2023}.

\begin{figure}
\centering
\subfigure[3D view of the initial configuration.\label{fig:bend_3D}]
{\includegraphics[width=0.45\textwidth]{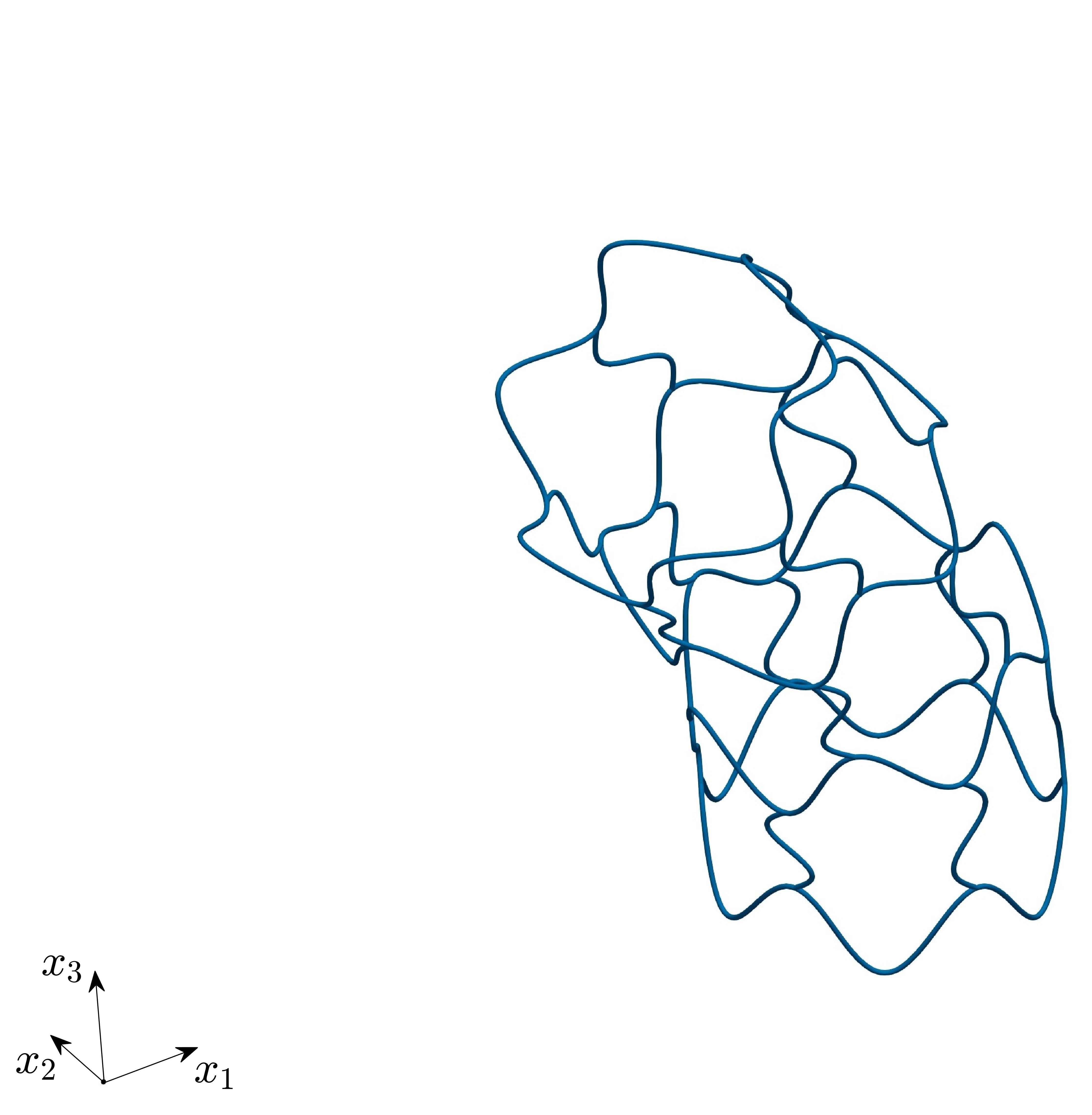}}
\hspace{0.04\textwidth}
\subfigure[3D view of the initial (shaded grey) and temporary, miniaturized (solid blue), configurations.\label{fig:bend_xy}]
{\includegraphics[width=0.45\textwidth]{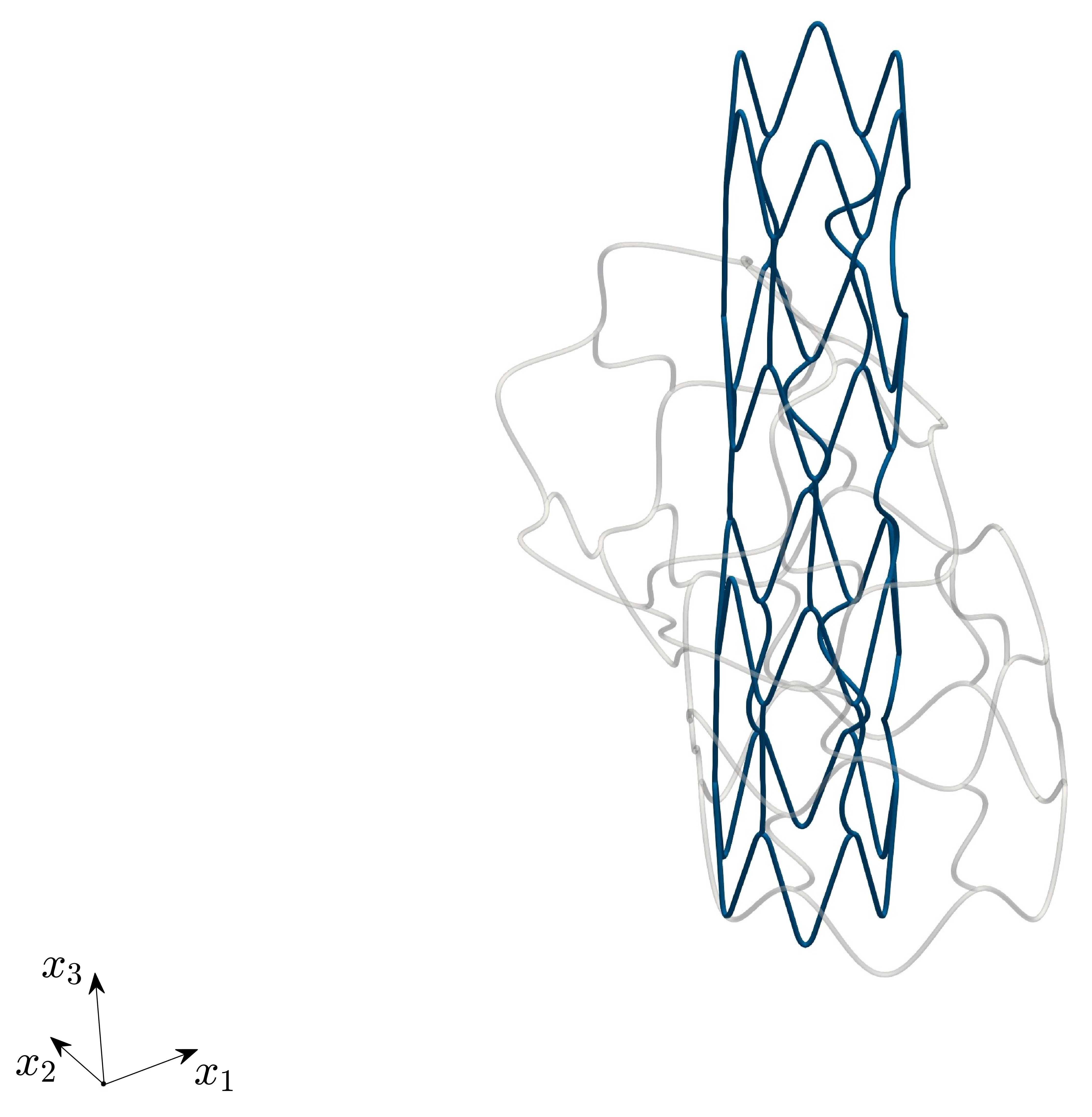}}
\subfigure[Time histories of temperature (red solid line) and displacement (blue solid line).\label{fig:load_bend_stent}]
{\includegraphics[width=1\textwidth]{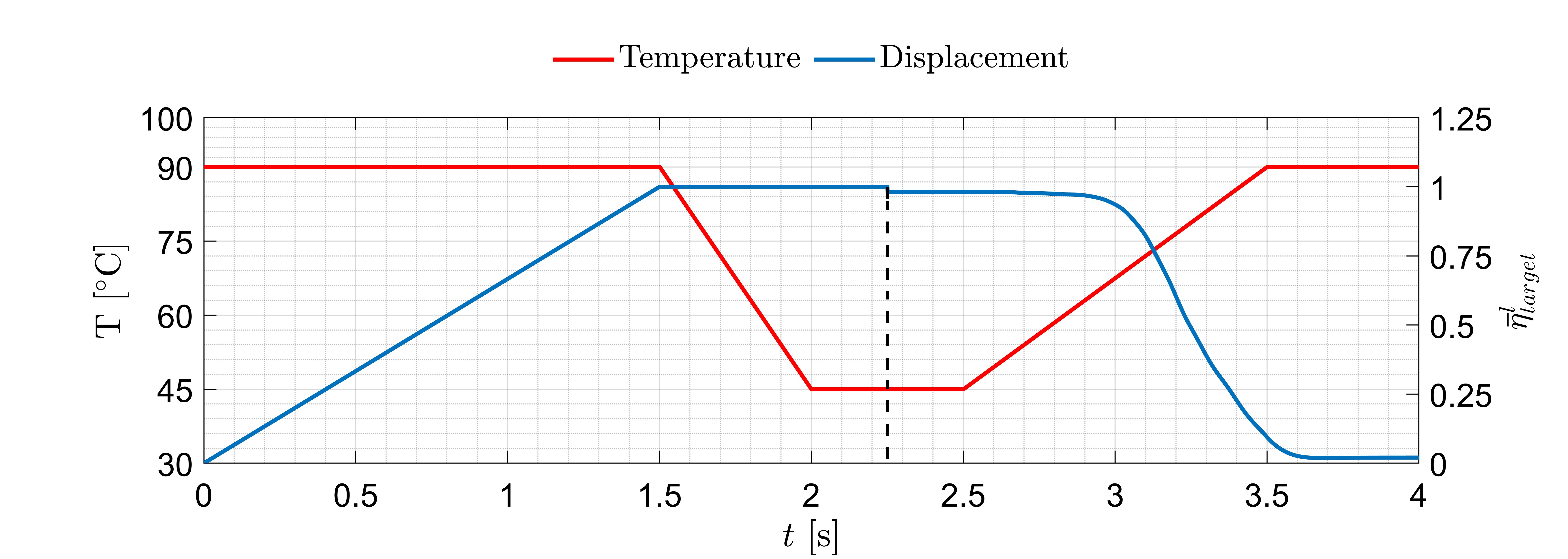}}
\caption{Complex morphing and shape recovery of a curved device: initial and temporary configurations, temperature and displacement time histories.}\label{fig:bend_stent_undef_target}
\end{figure}

We consider a four-crown structure characterized by a $45^\circ$ circular axis of radius $R_{stent}=\SI{100}{mm}$ and centre $\f O = [0\,0\,-5]\Tra\SI{}{mm}$. 
Since the shape-fixing (miniaturization) process involves both radial contraction and straightening of the device, we design sinusoidal bridges connecting subsequent crowns, to allow large deflections of the structure during the straightening of the stent axis. Figures~\ref{fig:bend_3D} and \ref{fig:bend_xy} show initial (permanent) and temporary (miniaturized) shapes (straightened and compacted) of the structure, respectively.

\begin{figure}
\centering
\subfigure[$t=\SI{0.0}{s}$, $\nm{T}=\SI{90}{^\circ \nm{C}}$.\label{fig:bend00_2d}]
{\includegraphics[width=0.24\textwidth]{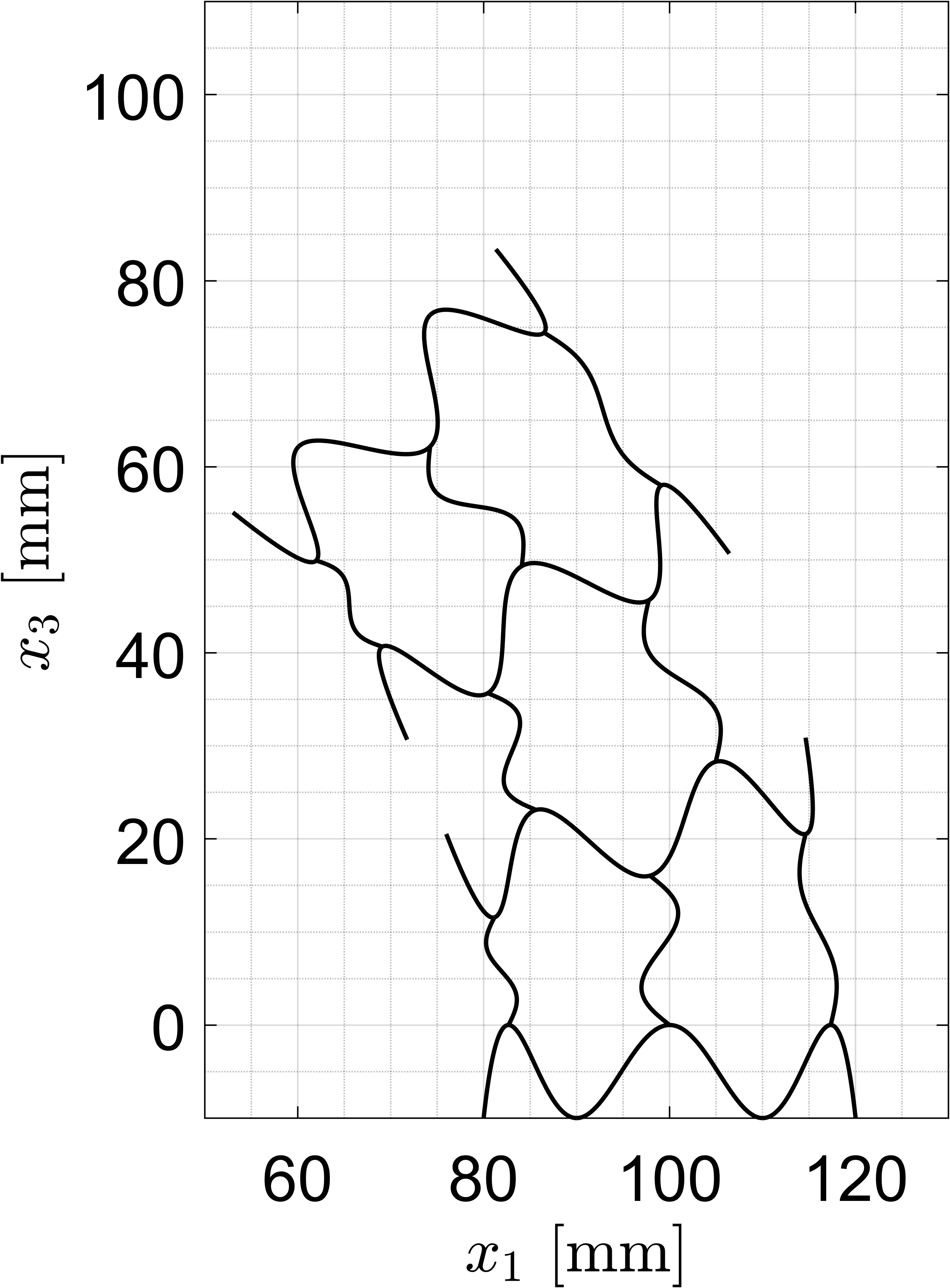}}
\subfigure[$t=\SI{0.5}{s}$, $\nm{T}=\SI{90}{^\circ \nm{C}}$.\label{fig:bend05_2d}]
{\includegraphics[width=0.24\textwidth]{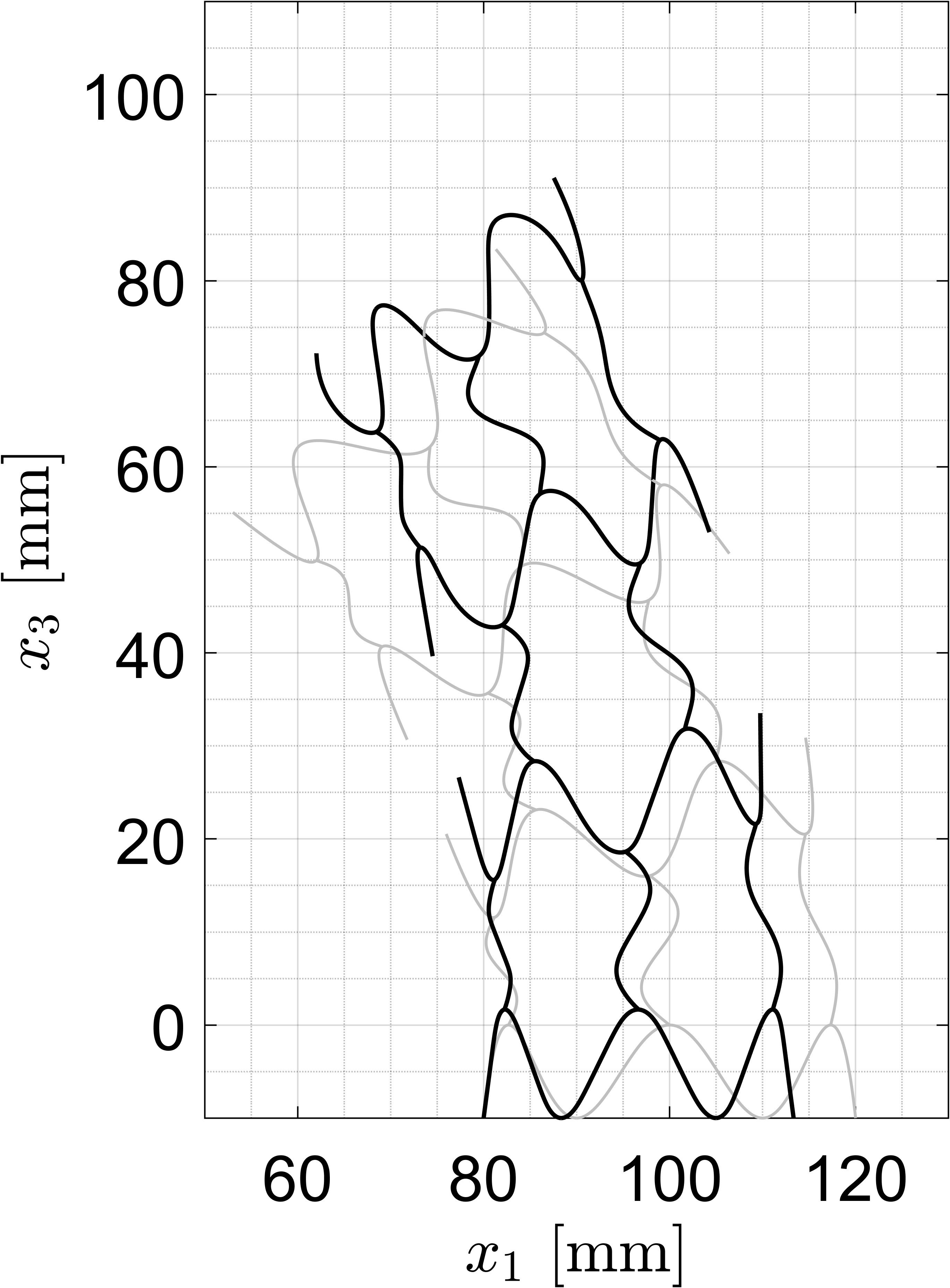}}
\subfigure[$t=\SI{1.5}{s}$, $\nm{T}=\SI{90}{^\circ \nm{C}}$.\label{fig:bend15_2d}]
{\includegraphics[width=0.24\textwidth]{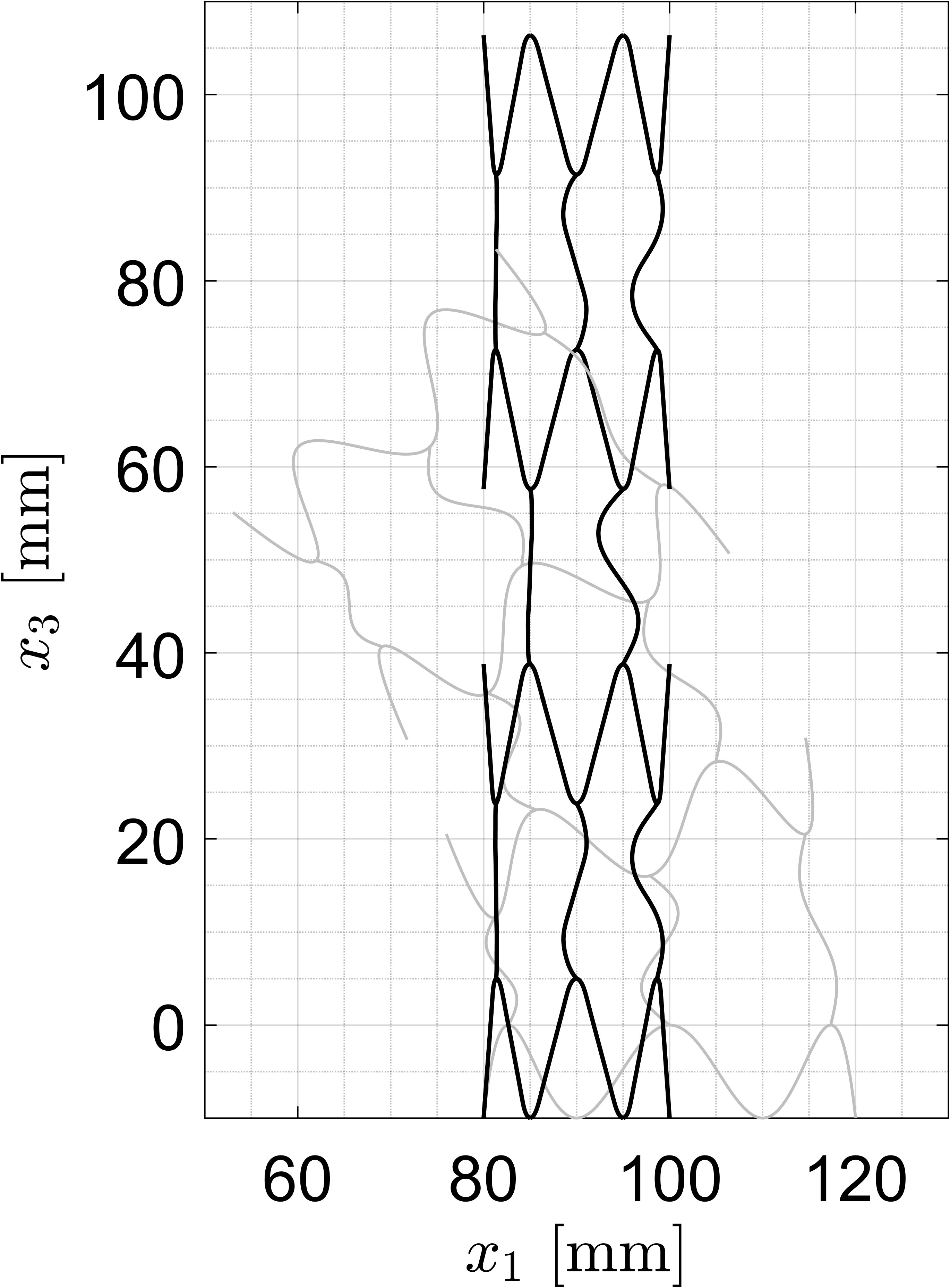}}
\subfigure[$t=\SI{2.25}{s}$, $\nm{T}=\SI{45}{^\circ \nm{C}}$.\label{fig:bend225_2d}]
{\includegraphics[width=0.24\textwidth]{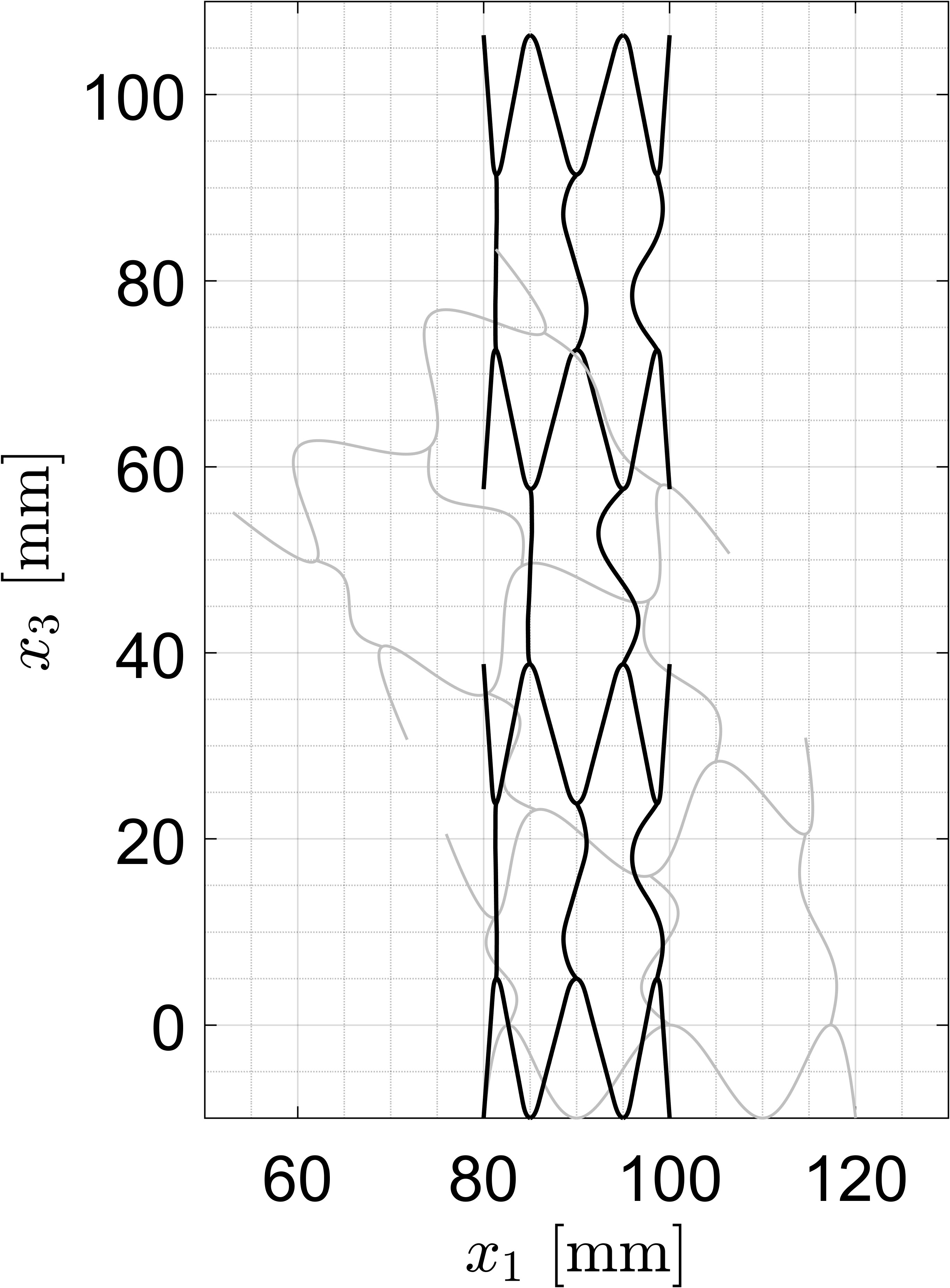}}
\subfigure[$t=\SI{3}{s}$, $\nm{T}=\SI{67.5}{^\circ \nm{C}}$.\label{fig:bend2251_2d}]
{\includegraphics[width=0.24\textwidth]{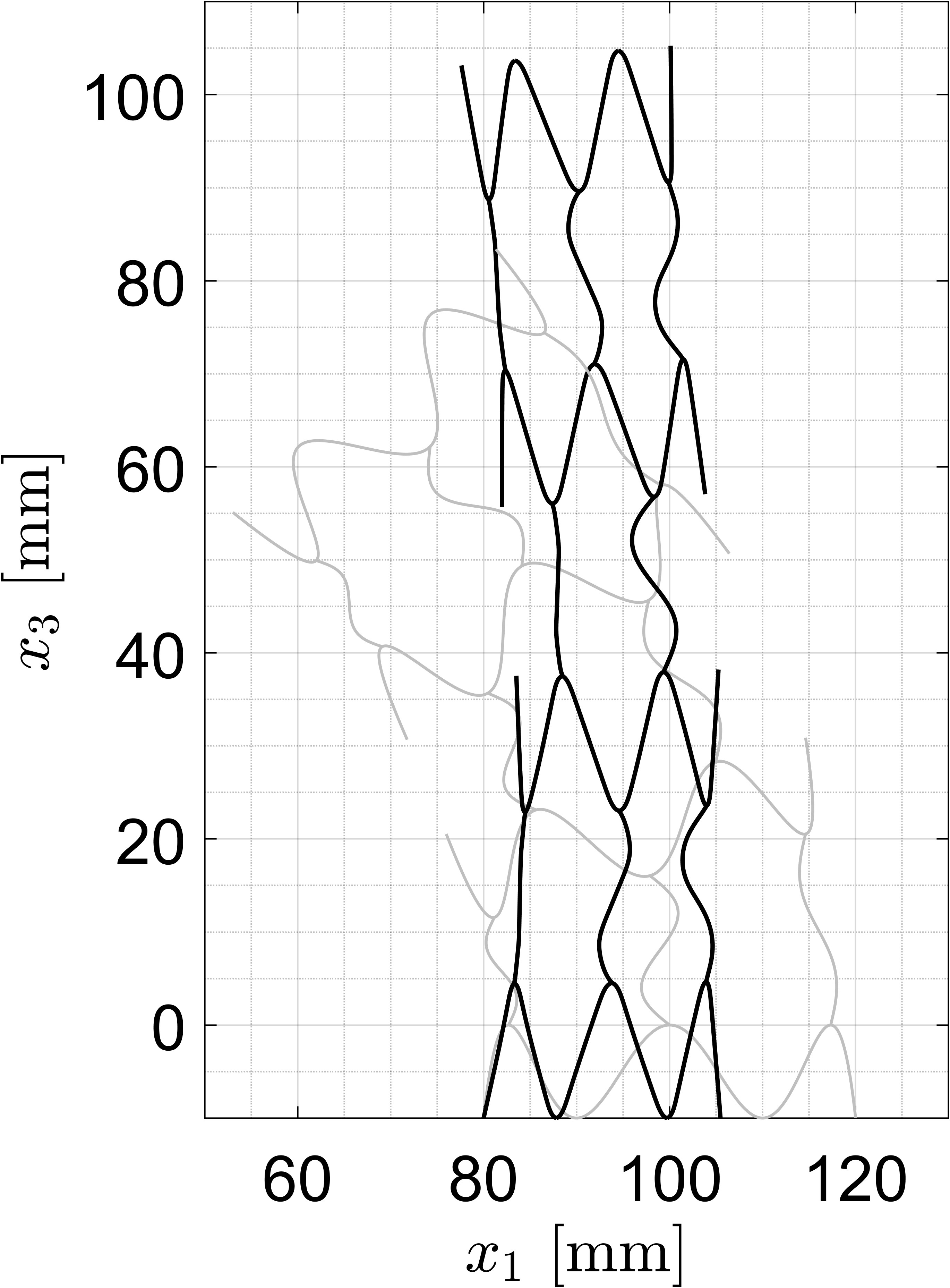}}
\subfigure[$t=\SI{3.1}{s}$, $\nm{T}=\SI{72}{^\circ \nm{C}}$.\label{fig:bend31_2d}]
{\includegraphics[width=0.24\textwidth]{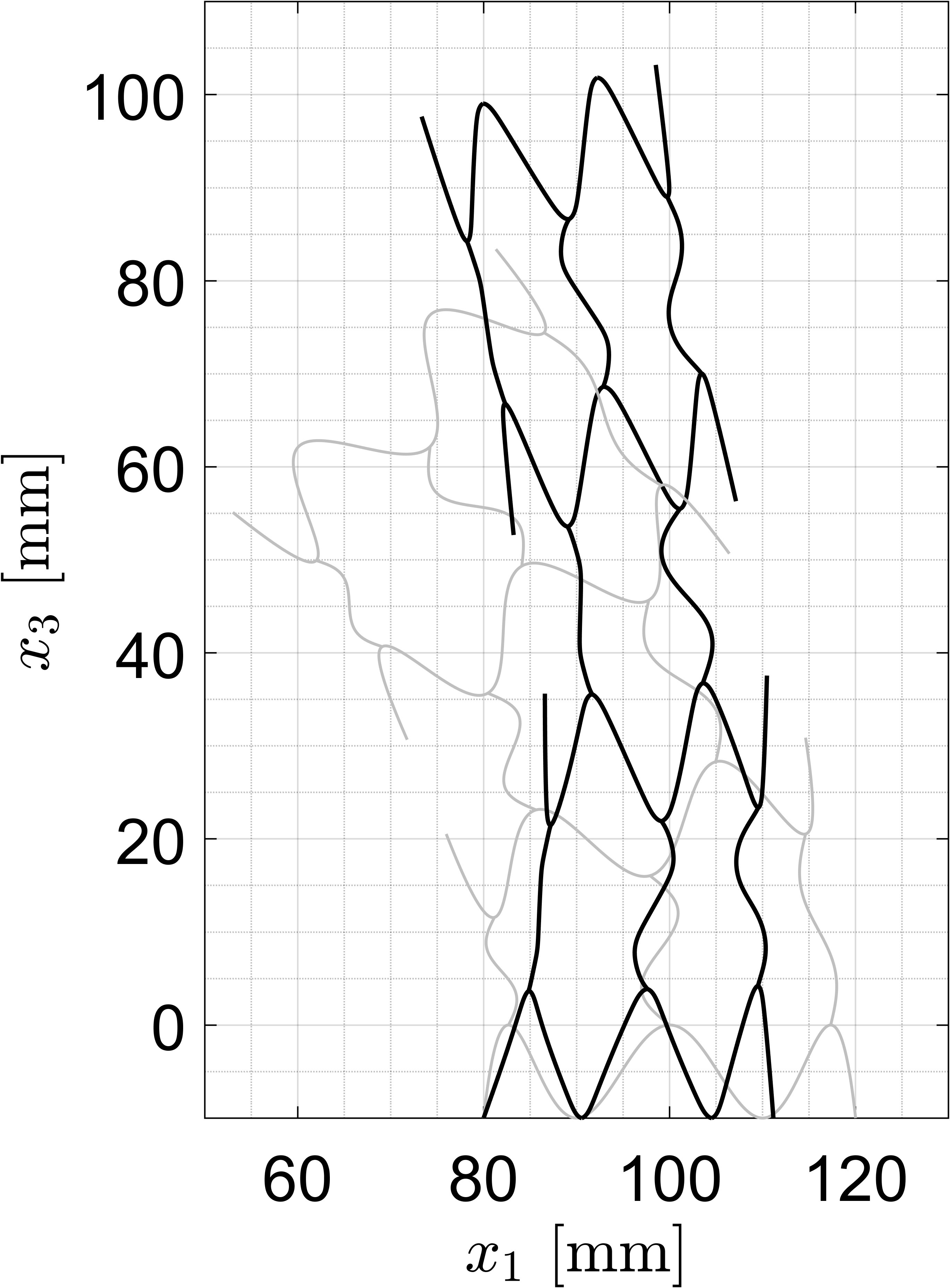}}
\subfigure[$t=\SI{3.4}{s}$, $\nm{T}=\SI{85.5}{^\circ \nm{C}}$.\label{fig:bend34_2d}]
{\includegraphics[width=0.24\textwidth]{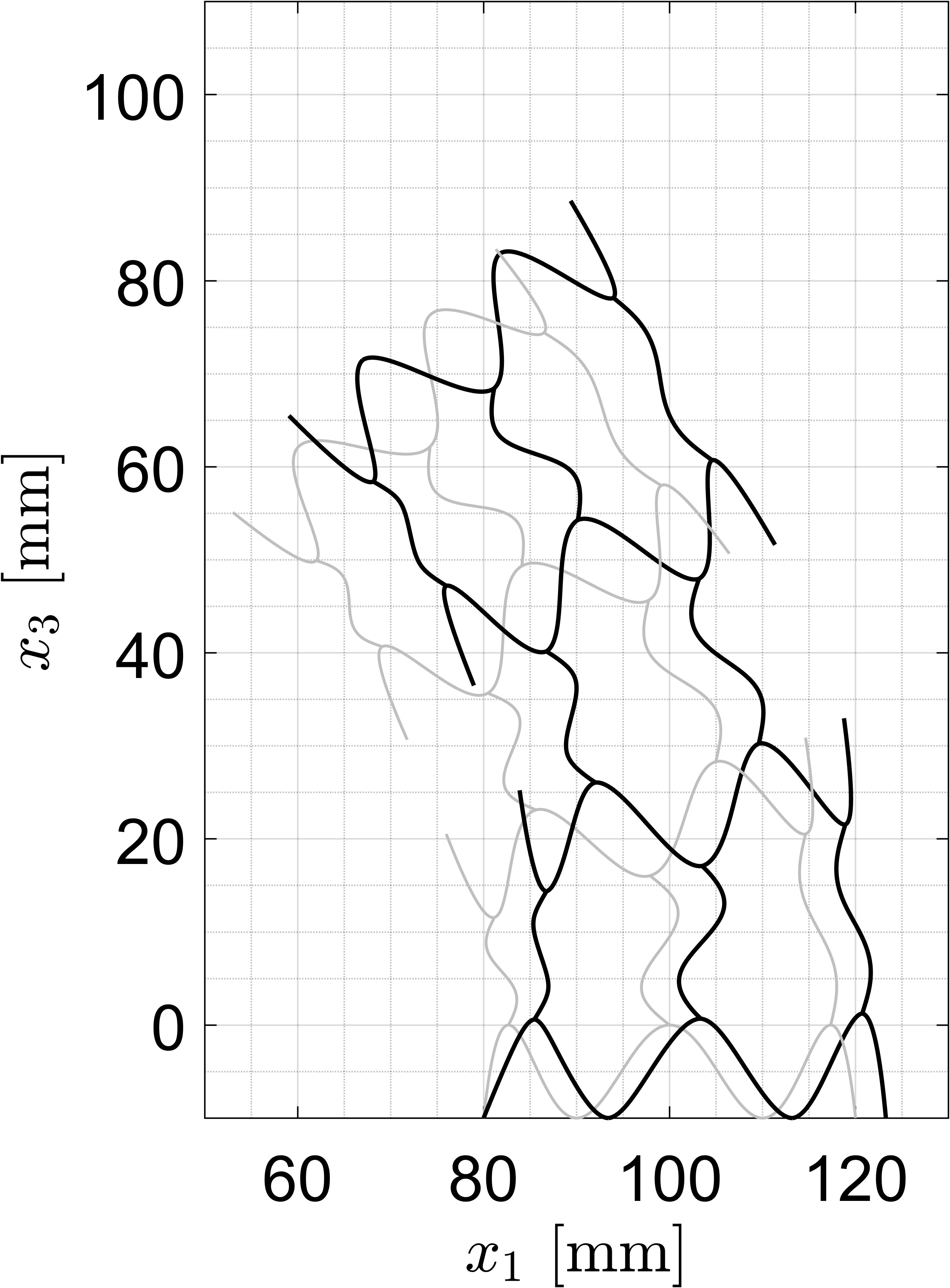}}
\subfigure[$t=\SI{4}{s}$, $\nm{T}=\SI{90}{^\circ \nm{C}}$.\label{fig:bend4_2d}]
{\includegraphics[width=0.24\textwidth]{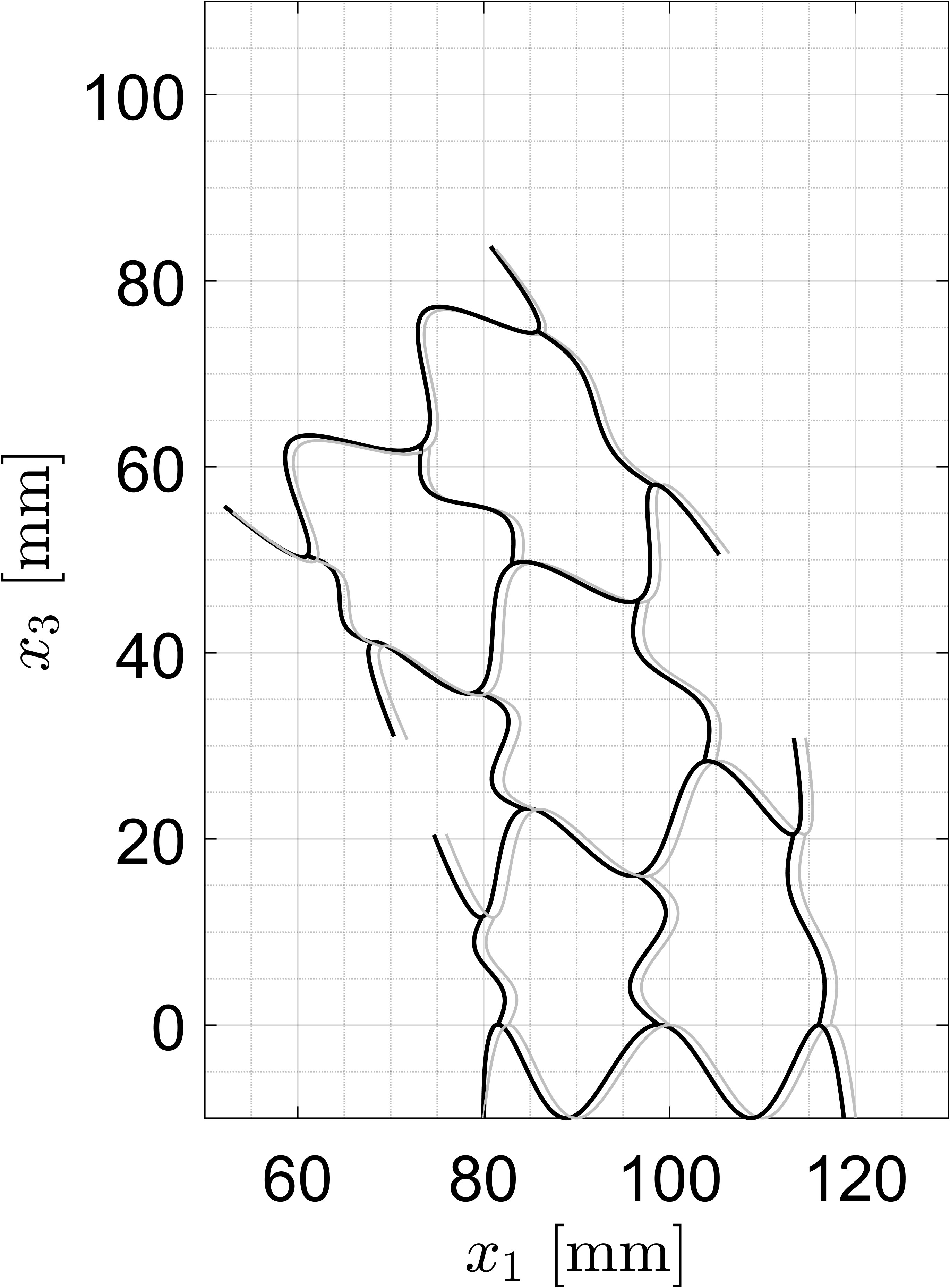}}
\caption{Complex morphing and shape recovery of a curved device: plane $(x_1-x_3)$ view of significant deformation snapshots.\label{fig:bend_stent_plane_config}}
\end{figure}

\begin{figure}
\centering
\subfigure[$t=\SI{0.0}{s}$, $||\f \eta^{l}||/||\baf \eta_{temp}^{l}||=0$, $\nm{T}=\SI{90}{^\circ \nm{C}}$. \label{fig:00_bend}]{\includegraphics[width=0.29\textwidth]{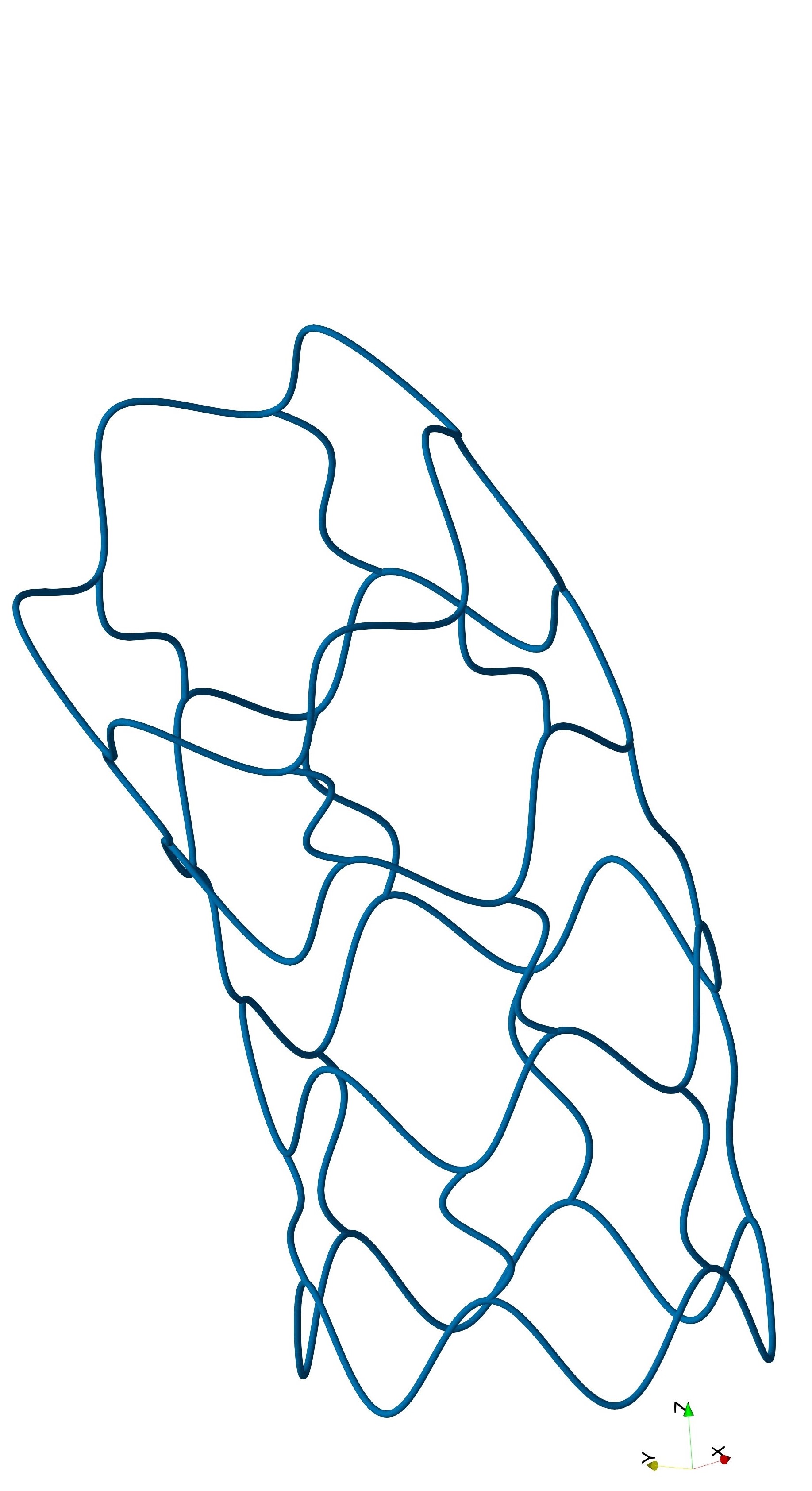}}
\hspace{0.02\textwidth}
\subfigure[$t=\SI{0.5}{s}$, $||\f \eta^{l}||/||\baf \eta_{temp}^{l}||=1/3$, $\nm{T}=\SI{90}{^\circ \nm{C}}$.\label{fig:05_bend}]{\includegraphics[width=0.29\textwidth]{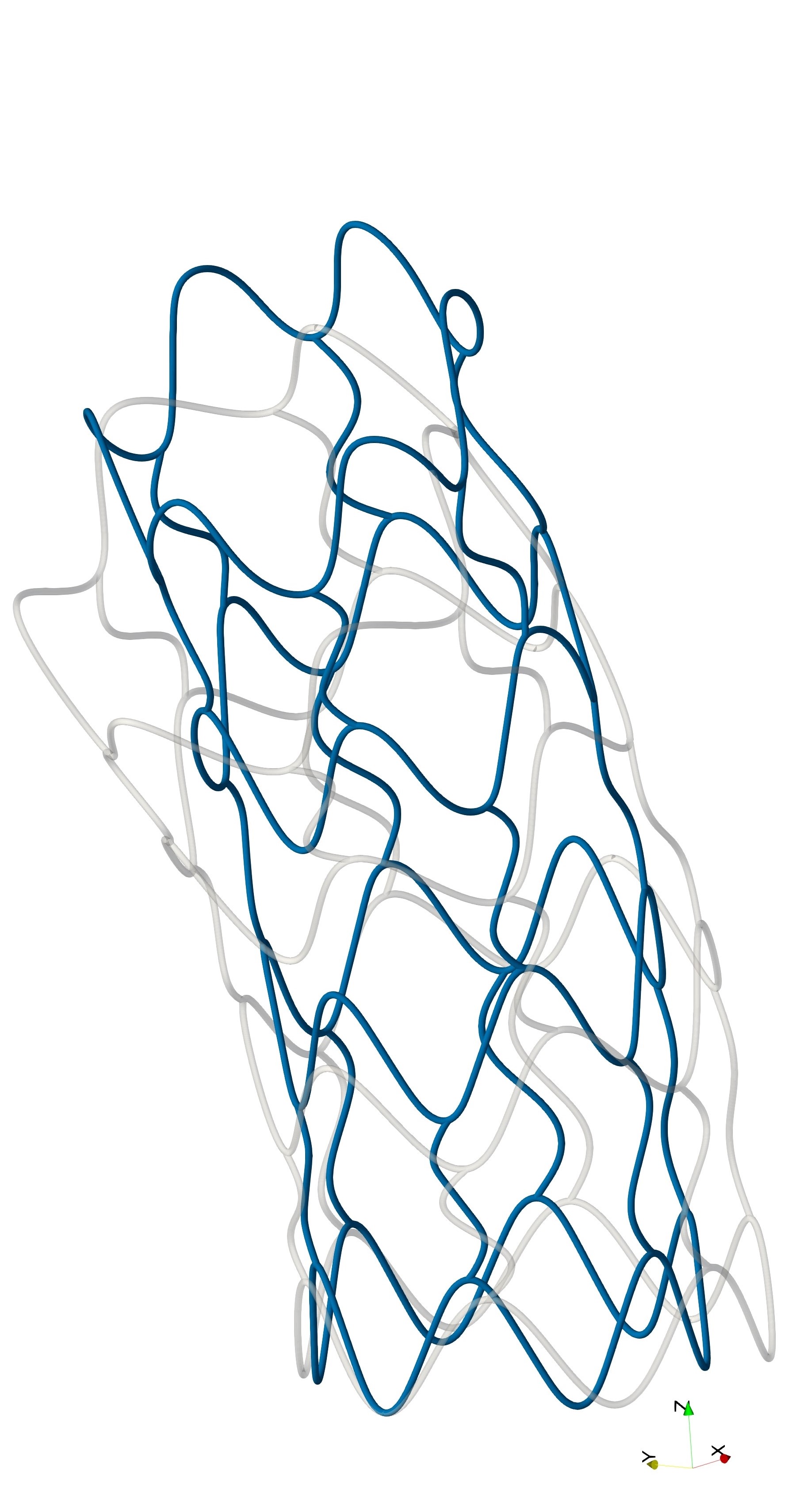}}
\hspace{0.02\textwidth}
\subfigure[$t=\SI{2.25}{s}$, $||\f \eta^{l}||/||\baf \eta_{temp}^{l}||=1$, $\nm{T}=\SI{45}{^\circ \nm{C}}$. \label{fig:20_bend}]{\includegraphics[width=0.29\textwidth]{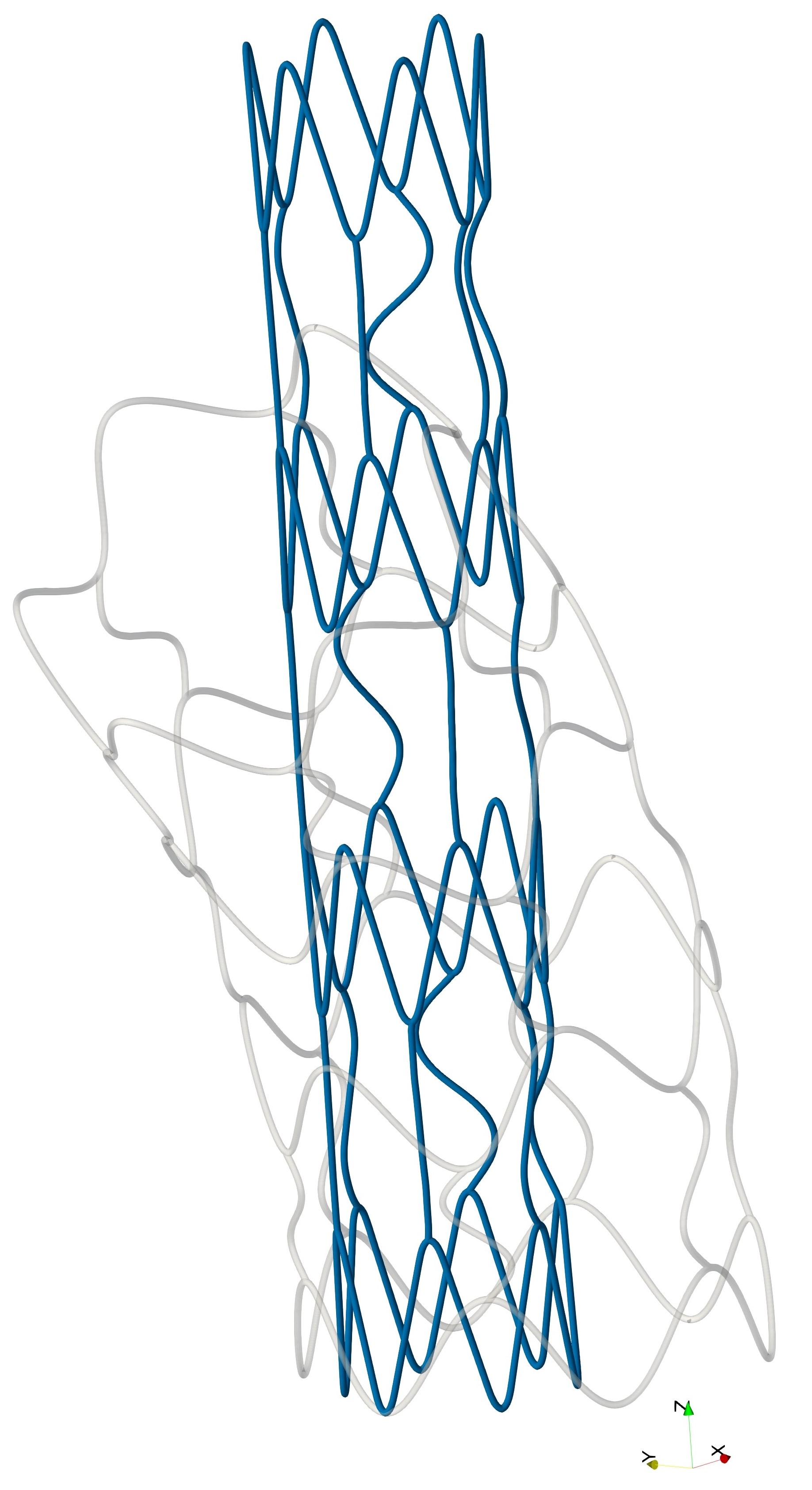}}
\subfigure[$t=\SI{3.1}{s}$, $||\f \eta^{l}||/||\baf \eta_{temp}^{l}||=0.82$, $\nm{T}=\SI{72}{^\circ \nm{C}}$.\label{fig:31_bend}]{\includegraphics[width=0.29\textwidth]{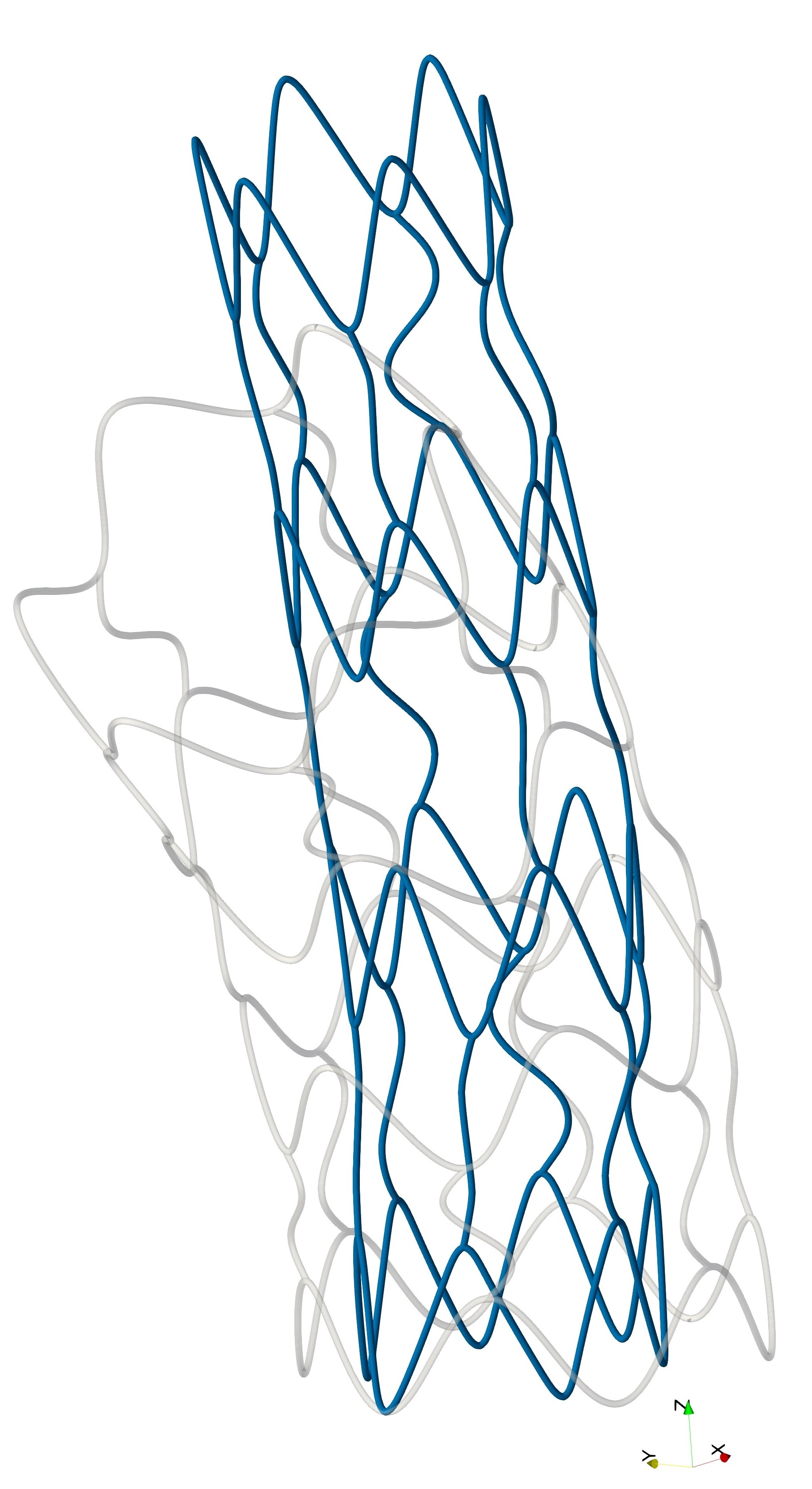}}
\hspace{0.02\textwidth}
\subfigure[$t=\SI{3.4}{s}$, $||\f \eta^{l}||/||\baf \eta_{temp}^{l}||=0.23$, $\nm{T}=\SI{85.5}{^\circ \nm{C}}$.\label{fig:34_bend}]{\includegraphics[width=0.29\textwidth]{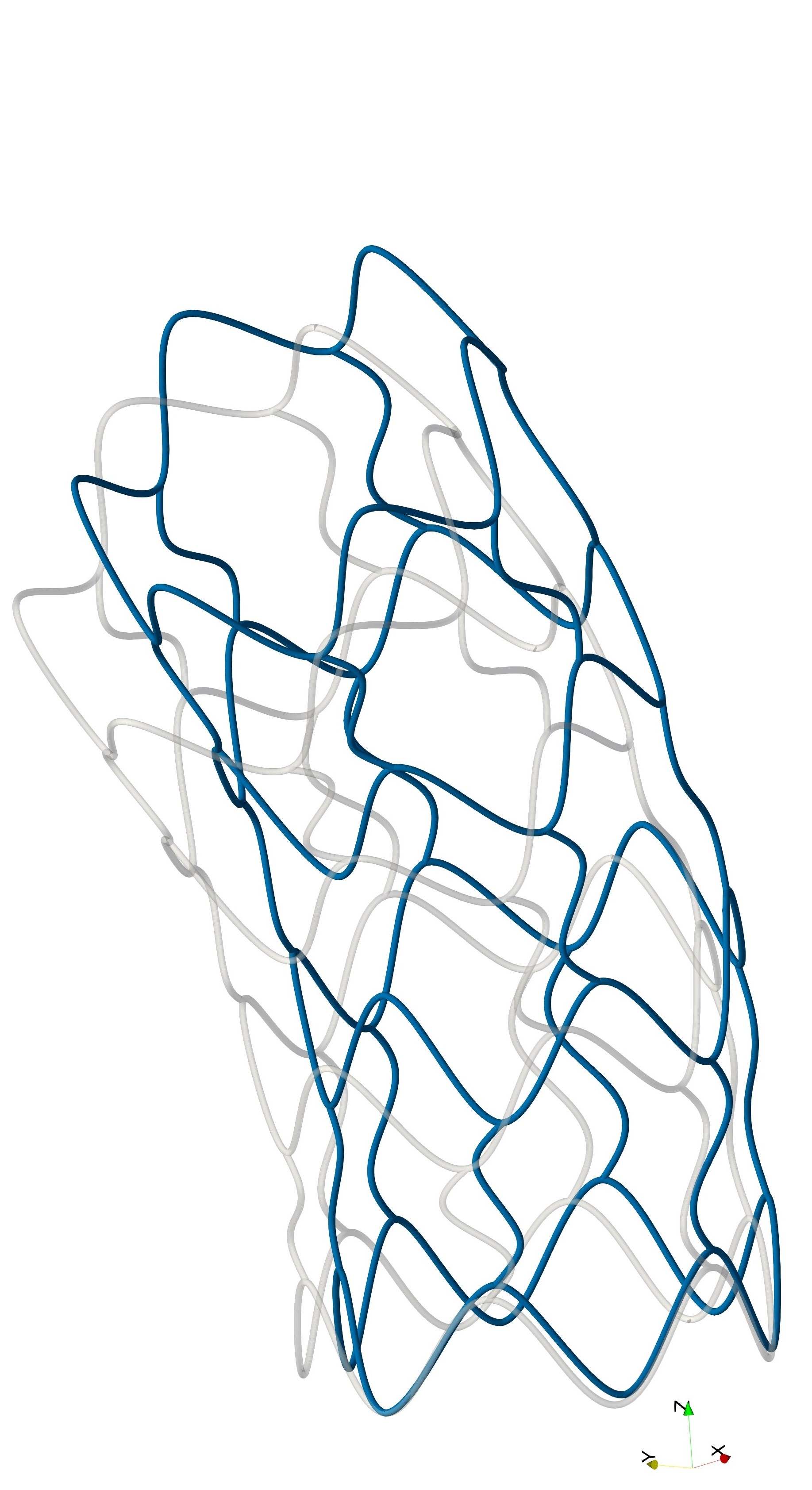}}
\hspace{0.02\textwidth}
\subfigure[$t=\SI{4}{s}$, $||\f \eta^{l}||/||\baf \eta_{temp}^{l}||=0.02$, $\nm{T}=\SI{90}{^\circ \nm{C}}$.\label{fig:40_bend}]{\includegraphics[width=0.29\textwidth]{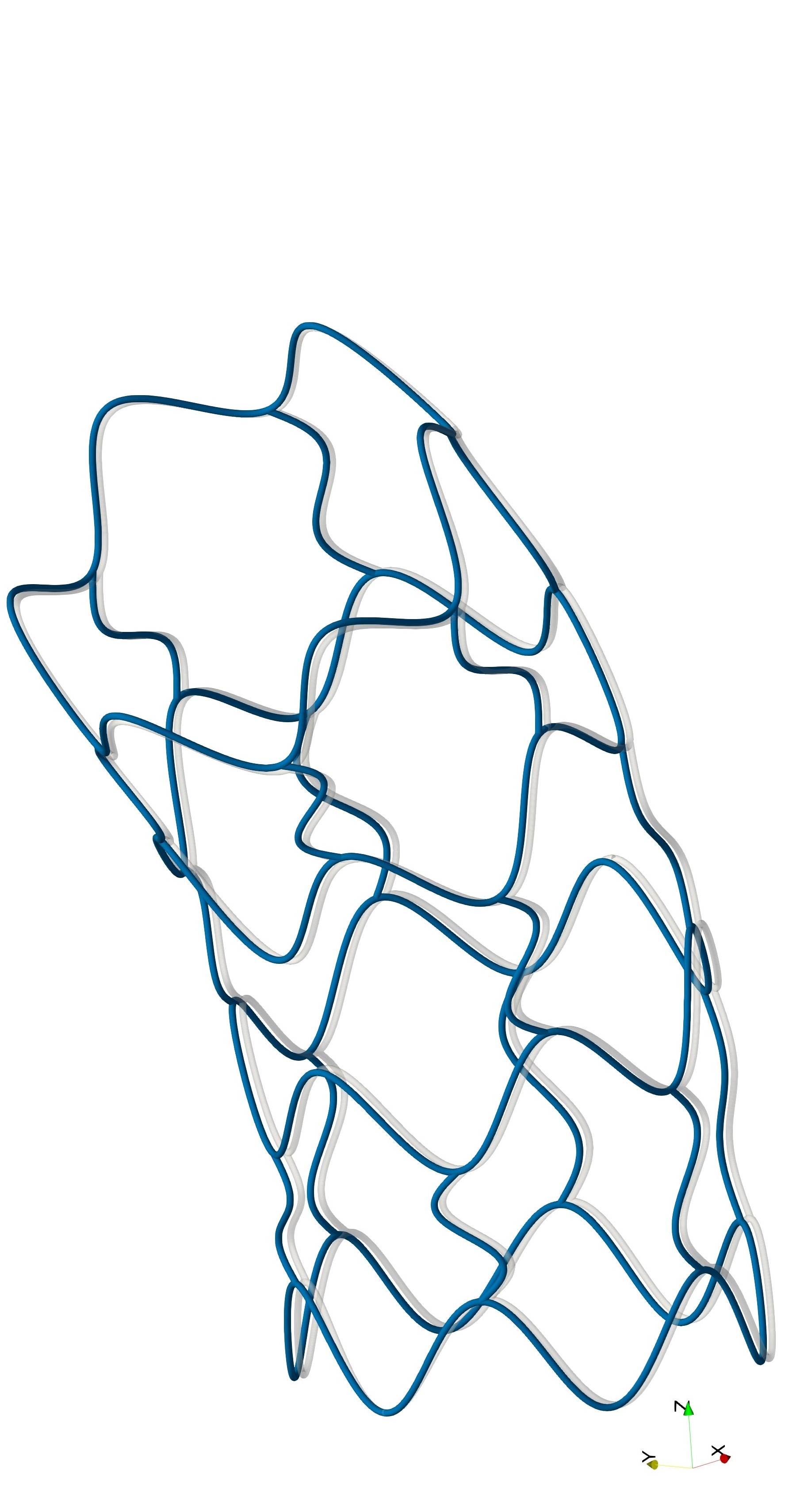}}
\caption{Complex morphing and shape recovery of a curved device: three-dimensional views of the deformed configurations ($\f \eta^{l}$ and $\baf \eta_ {temp}^{l}$ denote the current and target displacement of the $l$th node introduced in Eq.~\eqref{eq:target_disp}), respectively.}\label{fig:bend_stent_ext}
\end{figure}

We enforce Dirichlet BCs to the interface nodes of each beam. The total  displacement of the $l$-th node, which is imposed to reach the ``closed'' configuration is computed as
\bEq\label{eq:target_disp}
\baf \eta_ {temp}^{l}= \f P_{temp}^{l}-\f P_0^{l}\,,
\eEq
where $\f P_{temp}^l$ and $\f P_0^l$ denote the positions of the $l$th interface node in the temporary and the initial (permanent) configurations, respectively. 
The displacements $\baf \eta_ {temp}^{l}$ are enforced with a linear ramp from $t=\SI{0}{s}$ to $t=\SI{1.5}{s}$, and kept constant until $t=\SI{2.25}{s}$. 
Then, BCs are switched from Dirichlet to free Neumann. 
Figure~\ref{fig:load_bend_stent} shows the imposed displacement time history, as well as the evolution of the temperature during the entire process. 
The simulation time is set to $\SI{4}{s}$, with a time step size $h=\SI{0.001}{s}$. 
Each patch of the structure is discretized with $p=6$ basis functions and $\nm n=81$ collocation points. More details on the geometric modelling and patch subdivision are reported in the Appendix. 
Plane views and extruded three-dimensional views of the deformed configurations are plotted in Figure~\ref{fig:bend_stent_plane_config} and Figure~\ref{fig:bend_stent_ext}, respectively. 
During the shape-fixing process, occurring at $t=\SI{2.251}{s}$, a slight elastic snap is still noticeable (see the small step in the blue line at $t=\SI{2.25}{s}$ in Figure~\ref{fig:load_bend_stent}).
Then, as the temperature increases, the shape-recovery begins until the initial shape is almost fully recovered at $t=\SI{4}{s}$. 

\section{Conclusions}
In this paper, an efficient temperature-dependent viscoelastic model is proposed for the simulation of beams and beam structures made of SMPs. The formulation extends to the geometrically exact beam problem the temperature-dependent generalized Maxwell model for viscoelastic materials. 
Temperature is considered as an external action and the relaxation times, directly influencing the stiffness of the material, are expressed as a function of temperature through the shift factor exploiting the Time-Temperature Superimposition Principle (TTSP). With this material model, we demonstrate the possibility to reproduce the Shape Memory Effect of complex assemblies of curved beams made of polymeric materials. 
High efficiency is achieved discretizing the differential problem in space through the isogeometric collocation (IGA-C) method. IGA-C, apart from the well-known attributes of isogeometric analysis (IGA) in terms of exactness of the geometric reconstruction of complex shapes and high-order accuracy, does not require numerical integration over the elements since the method is based on the strong form of the problem. In addition to the discretization technique, other distinguishing features of our formulation are the $\SO3$-consistent linearization of the problem and its and time stepping, the use of minimal parameters for the parametrization of (finite) rotations, and the ability to not increase the number of unknowns compared to the purely elastic case to account for the material rate-dependency.
Through a series of numerical tests, we demonstrate that the proposed formulation preserves higher-order spatial convergence, is able to straightforwardly model complex assemblies of curved elements, and can simulate the full shape memory effect: from programming to shape recovery. 
Future works will be dedicated to validate the proposed method with experiments as well as to solve the thermo-mechanical coupled problem. Furthermore, new materials will be investigated in terms of biocompatibility and stiffness towards the use of programmable 4D-printed devices for biomedical applications.

\section*{Acknowledgements}
EM was partially supported  by the European Union - Next Generation EU, in the context of The National Recovery and Resilience Plan, Investment 1.5 Ecosystems of Innovation, Project Tuscany Health Ecosystem (THE). (CUP: B83C22003920001).

EM was also partially supported by the National Centre for HPC, Big Data and Quantum Computing funded by the European Union within the Next Generation EU recovery plan. (CUP B83C22002830001). 

EM and GF were partially supported  by the UniFI project IGA4Stent - ``Patient-tailored stents: an innovative computational isogeoemtric analysis approach for 4D printed shape changing devices''. (CUP B55F21007810001).

\appendix 
\section*{Appendix}
As described in section~\ref{curved_stent}, the curved device is build deploying the reference crown along the axis of the structure. For this test, we have subdivided the reference crown in $n_w=12$ wires. Each of them is built in the $(\vartheta,x_3)$ plane (see Figure~\ref{fig:straight_crown}) such that the $q$th wire centroid equation in  the $(x_1,x_2,x_3)$ is 
\bEq
\f c_w (s)=[R\cos{(\vartheta)}, R\sin{(\vartheta)}, h_c\sin{(\fr {n_w\vartheta}{2})}]\Tra, \sepr{} \vartheta\in[(q-1)\pi/6,\, q\pi/6],\,q=1,2,...,n_w   \,.
\eEq

\begin{figure}
\centering
\includegraphics[width=0.8\textwidth]{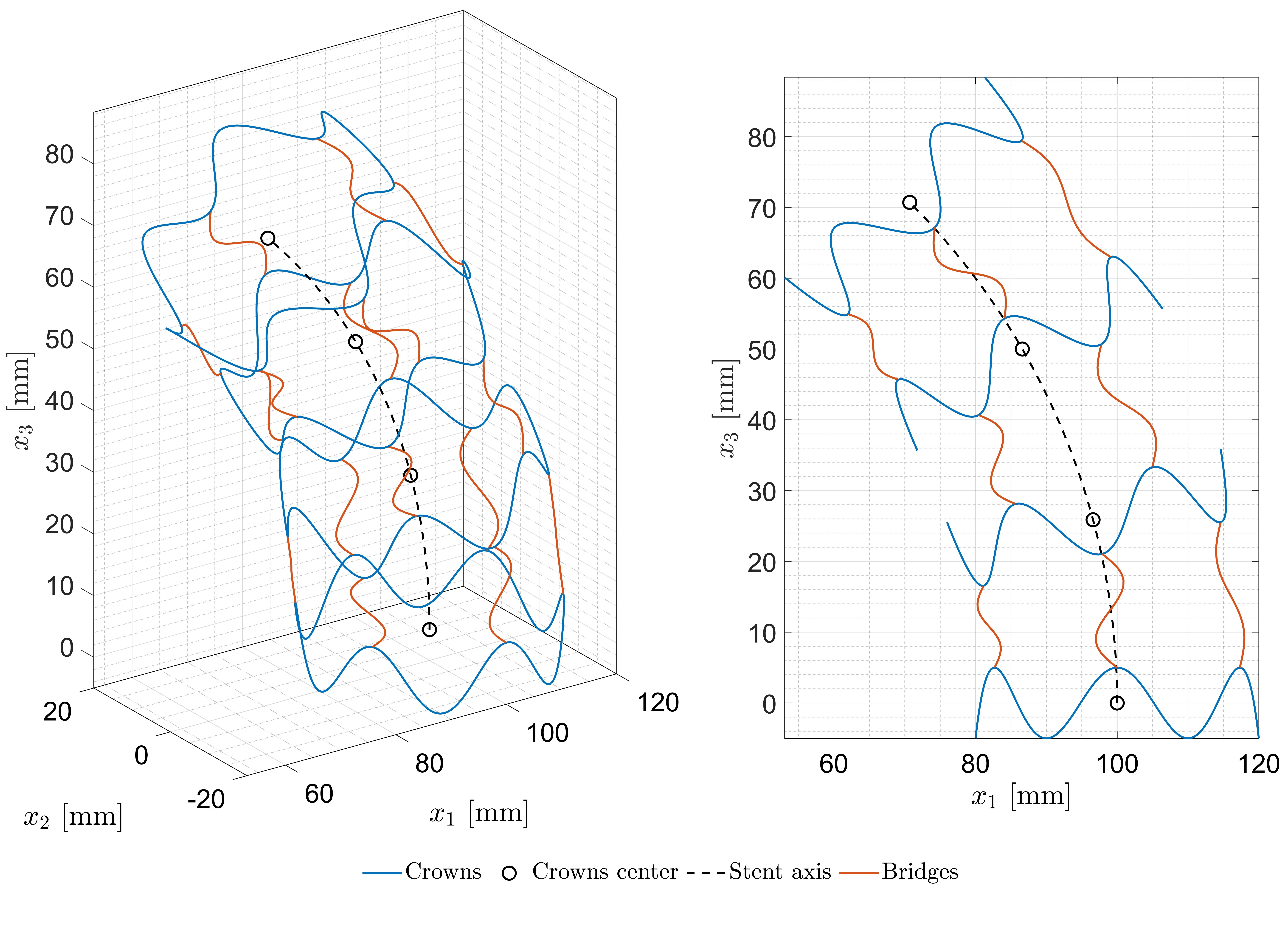}
\caption{Three-dimensional view (left) and planar ($x_1,x_3$) view of the initialised geometry of the curved device: blue solid lines refers to the crowns that are translated at the end points (black circles) of each stent axis ( black dashed lines) subdivision and aligned by means of $\fR R_{stent}$.\label{fig:Stent_appendix}}
\end{figure}

Given the curved centreline of the device as 
\bEq
\f c_{stent} (s)=[R_{stent}\cos{(\psi)},0, R_{stent}\sin{(\psi)}]\Tra, \sepr{} \psi\in[0,\, \pi/4]   \,, 
\eEq
we subdivide it in three equal circular sectors of amplitude $\pi/12$. Then, crowns are translated at the end points of each sector, and align along the stent axis making use of the rotation operator $\fR R_{stent}$ (see Figure~\ref{fig:Stent_appendix}). Crowns located at $\psi=\pi/12$ and $\psi=\pi/4$ are preliminary rotated of $\pi$ with respect to the $x_2$-axis.
This allows to align crests and trough of subsequent crowns and permits a better interconnection with bridges (see orange solid lines in Figure~\ref{fig:Stent_appendix}). These elements are radially oriented along the stent. They feature a sinusoidal shape of height $h_b=h_c/2$ and period equal to the distance between crests and troughs.  

Given the symmetry of the structure with respect to the plane $x_2=0$, only one half of the structure is studied, imposing the symmetry constraints at those interface nodes located at $x_2=0$. 

\clearpage
\bibliographystyle{elsarticle-num}
\bibliography{my_thermoviscobeam_bib_merge}

\end{document}